\documentstyle[a4wide]{article} 
\input epsf 
\begin{document}


\title{Critical phenomena in gravitational collapse}

\author{Carsten Gundlach \\ 
Faculty of Mathematical Studies, University of Southampton, \\
Southampton SO17 1BJ, UK}

\date{October 2002}

\maketitle


\begin{abstract}

In general relativity black holes can be formed from regular initial
data that do not contain a black hole already. The space of regular
initial data for general relativity therefore splits naturally into
two halves: data that form a black hole in the evolution and data that
do not. The spacetimes that are evolved from initial data near the
black hole threshold have many properties that are mathematically
analogous to a critical phase transition in statistical mechanics.

Solutions near the black hole threshold go through an intermediate
attractor, called the critical solution. The critical solution is
either time-independent (static) or scale-independent
(self-similar). In the latter case, the final black hole mass scales
as $(p-p_*)^\gamma$ along any one-parameter family of data with a
regular parameter $p$ such that $p=p_*$ is the black hole threshold in
that family. The critical solution and the critical exponent $\gamma$
are universal near the black hole threshold for a given type of
matter.

We show how the essence of these phenomena can be understood using
dynamical systems theory and dimensional analysis. We then review
separately the analogy with critical phase transitions in statistical
mechanics, and aspects specific to general relativity, such as
spacetime singularities. We examine the evidence that critical
phenomena in gravitational collapse are generic, and give an overview
of their rich phenomenology.

\end{abstract}


\newpage
\tableofcontents
\newpage


\section{Introduction}


Gravity is described by Einstein's theory of general relativity (from
now on GR). In the limit of weak fields and slowly moving objects it
reduces to Newtonian gravity. But it also has completely new
features. One of these is the creation of black holes in gravitational
collapse.

The Schwarzschild solution that describes a spherically symmetric
black hole has been known since 1917. It describes the spacetime
outside a spherical star. It took some time to understand the nature
of the surface $r=2M$ as the event horizon of a black hole when the
star is not there. In Painlev\'e-G\"ullstrand coordinates
$r,t,\theta,\varphi$ the Schwarzschild metric can be written as
\begin{equation}
ds^2=-dt^2 + \left(dr+\sqrt{2M\over r}dt \right)^2 + r^2
\left(d\theta^2 + \sin^2\theta \,d\varphi^2\right).
\end{equation}
As $r\gg 2M$, gravity is weak, and slowly moving bodies approximately
obey Newton's laws. This limit shows that the gravitational mass of
the black hole is indeed $M$. The metric itself shows that the area of
any 2-dimensional surface of constant $r$ and $t$ is given by $4\pi
r^2$. ($r$ is therefore called the area radius.)  In looking for
radial light rays, one solves for $ds^2=0$ subject to
$d\theta=d\varphi=0$. This gives two solutions
\begin{equation}
{dr\over dt}=-\sqrt{2M\over r}\pm 1.
\end{equation}
For $r\gg 2M$ these reduce to $dr/dt=\pm 1$. (We use units in which
the speed of light is 1.) At $r=2M$, the solutions are $dr/dt=0,-2$,
and for $0<r<2M$, both solutions are negative: both ``ingoing'' and
``outgoing'' radial light rays approach $r=0$. A more careful analysis
shows that this is true for all light rays and all material objects
(even when they are accelerated outwards by rockets). The tidal forces
on any freely falling extended object are finite at $r=2M$, and become
infinite only at $r=0$, where the spacetime ends in a singularity.

The 1939 paper of Oppenheimer and Snyder \cite{OppenheimerSnyder}
explicitly constructed a solution where a collapsing spherical star
disappears inside the surface $r=2M$ in a finite time (as measured by
an observer moving with the star). This established that a
Schwarzschild black hole can be created dynamically, as the end state
of gravitational collapse.

Some doubt was left that this scenario might be restricted to exact
spherical symmetry.  The singularity theorems of the 1960s
\cite{HawkingEllis} proved, independently of symmetry or the type of
matter, that once a closed trapped surface has formed, a singularity
must occur. A closed trapped surface is a smooth 2-dimensional closed
spacelike surface (that is, a surface in the everyday sense of the
word, such as a soap bubble) with the property that not only light
rays going into it but also those coming out of it are momentarily
converging. In a spherical situation this is equivalent to having a
mass $M$ inside a sphere of area radius $r=2M$.

The cosmic censorship hypothesis affirms that such a singularity is
always inside a black hole.  Some version of cosmic censorship is
generally believed to hold, but we shall see that critical phenomena
in gravitational collapse throw an interesting light on what this
version can be. 

The singularity theorems and the cosmic censorship hypothesis together
imply that a black hole is formed generically whenever mass is
concentrated enough. Thorne has suggested the rule of thumb that a
mass $M$ must be enclosed in a circumference of $4\pi M$. It should be
stressed, however, that no sufficient criterion for black hole
formation is known, other than the existence of a closed trapped
surface. But the singularity theorems also imply that the closed
trapped surface is already inside the black hole, so that when a
closed trapped surface is present, the black hole has already formed.

On the other hand a sufficiently weak configuration of matter or
gravitational waves will never become singular, at least for
``reasonable'' types of matter which do not become singular in the
absence of gravity. Rigorous proofs exist for certain
kinds of matter \cite{Rendall}, and in particular for pure
gravitational waves \cite{ChristodoulouKlainermann}.

The middle regime between certain collapse and certain dispersion was
not explored systematically until the 1990s. In this middle regime,
one cannot decide from the initial data if they will or will not form
a black hole. Choptuik \cite{Choptuik92} pioneered the use of
numerical experiments in this situation. He concentrated on the
simplest possible model in which black holes can form: a spherically
symmetric scalar field playing the role of matter is coupled to GR.

Even in this simple model, the space of initial data is a function
space, and therefore infinite-dimensional. To map it out, Choptuik
relied on 1-parameter families of initially data. Tens of families
were considered, and within each family hundreds of data sets. Each
family contained both data that formed a black holes, and data that
did not. The parameter $p$ of the family can therefore be interpreted
as a measure of the strength, or gravitational self-interaction, of
the initial data: strong data (say with large $p$) are those that form
black holes, weak data (say with small $p$) are those that do not. The
critical value $p_*$ of the parameter $p$ was then found empirically
for each family.

One open question at the time was this: do black holes at the
threshold have a minimum mass, or does the black hole mass go to zero
as the initial data approach the threshold? Choptuik gave highly
convincing numerical evidence that by fine-tuning the initial data to
the threshold along any 1-parameter family, one can make arbitrarily
small black holes. In the process he found three unexpected
phenomena. The first is the now famous scaling relation
\begin{equation}
\label{power_law}
M \simeq C \, (p-p_*)^\gamma
\end{equation}
for the black hole mass $M$ in the limit $p\to p_*$ from above. While
$p_*$ and $C$ depend on the particular 1-parameter family of data, the
exponent $\gamma$ has a universal value $\gamma\simeq 0.374$ for all
1-parameter families of scalar field data.

The second unexpected phenomenon is universality. For a finite time in
a finite region of space, the spacetime generated by all near-critical
data approaches one and the same solution. This universal phase ends
when the evolution decides between black hole formation and
dispersion. The universal critical solution is approached by any
initial data that are sufficiently close to the black hole threshold,
on either side, and from any 1-parameter family. (Only an overall
scale depending on the family needs to be adjusted.)

The third phenomenon is scale-echoing. Let $\phi(r,t)$ be the
spherically symmetric scalar field, where $r$ is radius and $t$ is
time. The critical solution $\phi_*(r,t)$ is the same when we rescale
space and time by a factor $e^\Delta$:
\begin{equation}
\phi_*(r,t) = \phi_*\left(e^{\Delta}r,e^{\Delta}t\right),
\end{equation}
with $\Delta\simeq 3.44$, so that $e^{\Delta}\simeq 30$. The scale
period $\Delta$ is a second dimensionless number that comes out of the
blue.

Choptuik's results created great excitement. Similar phenomena were
quickly discovered in many other types of matter coupled to gravity,
and even in gravitational waves (which can form black holes even in
the absence of matter) \cite{AbrahamsEvans}. The echoing period
$\Delta$ and critical exponent $\gamma$ depend on the type of matter,
but the existence of the phenomena appears to be generic.

There is also another kind of critical behavior at the black hole
threshold. Here, too, the evolution goes through a universal critical
solution, but it is static, rather than scale-invariant. As a
consequence, the mass of black holes near the threshold takes a finite
universal value, instead of showing power-law scaling. In an analogy
with first and second order phase transitions in statistical
mechanics, the critical phenomena with a finite mass at the black hole
threshold are called type I, and the critical phenomena with power-law
scaling of the mass are called type II. 

The existence of a threshold where a qualitative change takes place,
universality, scale-invariance, and critical exponents suggest that
there is a connection between type II critical phenomena and critical
phase transitions in statistical mechanics. The appearance of a
complicated structure and two mysterious dimensionless numbers out of
generic initial data and simple field equations is also
remarkable. The essence of this is now so well understood that one can
speak of a standard model.  But critical phenomena also have a deeper
significance for our understanding of GR.

Black holes are among the most important solutions of GR because of
their universality: Whatever the collapsing object that forms a black
hole, the final black hole state is characterized completely by its
mass, angular momentum and charge \cite{Heusler}. (For Yang-Mills
matter fields, there are exceptions to the letter, but not the spirit,
of black hole uniqueness.) All other information about the initial
state before collapse must be radiated away \cite{Price}. 

Critical solutions have a similar importance because they are generic
intermediate states of the evolution that are also independent of the
initial data, up to the requirement that they are fairly close to the
black hole threshold. Type II critical solutions are significant for
GR also because they contain a naked singularity, that is a point of
infinite spacetime curvature from which information can reach a
distant observer. (By contrast, the singularity inside a black hole is
hidden from distant observers.) It has been conjectured that naked
singularities do not arise from suitably generic initial data for
suitably well-behaved matter (cosmic censorship). The existence of
critical solutions sheds some light on this: because they have naked
singularities, and because they are the end states for all initial
data that are {\it exactly} on the black hole threshold, all initial
data on the black hole threshold form a naked singularity. These data
are not generic, but they are close, having co-dimension one.  For
these reasons, type II phenomena are more fundamentally important than
type I phenomena, and this review focuses largely on them. Beyond type
I and type II critical phenomena, one can also take a larger view of
the subject: it is the study of the boundaries between basins of
attraction in phase space.

This review consists of an overview part, which deliberately skates
over technical aspects of GR, and a details part. The overview part
begins with a fairly complete presentation of Choptuik's work in
Section \ref{section:Choptuik}, in order to give the reader one
concrete example of a system in which type II critical phenomena are
observed. The core ideas required in a derivation of the mass scaling
law (\ref{power_law}) come from dynamical systems theory and
dimensional analysis, and these are presented in Section
\ref{section:standardmodel}. The analogy with critical phase
transitions is then summarized in Section \ref{section:SM}.

The details part begins with Section \ref{section:GR}, where I have
grouped together those aspects of critical collapse that make it, in
my opinion, an interesting research area within GR. In Section
\ref{section:generic} we ask how generic critical phenomena really
are: are they restricted to specific types of matter, or specific
symmetries?  Given that black holes can have angular momentum and
charge, how do these scale at the black hole threshold, and what is
effect of charge and angular momentum in the initial data? (Section
\ref{section:angmom}, on angular momentum, also extends the analogy
with critical phase transitions.) The final Section
\ref{section:phenomenology} has been loosely titled ``phenomenology'',
but could fairly be called ``odds and ends''.

Other review papers on critical phenomena in gravitational collapse
include
\cite{Bizon,Choptuik_review,Choptuik_review2,Gundlach_Banach,Gundlach_critreview1,Gundlach_critreview2,Horne_MOG}.


\section{Case study: the spherically symmetric scalar field} 
\label{section:Choptuik}



\subsection{Field equations}


The system in which Choptuik first studied the threshold of black hole
formation is the spherically symmetric massless, minimally coupled,
scalar field. Spherical symmetry means that all fields depend only on
radius $r$ and time $t$ (1+1 effective spacetime dimensions). This
keeps the computational requirements low. Conversely, for given
computing power, it permits much higher numerical precision than a
numerical calculation in 2+1 or 3+1 dimensions. Scalar field matter is
in many ways the most simple and well-behaved matter. As it propagates
at the speed of light, it can also be used as a toy model for
gravitational radiation, which does not exist in spherical symmetry.

Consider therefore a spherically symmetric, massless scalar field $\phi$
minimally coupled to GR. The Einstein equations are
\begin{equation}
\label{scalar_stress_energy}
G_{ab} = 8 \pi \left(\nabla_a \phi \nabla_b \phi - {1\over 2} g_{ab}
\nabla_c \phi \nabla^c \phi\right)
\end{equation}
and the matter equation is
\begin{equation}
\nabla_a \nabla^a \phi = 0.
\end{equation}
Note that the matter equation of motion follows from stress-energy
conservation alone. 

Choptuik chose Schwarzschild-like coordinates (also called
polar-radial coordinates), in terms of which the spacetime line
element is
\begin{equation}
\label{tr_metric}
ds^2 = - \alpha^2(r,t) \, dt^2 + a^2(r,t) \, dr^2 + r^2 \, d\Omega^2.
\end{equation}
Here $d\Omega^2 = d\theta^2 + \sin^2\theta \, d\varphi^2$ is the
metric on the unit 2-sphere. This choice of coordinates is defined by
two properties: the surface area of the 2-spheres of constant $t$ and
$r$ is $4\pi r^2$, and $t$ is orthogonal to $r$, so that
there is no $dt\,dr$ cross term in the line element. One more
condition is required to fix the coordinate $t$ completely. Choptuik
chose $\alpha=1$ at $r=0$, so that $t$ is the proper time of the
central observer at $r=0$. An important geometric diagnostic is the
Hawking mass $m$ which in spherical symmetry is defined by
\begin{equation}
\label{mdef}
1-{2m(r,t)\over r}\equiv \nabla_a r \nabla^a r = a^{-2}.
\end{equation}
In particular, the limit $r\to\infty$ of $m$ is the ADM, or total mass
of the spacetime, and $r=2m$ signals an apparent horizon.

In the auxiliary variables
\begin{equation}
\Phi = \phi_{,r}, \qquad \Pi={a\over\alpha} \phi_{,t},
\end{equation}
the wave equation becomes a first-order system,
\begin{eqnarray}
\label{wave}
\Phi_{,t} & = &  \left({\alpha\over a}\Pi\right)_{,r}, \\
\Pi_{,t} & = & {1\over r^2} \left(r^2{\alpha\over a} \Phi\right)_{,r}.
\end{eqnarray}
In spherical symmetry there are four algebraically independent
components of the Einstein equations. Of these, one is a linear
combination of derivatives of the other and can be disregarded. In
Schwarzschild-like coordinates, the other three contain only first
derivatives of the metric, namely $a_{,t}$, $a_{,r}$ and
$\alpha_{,r}$. These equations are
\begin{eqnarray} 
\label{da_dr}
{a_{,r}\over a}  + {a^2 -1 \over 2r} - 2\pi r (\Pi^2 + \Phi^2) & = &
0, \\
\label{dalpha_dr}
{\alpha_{,r}\over \alpha}  - {a_{,r}\over a}  - {a^2 -1 \over r} &
= & 0, \\
\label{da_dt}
{a_{,t}\over \alpha}  -  4\pi r \Phi \Pi &=& 0.
\end{eqnarray}
Choptuik chose to use the first two equations, which contain $a_{,r}$
and $\alpha_{,r}$, but no $t$-derivatives, for his numerical
scheme. Only the scalar field $\phi$ is evolved forward in time, while
the two metric coefficients $a$ and $\alpha$ are calculated from the
matter at each new time step, by an explicit integration over $r$
starting at $r=0$. The third equation is then obeyed automatically. We
have already mentioned the gauge condition $\alpha=1$ at $r=0$. The
other boundary condition that we need at $r=0$ is $a=1$, which simply
means that the spacetime is regular there. The main advantage of such
a ``fully constrained'' numerical scheme is its stability. We have now
stated all the equations that are needed to repeat Choptuik's results.
Note that the field equations do not contain an intrinsic
scale. Therefore the rescaling 
\begin{equation}
\label{rescaling}
t\to kt, \quad r\to kr, \quad \phi\to\phi, \quad
\Pi\to k^{-1}\Pi, \quad\Psi\to k^{-1}\Psi
\end{equation}
transforms one solution into another for any positive constant $k$.


\subsection{The black hole threshold}


In spherical symmetry, the gravitational field has no degrees of
freedom independently of the matter -- there is no gravitational
radiation. This is fortunate for the purposes of this review in that
it hides some of the complications specific to GR: the reader not
familiar with GR can form a correct picture of many aspects of this
system using flat spacetime intuition. For example, the free data for
the system, in Choptuik's choice of variables, are the two functions
$\Pi(r,0)$ and $\Phi(r,0)$, just as they would be in the absence of
gravity.

Choptuik investigated 1-parameter families of such data by evolving
the data for many values each of the parameter, say $p$. He examined a
number of families in this way. A simple example of such a family is
$\Phi(r,0)=0$ and a Gaussian for $\Pi(r,0)$, with the parameter $p$
taken to be the amplitude of the Gaussian. For a sufficiently small
amplitude the scalar field will disperse, and for a sufficiently large
amplitude it will form a black hole. It is not difficult to construct
other 1-parameter families that cross the black hole threshold in this
way (for example by varying the width or the center of a Gaussian
profile).

Christodoulou has proved for the spherically symmetric scalar field
system that data sufficiently weak in a well-defined way evolve to a
Minkowski-like spacetime \cite{Christodoulou0a,Christodoulou3}, and
that a class of sufficiently strong data forms a black hole
\cite{Christodoulou2}. (See also
\cite{Christodoulou1,Christodoulou4}.) But what happens in between,
where the conditions of neither theorem apply?

Choptuik found that in all 1-parameter families of initial data that
he investigated he could make arbitrarily small black holes by
fine-tuning the parameter $p$ ever closer to the black hole
threshold. One must keep in mind that nothing singles out the black
hole threshold in the initial data. One simply cannot tell that one
given data set will form a black hole and another one that is
infinitesimally close will not, short of evolving both for a
sufficiently long time. ``Fine-tuning'' of $p$ to the black hole
threshold must therefore proceed numerically, for example by
bisection.

With $p$ closer to $p_*$, the spacetime varies on ever smaller
scales. The only limit was numerical resolution, and in order to push
that limitation further away, Choptuik developed numerical
techniques that recursively refine the numerical grid in spacetime
regions where details arise on scales too small to be resolved
properly. The finest grid will be many orders of magnitude finer than
the initial grid, but will also cover a much smaller area. The total
number of grid points (and computer memory) does not diverge. Covering
all of space in the finest grid would be quite impossible. 

In the end, Choptuik could determine $p_*$ up to a relative precision
of $10^{-15}$, limited only by finite precision arithmetic, and make
black holes as small as $10^{-6}$ times the ADM mass of the
spacetime. The power-law scaling (\ref{power_law}) was obeyed from
those smallest masses up to black hole masses of, for some families,
$0.9$ of the ADM mass, that is, over six orders of magnitude
\cite{Choptuik94}. There were no families of initial data which did
not show the universal critical solution and critical
exponent. Choptuik therefore conjectured that $\gamma$ is the same for
all 1-parameter families of smooth, asymptotically flat initial data
that depend smoothly on the parameter, and that the approximate
scaling law holds ever better for arbitrarily small $p-p_*$.

Clearly a collapse spacetime which has ADM mass 1, but settles down to
a black hole of mass $10^{-6}$ (for example) has to show structure on
very different scales. The same is true for a spacetime which is as
close to the black hole threshold, but on the other side: the scalar
wave contracts until curvature values of order $10^{12}$ are reached
in a spacetime region of size $10^{-6}$ before it starts to
disperse. Choptuik found that all near-critical spacetimes, for all
families of initial data, look the same in an intermediate region,
that is they approximate one universal spacetime, which is also called
the critical solution:
\begin{equation}
\label{universality}
Z(r,t)\simeq Z_*\left(kr,k(t-t_*)\right)
\end{equation}
The accumulation point $t_*$ and the factor $k$ depend on the family,
but the scale-periodic part $Z_*$ of the near-critical solutions does
not. The universal solution $Z_*$ itself has the property that
\begin{equation}
\label{tr_scaling}
Z_*(r,t)= Z_*\left(e^{n\Delta}r,e^{n\Delta}\right)
\end{equation}
for all integer $n$ and for $\Delta\simeq 3.44$, and where $Z$ stands
for either of the metric coefficients $a(r,t)$ and $\alpha(r,t)$, or
the matter field $\phi(r,t)$. One easily finds quantities derived from
these, such as $r\Pi$ or $r\Phi$, or $r^2R$ (where $R$ is the Ricci
scalar), that are also periodic. Choptuik called this phenomenon
``scale-echoing''. [The critical solution is really determined only up
to rescalings of the form (\ref{rescaling}). However if one
arbitrarily fixes $Z_*(r,t)$ to be just one member of this family,
then $k$ must be adjusted as a family-dependent constant in order to
obtain (\ref{universality}).]


\section{The standard model}
\label{section:standardmodel}


In this section we review the basic ideas underlying critical
phenomena in gravitational collapse. In a first step, we apply ideas
from dynamical systems theory to gravitational collapse. The critical
solution is identified with a critical fixed point of the dynamical
system. This explains universality at the black hole threshold. In a
second step we look at the nature of the critical fixed point,
separately for type I and type II phenomena. This leads to a
derivation of the black hole mass law. 


\subsection{The phase space picture}
\label{section:phasespace}


We now discuss GR as an infinite-dimensional continuous dynamical
system. A continuous dynamical system consists of a manifold with a
vector field on it. The manifold, also called the phase space, is the
set of all possible initial data. The vector field gives the direction
of the time evolution, and so can formally be called
$\partial/\partial t$. The integral curves of the vector field are
solutions. When we consider GR as a dynamical system, points in the
phase space are initial data sets on a 3-dimensional manifold. An
integral curve is a spacetime that is a solution of the Einstein
equations. Points along the curve are Cauchy surfaces in the
spacetime, which can be thought of as moments of time $t$. There are
many important technical problems associated with this simple picture,
but we postpone discussing them to Section \ref{section:dynsim}.

Before we consider the time evolution, we clarify the nature of the
black hole threshold. All numerical evidence collected for individual
1-parameter families of data suggests that the black hole threshold is
a hypersurface in the infinite-dimensional phase space of smooth,
asymptotically flat initial data. There is no evidence that the
threshold is fractal. A 1-parameter family that depends smoothly on
its parameter locally crosses the threshold only once. $p-p_*$, for
any family, is then just a measure of distance from that
threshold. The mass scaling law can therefore be stated without any
reference to 1-parameter families. Let $P$ be a function on phase
space such that data sets with $P>0$ form black holes, and data with
$P<0$ do not. Let $P$ be analytic in a neighborhood of the black hole
threshold $P=0$.  Along any 1-parameter family of data that depends
analytically on its parameter $p$, $P$ then depends on $p$ as
\begin{equation}
\label{Pdef}
P(p)={1\over C}(p-p_*)+O\Big((p-p_*)^2\Big)
\end{equation}
with family-dependent constants $C$ and $p_*$. In terms of $P$, the
black hole mass as a function on phase space is to leading order
\begin{equation}
\label{PM}
M \simeq \theta(P)\ P^\gamma.
\end{equation}
In the following we go back to 1-parameter families because that
notation is customary in the literature, but all statements we make in
terms of such families are really statements about functions on phase
space.

We now consider some qualitative aspects of the time evolution in our
dynamical system. Typically, an isolated system in GR ends up in one
of three final states. It either collapses to a black hole, forms a
star, or disperses completely. The phase space of isolated gravitating
systems is therefore divided into basins of attraction. The boundaries
are called critical surfaces. We shall focus initially on the critical
surface between black hole formation and dispersion. (For the
spherically symmetric massless scalar field a black hole or complete
dispersion of the field are in fact the only possible end states: no
stars can be formed from this kind of matter.)

A phase space trajectory that starts out in a critical surface by
definition never leaves it. A critical surface is therefore a
dynamical system in its own right, with one dimension fewer. If it has
an attracting fixed point, such a point is called a critical point. It
is an attractor of codimension one, and the critical surface is its
basin of attraction. The fact that the critical solution is an
attractor of codimension one is visible in its linear perturbations:
it has an infinite number of decaying perturbation modes tangential to
(and spanning) the critical surface, and a single growing mode
not tangential to the critical surface.

\begin{figure}
\epsfysize=10cm
\centerline{\epsffile{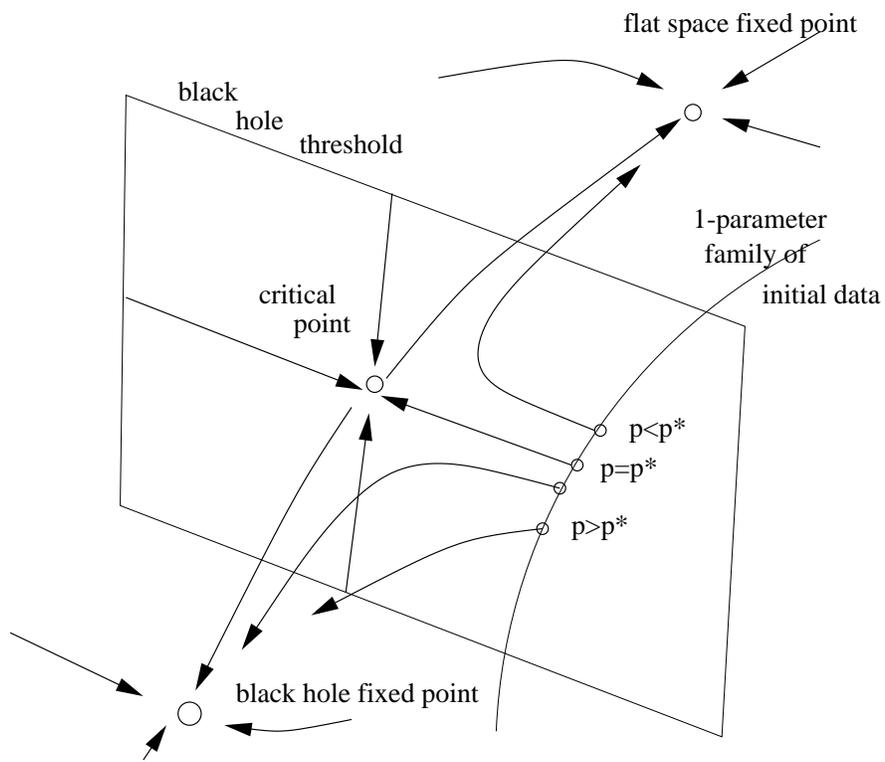}}
\caption{
\label{fig:dynsim}
The phase space picture for the black hole threshold in the
presence of a critical point. The arrow lines are time evolutions,
corresponding to spacetimes. The line without an arrow is not a time
evolution, but a 1-parameter family of initial data that crosses the
black hole threshold at $p=p_*$.}
\end{figure}

Fig. \ref{fig:dynsim} now illustrates the following qualitative
considerations: any trajectory beginning near the critical surface,
but not necessarily near the critical point, moves almost parallel to
the critical surface towards the critical point. As the phase point
approaches the critical point, its movement parallel to the surface
slows down, while its distance and velocity out of the critical
surface are still small. The phase point spends some time moving
slowly near the critical point. Eventually it moves away from the
critical point in the direction of the growing mode, and ends up on an
attracting fixed point.

This is the origin of universality: any initial data set that is close
to the black hole threshold (on either side) evolves to a spacetime
that approximates the critical spacetime for some time. When it
finally approaches either the dispersion fixed point or the black hole
fixed point it does so on a trajectory that appears to be coming from
the critical point itself. All near-critical solutions are passing
through one of these two funnels. All details of the initial data have
been forgotten, except for the distance from the black hole threshold:
The closer the initial phase point is to the critical surface, the
more the solution curve approaches the critical point, and the longer
it will remain close to it. (We shall see how this determines the
black hole mass in Section \ref{section:massscaling}.

\begin{figure}
\epsfysize=10cm
\centerline{\epsffile{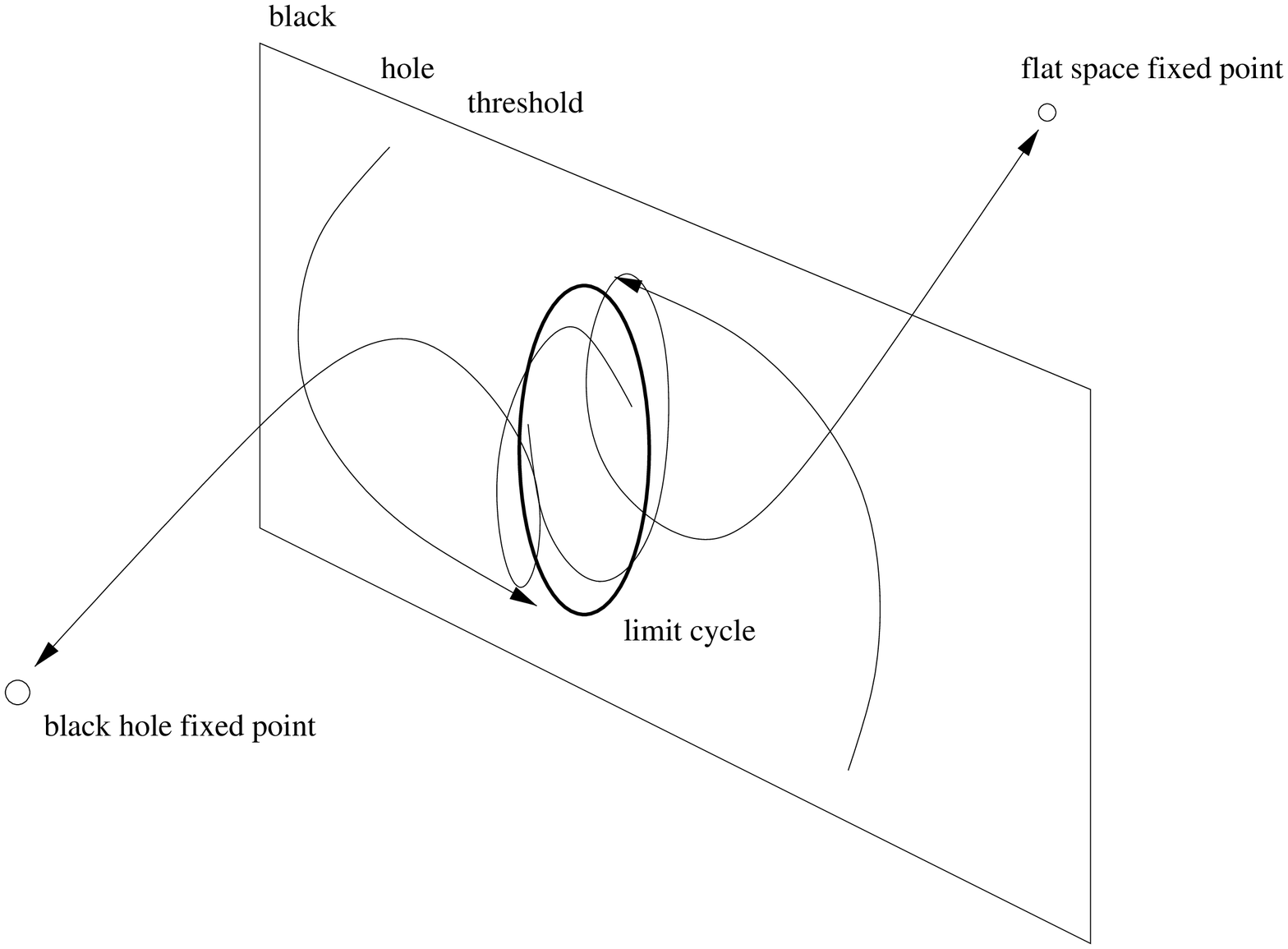}}
\caption{
\label{fig:limitcycle}
The phase space picture in the presence of a limit cycle. The
plane represents the critical surface. 
The circle (fat unbroken line) is the limit cycle representing the
critical solution. Shown also are two trajectories in the critical
surfaces and therefore attracted to the limit cycle, and two
trajectories out of the critical surface and repelled by it.}
\end{figure}

The black hole threshold in all toy models that have been examined
contains at least one critical point. In the GR context, a fixed
point, which is a solution that is independent of the time parameter
$t$, is a spacetime that has an additional continuous symmetry that
generic solutions do not have. As we shall see this can mean either
that the spacetime is time-independent in the usual sense, or else
that it is scale-invariant. The attractor within the critical surface
may also be a limit cycle, rather than a fixed point. The critical
solution is then periodic in the time parameter $t$. In spacetime
terms this corresponds to a discrete symmetry. Our qualitative remarks
are not affected by this variation. The phase space picture in the
presence of a limit cycle critical solution is sketched in
Fig. \ref{fig:limitcycle}. (Recall that the scalar field critical
solution has a discrete symmetry.)


\subsection{Type I critical phenomena}
\label{section:typeI}


We now look at the nature of the critical point as a spacetime. Recall
that a critical point is really equivalent to an entire integral curve,
and therefore, in GR, to a spacetime. All critical points that have
been found in black hole thresholds so far have an additional
spacetime symmetry beyond those present in a generic phase point. The
additional symmetry is either time-independence or
scale-independence. Each of these exists in a continuous and a
discrete version. 

Here our exposition branches. We first consider type I critical
phenomena, due to a time-independent critical solution. Type I
phenomena are not very spectacular, but they have a scaling law of
their own, which provides a simpler model for the power-law scaling of
the mass in type II phenomena.

A static solution is invariant under an infinitesimal displacement in
time. In suitable coordinates, its dynamical variables are all
independent of time. By contrast, an oscillating, or periodic,
solution is invariant under a discrete displacement in time (or its
integer multiples). All its variables are periodic in a suitable time
coordinate. For simplicity we focus on the continuous symmetry. The
generalization to the discrete symmetry is straightforward. Assume
therefore that the critical solution is time-independent. If the time
coordinate is $t$, and $r$ is radius (in spherical symmetry), then ñwe
can write the critical solution formally as
\begin{equation}
Z(t,r)=Z_*(r)
\end{equation}
(In technical GR terms, the critical solution is stationary, which
means that it has a timelike Killing vector $\partial/\partial t$. All
known stationary critical solutions are in fact spherically symmetric,
and therefore static, meaning that $\partial/\partial t$ is normal to
the surfaces of constant $t$). Because the critical solution is
independent of $t$, its general linear perturbation can be written as
a sum of modes of the form
\begin{equation}
\delta Z(t,r)=\sum_i C_i(p) e^{\lambda_i t} Z_i(r).
\end{equation} 
$\lambda_i$ and $Z_i(r)$ are in general complex but come in complex
conjugate pairs. The amplitudes $C_i$ depend on the initial data, and
in particular on the parameter $p$ of a one-parameter family of
data. The critical solution is by definition an attractor of
co-dimension one. This means that it has precisely one unstable
perturbation, with $\lambda_0 >0$, while all other perturbations
decay, ${\rm Re}\lambda_i<0$. For initial data that are sufficiently
close to the critical surface, the solution curve passes so much time
near the critical point that the decaying perturbations can eventually
be neglected. We are left with the one growing mode. By definition, if
the initial data are exactly in the critical surface, the solution
curve will run into the critical point. In perturbative terms this
means that the amplitude $C_0$ of the one growing mode is exactly
zero. This happens for $p=p_*$, where $p$ is the parameter of a family
of initial data. To leading order $C_0$ is then proportional to
$p-p_*$. Putting all this together, we find that there is an
intermediate regime where we have the approximation
\begin{equation}
\label{typeIintermediate}
Z(r,t) \simeq Z_*(r) + {d C_0\over dp} (p_*)(p-p_*) e^{\lambda_0 t} Z_0(r) +
\hbox{decaying modes}.
\end{equation}
We define a time $t_p$ by
\begin{equation}
{d C_0\over dp} |p-p_*| e^{\lambda_0 t_p} \equiv \epsilon,
\end{equation}
where $\epsilon$ is an arbitrary small positive constant. (We assume $
dC_0/dp>0$.)  The initial data at $t_p$ are
\begin{equation}
Z(r,t_p) \simeq Z_*\left(r\right) \pm \epsilon \ Z_0\left(r \right),
\end{equation}
where the sign in front of $\epsilon$ is the sign of $p-p_*$. Again by
definition, these initial data must form a black hole for one sign,
and disperse for the other. The data at $t=t_p$ are independent of
$|p-p_*|$, and so the final black hole mass for $p>p_*$ is independent
of $p-p_*$. This mass is some fraction of the mass of the critical
solution $Z_*$. The magnitude of $|p-p_*|$ survives only in $t_p$,
which is the time interval during which the solution is approximately
equal to the critical solution: the ``lifetime'' of the critical
solution. From its definition, we see that this time scales as
\begin{equation}
\label{typeIscaling}
t_p = - {1\over \lambda_0} \ln|p-p_*| + {\rm const.}
\end{equation}

Intuitively, type I critical solutions can be thought of as metastable
stars. Therefore, type I critical phenomena typically occur when the
field equations set a mass scale in the problem (although there could
be a family of critical solutions related by scale). Conversely, as
the type II power law is scale-invariant, type II phenomena occur in
situations where either the field equations do not contain a scale, or
this scale is dynamically irrelevant. Many systems, such as the
massive scalar field, show both type I and type II critical phenomena,
in different regions of the space of initial data.


\subsection{Type II critical phenomena}
\label{section:typeII}



\subsubsection{The critical solution}


Type II critical phenomena occur where the critical solution is
scale-invariant, or self-similar. Like time-independence in type I,
this symmetry comes in a continuous and a discrete version: A
continuously self-similar (CSS) solution is invariant under an
infinitesimal (by a factor of $1+\epsilon$) rescaling of both space
and time (and therefore invariant under an arbitrary finite
rescaling). A discretely self-similar (DSS) solution is invariant only
under rescaling by a particular finite factor, or its integer
powers. These solutions are independent of (or periodic in) a suitable
scale coordinate. They give rise to power-law scaling of the black
hole mass at the threshold, and the other phenomena discovered by
Choptuik. These are referred to as type II critical phenomena, in
analogy with a second-order phase transition, where the order
parameter is continuous.

In order to construct a scale coordinate, we formulate the
scale-periodicity (\ref{tr_scaling}) observed by Choptuik in a
slightly different manner. We replace $r$ and $t$ by new coordinates
such that one of them is the logarithm of an overall spacetime
scale. A simple example is
\begin{equation}
\label{x_tau}
x = -{r \over t}, \quad \tau = - \ln\left(-{t\over l}\right),
\quad t<0.
\end{equation}
We have again shifted the origin of $t$ to the accumulation point
$t_*$ of echos, which is why $t<0$. $\tau$ has been defined with the
two minus signs so that it increases as $t$ increases and approaches
$t_*$ from below. It is useful to think of $r$, $t$ and $l$ as having
dimension length in units $c=G=1$, and of $x$ and $\tau$ as
dimensionless.  Choptuik's observation, expressed in these
coordinates, is that in any near-critical solution there is a
space-time region where the fields $a$, $\alpha$ and $\phi$ are well
approximated by their values in a universal solution, or
\begin{equation}
Z(x,\tau) \simeq Z_*(x,\tau),
\end{equation}
where the fields $Z=\{a,\alpha,\phi\}$ of the critical solution 
have the periodicity
\begin{equation}
Z_*(x,\tau+\Delta) = Z_*(x,\tau).
\end{equation}
The dimensionful constants $t_*$ and $l$ depend on the particular
1-parameter family of solutions, but the dimensionless critical fields
$a_*$, $\alpha_*$ and $\phi_*$, and in particular their dimensionless
period $\Delta$, are universal. The critical solution for the massless
scalar field is DSS, but the critical solution for a perfect fluid,
for example, is CSS. In the CSS case suitable variables $Z$ exist so
that $Z_*(x)$ is independent of $\tau$. (Note that the time parameter
of the dynamical system must be chosen as $\tau$ if a CSS solution is
to be a fixed point.)

Self-similarity in GR will be defined rigorously in Section
\ref{section:selfsimilarity} below. Here we should stress the double
role of $\tau$: It is both a time coordinate (just the logarithm of
$t$), and also the logarithm of spacetime scale in the critical
solution: In going from $\tau$ to $\tau+\Delta$ at constant $x$, we
find the same solution, but on a space and time scale smaller by a
factor $e^{-\Delta}$.

We need to introduce a vector $Z(x)$ of variables such that they
specify the state of the system at one moment of time, and such that a
solution $Z(x,\tau)$ is CSS if and only if it is independent of
$\tau$, and DSS if it is periodic in $\tau$. This is done in general
below in Section \ref{section:dynsim}. Here we only define the $Z$ for
our example of the spherically symmetric scalar field. In spherical symmetry,
there are no independent degrees of freedom in the gravitational
field, and initial data are required only for the scalar field. As it
obeys the wave equation, Cauchy data are $\phi$ and $\phi_{,t}$. It is
easy to see that if the scalar field is CSS, so that
$\phi(r,t)=\phi(x)$, then $r\phi_{,t}$ is also a function of $x$
only. Similarly, in DSS, this variable is periodic in $\tau$. For $Z$
we can therefore choose $Z=\{\phi,r\phi_{,t}\}$.


\subsubsection{Mass scaling}
\label{section:massscaling}


The critical exponent $\gamma$ can be calculated from the linear
perturbations of the critical solution by an application of
dimensional analysis. This was suggested by Evans and Coleman
\cite{EvansColeman} and the details worked out independently by Koike,
Hara and Adachi \cite{KoikeHaraAdachi,KoikeHaraAdachi2} and Maison
\cite{Maison}. The calculation is simple enough, and sufficiently
independent of GR technicalities, that we summarize it here.  In order
to keep the notation simple, we restrict ourselves to a critical
solution that is spherically symmetric and CSS. We only state the
result in the more general DSS. The generalization away from spherical
symmetry is only a matter of notation.

Let us assume that we have fine-tuned the initial data to the black
hole threshold so that in a region of the resulting spacetime it is
well approximated by the CSS critical solution, $Z(r,t)\simeq
Z_*(x)$. This part of the spacetime corresponds to the section of the
phase space trajectory that lingers near the critical point. In this
region we can linearize around $Z_*$, while both before (near the
initial time) and afterwards (when a black hole forms or the solution
disperses) we cannot do this. Therefore we call this spacetime region
the ``intermediate linear regime''. As $Z_*$ does not depend on
$\tau$, its linear perturbations can depend on $\tau$ only
exponentially, so that the general linear perturbation has the form
\begin{equation}
\delta Z(x,\tau)=\sum_{i=0}^\infty C_i \, e^{\lambda_i \tau} Z_i(x),
\end{equation}
where the $C_i$ are free constants. The $\lambda_i$ and $Z_i(x)$ are
in general complex, but form complex conjugate pairs. We assume here
that the perturbation spectrum is discrete. This seems to be the case
for the critical solutions studied so far. Indeed, as the critical
solution has precisely one growing mode, the spectrum must be discrete
in the right half plane.

To linear order, the solution in the intermediate linear region is
then of the form
\begin{equation}
Z(x,\tau;p) \simeq Z_*(x) + \sum_{i=0}^\infty C_i(p) \, e^{\lambda_i
\tau} Z_i(x).
\end{equation} 
The coefficients $C_i$ depend in a complicated way on the initial
data, and hence on $p$. If $Z_*$ is a critical solution, by
definition there is exactly one $\lambda_i$ with positive real part
(in fact it is purely real), say $\lambda_0$. As $t\to t_*$ from below
and $\tau\to\infty$, all other perturbations vanish. In the following
we consider this limit, and retain only the one growing
perturbation. By definition the critical solution corresponds to
$p=p_*$, so we must have $C_0(p_*)=0$. Linearizing around $p_*$, we
obtain
\begin{equation}
\label{intermediatelinear}
\lim_{\tau\to\infty} Z(x,\tau) \simeq Z_*(x) + {dC_0\over dp} (p-p_*)
e^{\lambda_0\tau} Z_0(x).
\end{equation}
In the limit of perfect fine-tuning of $p$ to its critical value,
the growing mode would be completely suppressed and the solution would
approximate to $Z_*$ ever better as $\tau\to\infty$. For any finite
value of $p-p_*$, however, the growing mode will eventually become
large, and the solution leaves the intermediate linear regime.

The solution has the approximate form (\ref{intermediatelinear}) over a
range of $\tau$. Now we extract Cauchy data at one particular value of
$\tau$ within that range, namely $\tau_p$ defined by
\begin{equation}
{dC_0\over dp} |p-p_*| e^{-\lambda_0\tau_p} \equiv \epsilon,
\end{equation}
where $\epsilon$ is a fixed constant. Its exact value does not matter,
but it should be chosen small enough so that at this amplitude $Z_0$
can still be treated as a linear perturbation, and large enough so
that it becomes nonlinear soon after. With this choice, $\tau_p$ is
the amount of $\tau$-time spent in the intermediate linear regime (up
to an additive constant). We have
\begin{equation}
\tau_p = {1\over \lambda_0} \ln|p-p_*| + {\rm const.}
\end{equation}
Note that this holds for both supercritical and subcritical solutions.
With a DSS solution with period $\Delta$, the number of echos observed
in the intermediate linear regime is $N=\tau_p/\Delta$.

At sufficiently large $\tau$, the linear perturbation has grown so
much that the linear approximation breaks down. Later on either a
black hole forms, or the solution disperses, depending only on the
sign of $p-p_*$. We define signs so that a black hole is formed for
$p>p_*$. The crucial point is that we need not follow this evolution
in detail, nor does it matter at what amplitude $\epsilon$ we consider
the perturbation as becoming non-linear. It is sufficient to note that
the Cauchy data at $\tau=\tau_p$ depend on $r$ only through the
argument $x$, because by definition we have
\begin{equation}
Z(x,\tau_p) \simeq Z_*(x) \pm \epsilon \ Z_0(x).
\end{equation}
The $\pm$ sign is the sign of $p-p_*$, left behind because by
definition $\epsilon$ is positive.  Going back to coordinates $t$ and
$r$, and shifting the origin of $t$ once more so that it now coincides
with $\tau=\tau_p$, we have
\begin{equation}
Z(r,0) \simeq Z_*\left(-{r\over L_p}\right) \pm \epsilon \
Z_0\left(-{r\over L_p} \right), \qquad L_p \equiv Le^{-\tau_p}.
\end{equation}

These intermediate data at $t=0$ depend on the initial data at $t=0$
only through the overall scale $L_p$, and through the sign in front of
$\epsilon$.  The field equations themselves do not have an intrinsic
scale. It follows that the solution based on the data at $t=0$ must be
universal up to the overall scale. It is then of the form
\begin{equation}
Z(r,t) = f_\pm\left({r\over L_p}, {t\over L_p}\right),
\end{equation}
for two functions $f_\pm$ that are universal for all 1-parameter
families \cite{HE2}. This universal form of the solution applies for
all $t>0$, even after the approximation of linear perturbation theory
around the critical solution breaks down. Because the black hole mass
has dimension length, it must be proportional to $L_p$, the only
length scale in the solution $f_+$. Therefore
\begin{equation}
M \propto L_p \propto (p-p_*)^{1\over \lambda_0},
\end{equation}
and we have found the critical exponent $\gamma = 1/\lambda_0$.

When the critical solution is DSS, the scaling law is modified.  This
was predicted in \cite{Gundlach_Chop2} and predicted independently and
verified in collapse simulations by Hod and Piran
\cite{HodPiran_wiggle}. On the straight line relating $\ln M$ to
$\ln(p-p_*)$, a periodic ``wiggle'' or ``fine structure'' of small
amplitude is superimposed:
\begin{equation}
\label{wiggle}
\ln M = \gamma \ln (p-p_*) + c + f[\gamma \ln (p-p_*) + c],
\end{equation}
with $f(z)=f(z+\Delta)$.  The periodic function $f$ is again universal
with respect to families of initial data, and there is only one
parameter $c$ that depends on the family of initial data,
corresponding to a shift of the wiggly line in the $\ln(p-p_*)$
direction. (No separate adjustment in the $\ln M$ direction is
possible.)

In the notation of \cite{Sornette}, the result (\ref{wiggle}) can be
written as
\begin{equation}
M(p-p_*)=(p-p_*)^\gamma \sum_{n=-\infty}^\infty f_n
(p-p_*)^{in{2\pi\gamma\over \Delta}},
\end{equation}
where the numbers $f_n$ are the Fourier coefficients of $\exp
f(p-p_*)$. In this sense, one could speak of a family of complex
critical exponents
\begin{equation}
\gamma_n=\gamma+in{2\pi\gamma\over \Delta}.
\end{equation}
Keeping only the $n=0,1$ terms would be a sensible approach if DSS
was only a perturbation of CSS, but this is not the case for the known
critical solutions.

The maximal value of the scalar curvature, and similar quantities, for
near-critical solutions, scale just like the black hole mass, but with
a critical exponent $-2\gamma$ because they have dimension
(length$)^{-2}$ and are proportional to $L_p^{-2}$. Note that this is
true both for supercritical and subcritical data.  Technically this is
useful because it is easier to measure the maximum curvature in a
subcritical evolution than to measure the black hole mass in the
supercritical regime \cite{GarfinkleDuncan}.


\section{The analogy with statistical mechanics}
\label{section:SM}


Some basic aspects of critical phenomena in gravitational collapse,
such as fine-tuning, universality, scale-invariant physics, and
critical exponents for dimensionful quantities, can also be identified
in critical phase transitions in statistical mechanics. In the
following we are not trying to give a self-contained description of
statistical mechanics, but review only those aspects that will be
important to establish an analogy with critical collapse. (For a basic
textbook on critical phase transitions see \cite{Yeomans}.)

We shall not attempt to review the concepts of thermal equilibrium,
temperature, or entropy, as they are not central to our purpose. We
begin directly by noting that in equilibrium statistical mechanics
observable macroscopic quantities, such as the pressure of a fluid, or
the magnetization of a ferromagnetic material are derived as
statistical averages over micro-states of the system, which are not
observed because they contain too much information. The expected value
of an observable is
\begin{equation}
\langle A\rangle=\sum_{\rm microstates} A({\rm microstate}) \ e^{-{1\over
kT} H({\rm microstate},\mu,f)}.
\end{equation}
Here $H$ is the Hamiltonian of the system, and $T$ is the temperature
of the equilibrium distribution. $k$ is the Boltzmann constant. $f$
are macroscopic external forces on the system, such as the volume in
which a fluid sample is confined, or the external magnetic field
permeating a ferromagnetic material. $\mu$ are parameters inside the
Hamiltonian. We introduce them because it will be useful to think of
different fluids, or different ferromagnetic materials, as having the
same basic Hamiltonian, with different values of the parameters
$\mu$. Examples for such parameters would be the masses and
interaction energies of the constituent atoms or molecules of the
system. $kT$ is special in that it appears in expectation values only
as an overall factor multiplying the Hamiltonian. Nevertheless it will
be useful to reclassify it as one of the parameters $\mu$ of the
Hamiltonian, and to write the expectation values as
\begin{equation}
\langle A\rangle=\sum_{\rm microstates} A({\rm microstate}) \ e^{-
H({\rm microstate},\mu,f)}.
\end{equation}
The objective of statistical mechanics is to derive relations between
the macroscopic quantities $\langle A\rangle$ and $ f$ (for fixed
$\mu$). (Except for the simplest systems, the summation over
micro-states cannot be carried out explicitly, and one has to use
other methods.) For suitable observables $A$, the expectation values
$\langle A\rangle$ can be generated as partial derivatives of the
partition function
\begin{equation}
Z( \mu, f)=\sum_{\rm microstates} \ e^{-  H({\rm
microstate}, \mu, f)}
\end{equation}
with respect to its arguments. The separation of external variables
into $A$ and $f$ is somewhat arbitrary. For example, one could
classify the volume of a fluid as an $f$ and its pressure as an $A$,
or the other way around, depending on which of the two is being
controlled in an experiment.

Phase transitions in thermodynamics are thresholds in the space of
external forces $f$ at which the macroscopic observables $A$, or one
of their derivatives, change discontinuously. We consider two
examples: the liquid-gas transition in a fluid, and the ferromagnetic
phase transition.

The liquid-gas phase transition in a fluid occurs at the boiling curve
$p=p_b(T)$. In crossing this curve, the fluid density changes
discontinuously. However, with increasing temperature, the difference
between the liquid and gas density on the boiling curve decreases, and
the boiling curve ends at the critical point $(p_*,T_*)$ where liquid
and gas have the same density. By going around the boiling curve one
can bring a fluid from the liquid state to the gas state without
boiling it. More important for us are two other aspects. As a function
of temperature along the boiling curve, the density discontinuity
vanishes as a non-integer power:
\begin{equation}
\rho_{\rm liquid}-\rho_{\rm gas}\sim (T_*-T)^\gamma.
\end{equation}
Also, at the critical point an otherwise clear fluid becomes opaque,
due to density fluctuations appearing on all scales up to scales much
larger than the underlying atomic scale, and including the wavelength
of light. This indicates that the fluid near its critical point is
approximately scale-invariant.

In a ferromagnetic material at high temperatures, the magnetization
$\bf m$ of the material (alignment of atomic spins) is determined by
the external magnetic field $\bf B$. At low temperatures, the material
shows a spontaneous magnetization even at zero external field. In the
absence of an external field this breaks rotational symmetry: the
system makes a random choice of direction. With increasing
temperature, the spontaneous magnetization $\bf m$ decreases and
vanishes at the Curie temperature $T_*$ as
\begin{equation}
|{\bf m}|\sim (T_*-T)^\gamma.
\end{equation}
Again, the correlation length, or length scale of a typical
fluctuation, diverges at the critical point, indicating
scale-invariant physics. 

Quantities such as $|\bf m|$ or $\rho_{\rm liquid}-\rho_{\rm gas}$ are
called order parameters. In statistical mechanics, one distinguishes
between first-order phase transitions, where the order parameter
changes discontinuously, and second-order, or critical, ones, where it
goes to zero continuously. One should think of a critical phase
transition as the critical point where a line of first-order phase
transitions ends as the order parameter vanishes. This is already
clear in the fluid example. 

In the ferromagnet example, at first one seems to have only the one
parameter $T$ to adjust. But in the presence of a very weak external
field, the spontaneous magnetization aligns itself with the external
field $\bf B$, while its strength is to leading order independent of
b$\bf B$. The function ${\bf m}({\bf B},T)$ therefore changes
discontinuously at ${\bf B}=0$. The line ${\bf B}=0$ for $T<T_*$ is
therefore a line of first order phase transitions between directions
(if we consider one spatial dimension only, between $\bf m$ up and
$\bf m$ down). This line ends at the critical point $({\bf
B}=0,T=T_*)$ where the order parameter $|{\bf m}|$ vanishes. In this
interpretation both the ferromagnet and the fluid have a line of
first-order phase transitions that ends in a critical point, or
critical phase transition. At the critical point, the order parameter
vanishes as a power of distance along the first order line. The role
of ${\bf B}=0$ as the critical value of $\bf B$ is obscured by the
fact that ${\bf B}=0$ is singled out by symmetry: by contrast the
critical parameter values $p_c$ and $T_c$ of the fluid need to be
computed.

We have already stated that a critical phase transition involves
scale-invariant physics. Scale-invariance here means that fluctuations
appear on a large range of length scales between the underlying atomic
scale and the scale of the sample. In particular, the atomic scale,
and any dimensionful parameters associated with that scale, must
become irrelevant at the critical point. This is taken as the starting
point for obtaining properties of the system at the critical point.

One first defines a semi-group acting on micro-states: the
renormalization group. Its action is to group together a small number
of particles (for example, eight particles sitting on a cubic lattice)
as a single particle of a fictitious new system (a lattice with twice
the distance between particles), using some averaging procedure. This
can also be done in a more abstract way in Fourier space. One then
defines a dual action of the renormalization group on the space of
Hamiltonians by demanding that the partition function
is invariant under the renormalization group action:
\begin{equation}
\sum_{\rm microstates} e^{-H}=\sum_{\rm microstates'} e^{-H'}. 
\end{equation}
The renormalized Hamiltonian is in general more complicated than the
original one. In practice, one truncates the infinite-dimensional
space of Hamiltonians to a finite number of parameters $\mu$. The
temperature must be mixed in with the other parameters in order for
this truncation to work (which is why we made it one of the $f$
before), and the forces $f$ and coupling constants $\mu$ are also
mixed, and are therefore in practice lumped together. Fixed points of
the renormalization group correspond to Hamiltonians with the
parameters $\mu$ and $f$ at their critical values. The critical values
of many of these parameters will be zero (or infinity), meaning that
the dimensionful parameters $ \mu$ they were originally associated
with are irrelevant. Because a fixed point of the renormalization
group can not have a preferred length scale, the only parameters that
can have nontrivial values are dimensionless.

We make contact with critical phenomena in gravitational collapse when
we consider the renormalization group as a dynamical system. Consider
for now the ferromagnetic material in the absence of an external
magnetic field. (The action of the renormalization group will change
the value of an external field, but cannot generate a nonzero field
from a zero field because the zero field Hamiltonian has higher
symmetry. Therefore zero external field is a consistent
truncation.) With zero external field, we only need to fine-tune one
parameter, the temperature, to its critical value, in order to reach a
critical phase transition. This means that the critical surface is a
hypersurface of codimension one. The behavior of thermodynamical
quantities at the critical point is in general not trivial to
calculate. But the action of the renormalization group on length
scales is given by its definition. The blowup of the correlation
length $\xi$ at the critical point is therefore the easiest critical
exponent to calculate. But the same is true for the black hole mass,
which is just a length scale! 

We can immediately reinterpret the mathematics of Section
\ref{section:typeII} as a calculation of the critical exponent for
$\xi$, substituting the correlation length $\xi$ for the black hole
mass $M$, $T_*-T$ for $p-p_*$, and taking into account that the
$\tau$-evolution in critical collapse is towards smaller scales, while
the renormalization group flow goes towards larger scales: $\xi$
therefore diverges at the critical point, while $M$ vanishes.

In type II critical phenomena in gravitational collapse, we seem at
first to have infinitely many parameters in the initial data. But we
have shown in section \ref{section:phasespace} that we should think of
the black hole mass being controlled by the one global function $P$ on
phase space. Clearly, $P$ is the equivalent of the reduced temperature
$T-T_*$. What then is the second parameter, the equivalent of $\vec B$
or the pressure $p$? We shall suggest in Section \ref{section:angmom}
that in some situations the angular momentum of the initial data can
play this role. Note that like $\bf B$, angular momentum is a vector,
with a critical value that is just zero because all other values break
rotational symmetry. Furthermore, the final black hole can have
nonvanishing angular momentum, which must depend on the angular
momentum of the initial data. The former is analogous to the
magnetization $\bf m$, the latter to the external field $\bf B$.


\section{GR aspects of critical collapse}
\label{section:GR}


In this section we review the geometric definition of self-similarity
in GR, and review how self-similar solutions, and in particular
critical solutions, are obtained numerically. We review analytical
approaches to finding critical solutions, and the use of 2+1 spacetime
dimensions as a toy problem where analytical approaches appear to be
more promising. We focus on three aspects of type II critical collapse
that make them interesting specifically to relativists: the fact that
they force us to consider GR as a dynamical system, the surprising
fact that critical solutions exist when coupled to gravity but do not
have a flat spacetime limit, and the connection between critical
phenomena and naked singularities.


\subsection{Self-similarity in GR}
\label{section:selfsimilarity}


The critical solution found by Choptuik
\cite{Choptuik91,Choptuik92,Choptuik94} for the spherically symmetric
scalar field is scale-periodic, or discretely self-similar (DSS),
while other critical solutions, for example for a spherically
symmetric perfect fluid \cite{EvansColeman} are scale-invariant, or
continuously self-similar (CSS). Even without gravity, continuously
scale-invariant, or self-similar, solutions arise as intermediate
attractors in some fluid dynamics problems
\cite{BarenblattZeldovich,Barenblatt,Barenblatt2}. Discrete
self-similarity also arises in physics. (See \cite{Sornette} for a
review, but note that discrete scale-invariance is defined there only
as a perturbation of continuous scale-invariance.) We begin with the
continuous symmetry because it is simpler.

In Newtonian physics, a solution $Z$ is self-similar if it
is of the form
\begin{equation}
Z({\bf x}, t) = Z\left({{\bf x}\over f(t)}\right)
\end{equation}
If the function $f(t)$ is derived from dimensional analysis
alone, one speaks of self-similarity of the first kind. An example is
$f(t)=\sqrt{\lambda t}$ for the diffusion equation $Z_{,t}=\lambda
\Delta Z$. In more complicated equations, $f(t)$ may contain
additional dimensionful constants (which do not appear in the field
equation) in terms such as $(t/L)^\alpha$, where $\alpha$, called an
anomalous dimension, is not determined by dimensional considerations
but through the solution of an eigenvalue problem. This is called
self-similarity of the second kind \cite{Barenblatt}.

In GR, we can use the freedom to relabel either the space coordinates
${\bf x}$ or the time coordinate $t$ to make $f(t)$ anything we
like. Therefore the notions of self-similarity of the first and second
kinds cannot be applied straightforwardly. The most natural kind of
self-similarity in GR is homotheticity, which is also referred to as
continuous self-similarity (CSS) in the critical collapse literature:
A continuous self-similarity of the spacetime in GR corresponds to the
existence of a homothetic vector field $\xi$, defined by the property
\cite{CahillTaub}
\begin{equation}
\label{homothetic_metric}
{\cal L}_\xi g_{ab} = - 2 g_{ab}.
\end{equation}
This is a special type of conformal Killing vector, namely one with a
constant coefficient on the right-hand side. The value of this
constant coefficient is conventional, and can be set equal to $-2$ by
a constant rescaling of $\xi$.
From (\ref{homothetic_metric}) it follows that
\begin{equation}
\label{homothetic_curvature}
{\cal L}_\xi {R^a}_{bcd} = 0,
\end{equation}
and therefore
\begin{equation}
\label{homothetic_matter}
{\cal L}_\xi G_{ab} = 0,
\end{equation}
but the inverse does not hold: the Riemann tensor and the metric need
not satisfy (\ref{homothetic_curvature}) and (\ref{homothetic_metric})
if the Einstein tensor
obeys (\ref{homothetic_matter}). 

In coordinates $x^\mu=(\tau,x^i)$ adapted to the homothety, the metric
coefficients are of the form
\begin{equation}
\label{CSS_coordinates}
g_{\mu\nu}(\tau,x^i) = l^2 e^{-2\tau} \bar g_{\mu\nu}(x^i),
\end{equation}
where the constant $l$ has dimension length. In these coordinates,
the homothetic vector field is
\begin{equation}
\label{xi_in_coordinates}
\xi = {\partial\over\partial\tau}.
\end{equation}

If one replaces $\tau\equiv x^0$ by $t\equiv -le^{-\tau}$, the metric
becomes
\begin{equation}
\label{CSS_coordinates2}
ds^2=\bar g_{00}\, dt^2 + 2t\,\bar g_{0i} \,dt\,dx^i + t^2\, \bar
g_{ij}\,dx^i\,dx^j 
\end{equation}
If one also replaces the $x^i$ for $i=1,2,3$ by $r^i\equiv (-t)x^i$, the
metric becomes
\begin{equation}
\label{CSS_coordinates3}
ds^2=(\bar g_{00}+2\bar g_{0i}x^i+g_{ij}x^ix^j)\, dt^2 + 2(\bar
g_{0i}+\bar g_{ij}x^j)dt \,dr^i +\bar g_{ij}\,dr^i\,dr^j
\end{equation}
The coordinates $x^i$ and $\tau$ are dimensionless, while $t$ and
$r^i$ have dimension length. Note that all coordinates can be either
spacelike, null, or timelike, as we have not made any assumptions
about the sign of the metric coefficients. Note that in coordinates
$t$ and $r^i$ all metric coefficients still depend only on the
$x^i\equiv r^i/(-t)$.

Most of the literature on CSS solutions is restricted to spherically
symmetric solutions, and it may be worth writing down the three types
of metric in spherical symmetry. The form (\ref{CSS_coordinates})
reduces to
\begin{equation}
ds^2 = l^2e^{-2\tau} \left(A\, d\tau^2 + 2B\,d\tau dx + C\,dx^2 +
F^2\,d\Omega^2\right),
\end{equation}
where $A$, $B$, $C$ and $F$ are functions of $x$ only. The form
(\ref{CSS_coordinates2}) reduces to
\begin{equation}
ds^2 =  A\, dt^2 + 2B\,dt\, dx + t^2 C\,dx^2 +
t^2 F^2\,d\Omega^2,
\end{equation}
and the form (\ref{CSS_coordinates2}) reduces to
\begin{equation}
ds^2 =  (A+2xB+x^2C)\, dt^2 + 2(B+xC)\,dt\, dr + C\,dr^2 +
t^2 F^2\,d\Omega^2,
\end{equation}
where the coefficients in round brackets, $C$ and $F^2$ still only
depend on $x\equiv r/(-t)$. 

The generalization to a discrete self-similarity is obvious in the
coordinates (\ref{CSS_coordinates}), and was made in
\cite{Gundlach_Chop2}:
\begin{equation}
\label{DSS_coordinates}
g_{\mu\nu}(\tau,x^i) = e^{-2\tau} \bar g_{\mu\nu}(\tau,x^i), \quad
\hbox{where} \quad \bar g_{\mu\nu}(\tau,x^i) = \bar 
g_{\mu\nu}(\tau+\Delta,x^i).
\end{equation}
The conformal metric $\bar g_{\mu\nu}$ does now depend on $\tau$,
but only in a periodic manner. Like the continuous symmetry, the
discrete version has a geometric definition
\cite{Gibbonspc}: A spacetime is discretely self-similar if 
there exists a discrete diffeomorphism $\Phi$ and a real constant
$\Delta$ such that
\begin{equation} 
\label{DSS_geometric}
\Phi^* g_{ab} = e^{-2\Delta} g_{ab},
\end{equation}
where $\Phi^* g_{ab}$ is the pull-back of $g_{ab}$ under the
diffeomorphism $\Phi$. Note that this definition does not introduce a
vector field $\xi$. Note also that $\Delta$ is dimensionless and
coordinate-independent.

One simple coordinate transformation that brings the
Schwarzschild-like coordinates (\ref{tr_metric}) into the form
(\ref{DSS_coordinates}) was given above in Eq. (\ref{x_tau}), as one easily
verifies by substitution. The general spherically symmetric metric in
these coordinates becomes
\begin{equation}
\label{SchwarzschildDSS}
ds^2=l^2e^{-2\tau}\left[-\alpha^2\,d\tau^2+a^2(dx-x\,d\tau)^2
+x^2\,d\Omega^2\right].
\end{equation}
Note that that surfaces of constant $\tau$ are everywhere spacelike,
but that the vector $\partial/\partial \tau$ becomes spacelike at
large $x$ because of the presence of a shift. The spacetime is CSS if
$a$ and $\alpha$ depend only on $x$, and is DSS if they
are periodic functions of $\tau$.

It should be stressed here that the coordinate systems adapted to CSS
(\ref{CSS_coordinates}) or DSS (\ref{DSS_coordinates}) form large
classes, even in spherical symmetry. In CSS, the freedom is
parameterized by the choice of a surface $\tau=0$, and of coordinates
$x^i$ on it. In DSS, one can fix a surface $\tau=0$ arbitrarily. One
can then foliate the spacetime between $\tau=0$ and its image
$\tau=\Delta$ freely, and cross them with arbitrary coordinates $x^i$,
subject only to the requirement that they are the same on $\tau=0$ and
$\tau=\Delta$. The $\tau$-surfaces can be chosen to be spacelike, as
for example defined by (\ref{tr_metric}) and (\ref{x_tau}) above, and
in this case the coordinate system cannot be global.  Alternatively,
one can find global coordinate systems, where $\tau$-surfaces must
become spacelike at large $r$.

If the matter is a perfect fluid with stress-energy tensor
\begin{equation}
\label{fluid_stress_energy}
T_{ab} = (p+\rho) u_a u_b + p g_{ab},
\end{equation} 
it follows from (\ref{homothetic_metric}), (\ref{homothetic_matter})
and the Einstein equations that
\begin{equation}
{\cal L}_\xi u^a = u^a, \quad {\cal L}_\xi \rho = 2\rho, \quad
{\cal L}_\xi p = 2p.
\end{equation}
Therefore only the equation of state $p=k\rho$, where $k$ is a
constant, allows for CSS solutions. In CSS coordinates, the direction
of $u^a$ depends only on $x$, and the density is of the form
\begin{equation}
\rho(x,\tau) = e^{2\tau} \bar\rho(x).
\end{equation}
Similarly, if the matter is a massless scalar field $\phi$, with
stress-energy tensor (\ref{scalar_stress_energy}), it follows from the
Einstein equations that
\begin{equation}
\label{scalar_CSS}
{\cal L}_\xi \phi=\kappa, 
\end{equation}
where $\kappa$ is a constant. The most general CSS solution is therefore
\begin{equation}
\phi=f(x)+\kappa\tau.
\end{equation}
The most general DSS solution is
\begin{equation}
\label{scalar_DSS}
\phi=f(\tau,x^i) + \kappa \tau, \quad {\rm where} \quad
f(\tau,x^i)=f(\tau+\Delta,x^i).
\end{equation}
In the Choptuik critical solution, $\kappa=0$ for unknown reasons.

Finally, two remarks on terminology.  In a possible source of
confusion, Evans and Coleman \cite{EvansColeman} use the term
``self-similarity of the second kind'', because they define their
self-similar coordinate $x$ as $x=r/f(t)$, with
$f(t)=t^\alpha$. Nevertheless, the spacetime they calculate is
homothetic. The difference is only a coordinate transformation: the
$t$ of \cite{EvansColeman} is not proper time at the origin, but what
would be proper time at infinity if the spacetime was truncated at
finite radius and matched to an asymptotically flat exterior
\cite{Evanspc}.

In yet another possible source of confusion, Carter and Henriksen
\cite{CarterHenriksen,Coley} reserve the term self-similarity of the
first kind for homotheticity. By self-similarity of the second kind
they understand breaking homotheticity by introducing a preferred
time-slicing. One can then rescale proper space and proper time in the
preferred frame by different powers of the rescaling factor, as in the
Newtonian sense of the term self-similarity of the second kind. This
symmetry does not seem to occur in critical collapse, or to be
relevant in physical situations.


\subsection{Calculations of critical solutions}
\label{section:DSScalculations}


Critical phenomena were originally discovered by evolving initial
data, and fine-tuning those data to the black hole threshold. This
approach continues for new systems in spherical symmetry, but remains
a major challenge for numerical relativity in axisymmetry or the full
3D case. A different approach is to construct a candidate critical
solution by demanding that it be CSS or DSS and suitably regular, and
then to construct its perturbation spectrum in order to see if it has
only one growing mode. The assumption of CSS, by introducing a
continuous symmetry, reduces the number of coordinates on which the
fields depend by one.  The combination of CSS and spherical symmetry,
in particular, reduces the field equations to a system of ODEs. In
this ansatz, the requirement of analyticity at the center and at the
past matter characteristic of the singularity provides sufficient
boundary conditions for the ODE system. While some solutions of these
equations may be derived in closed form, all known CSS critical
solutions are only known as numerical solutions of the ODE boundary
value problem. (However, it may be possible to calculate a critical
solution in 2+1 spacetime dimensions in closed form, see Section
\ref{section:2+1}.)

The ansatz of CSS with regularity at both the center and the past
sound cone was pioneered by Evans and Coleman \cite{EvansColeman} for
the perfect fluid with the equation of state $p=\rho/3$. The solution
they found is ingoing near the center, and outgoing at large radius,
so that there is exactly one zero of the radial velocity (besides the
one at the center). Later, this solution was constructed for the range
$0<k<0.89$, and it was shown that it has again precisely one growing
mode \cite{Maison,KoikeHaraAdachi3}. It was claimed that the
generalized Evans-Coleman solution does not exist for $k>0.89$, but
this was found to be an artifact of the numerical method
\cite{NeilsenChoptuik}. Harada has found that the Evans-Coleman
solution for $k>0.89$ has an additional unstable mode in which the
density gradient is discontinuous at the sound cone, the ``kink
instability'' \cite{Harada2}. The perturbative result of Harada is
compatible with the perturbative result of Gundlach
\cite{Gundlach_critfluid2}, who considered only analytic perturbation
modes. At the same time, Neilsen and Choptuik found the Evans-Coleman
solution as the critical solution at the threshold of black hole
formation in numerical evolutions for the entire range $0<1<k$,
including $0.89<k<1$ \cite{NeilsenChoptuik} (see also the collapse
simulations announced in \cite{BradyCai}). Harada \cite{Harada2} has
speculated why the kink instability is not seen in time evolutions.

\begin{figure}
\epsfysize=14cm
\centerline{\epsffile{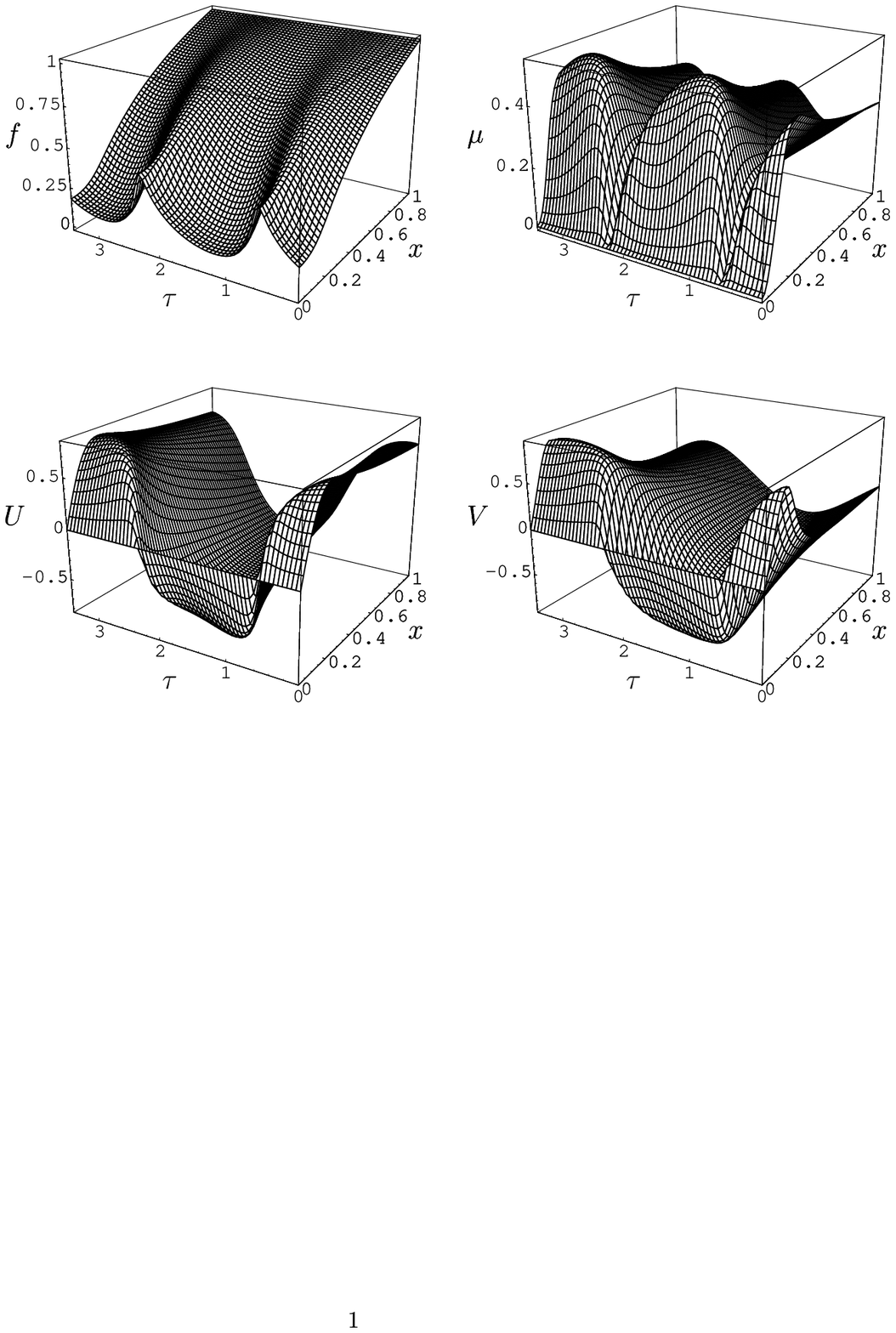}}
\caption{
\label{fig:choptuon}
Choptuik's critical solution in coordinates adapted to DSS.  The
first-order matter variables $U$ and $V$ are defined in
(\ref{UVdef}). $\mu=2m/r$ where $m$ is the Hawking mass defined in
(\ref{mdef}) and $r$ the area radius defined in
(\ref{tr_metric}). $\mu=1$ therefore signals an apparent
horizon. $f=\alpha/a$ where $\alpha$ and $a$ are defined in
(\ref{tr_metric}). The two axes are $0\le x\le 1$ and $0\le \tau\le
\Delta\simeq 3.44$, where $x$ and $\tau$ are defined in
(\ref{x_tau}). In particular $x=0$ is the center of spherical symmetry
and $x=1$ is the past lightcone of the singularity. Note that the
period of the matter variables is $\Delta$ with
$U(x,\tau+\Delta/2)=-U(x,\tau)$, while that of the metric variables is
$\Delta/2$.  This figure is taken from \cite{critcont}.
} 
\end{figure}

\begin{figure}
\epsfysize=14cm
\centerline{\epsffile{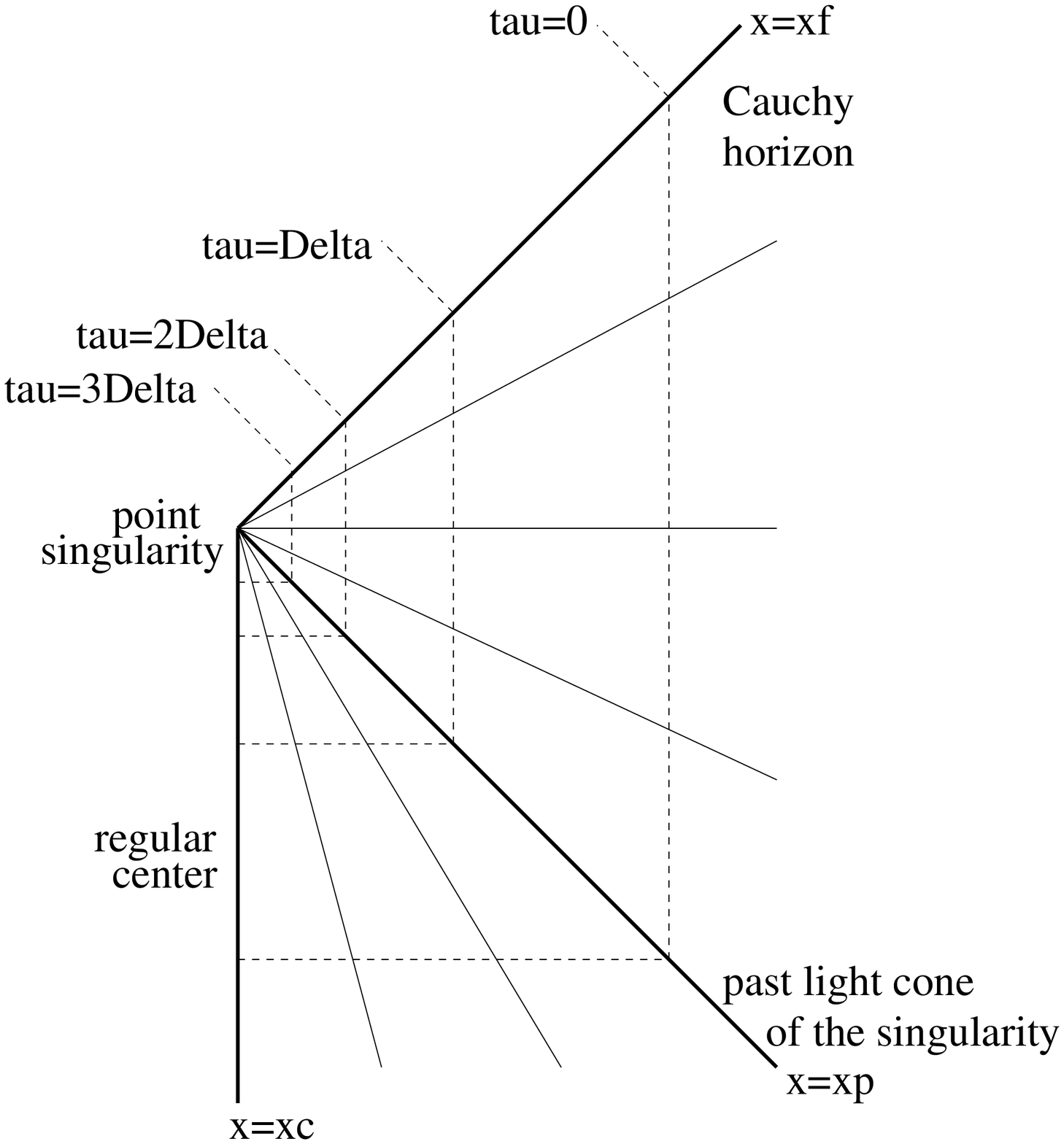}}
\caption{
\label{fig:preCH}
Spacetime diagram of spherically symmetric critical solution (for
example Choptuik's solution for the scalar field) with schematic
indication of coordinate patches. Here the spacetime from the regular
center $x=x_c$ to the past lightcone $x=x_p$ of the singularity has
been covered with a coordinate patch where the surfaces of constant
$\tau$ are spacelike (eg $\tau=-\ln(-t)$). The region between the past
lightcone and the future lightcone (Cauchy horizon) $x=x_f$ has been covered
with a patch where the $\tau$-surfaces are timelike (eg $\tau=-\ln
r$). A part of the region to the future lightcone has been covered
with a patch where the $\tau$-lines are ingoing null lines (eg
$\tau=-\ln v$). The spoke-like lines are $x$-lines. This figure is
taken from \cite{critcont}.  }
\end{figure}

The DSS scalar critical solution of scalar field collapse was
constructed as a boundary value problem by Gundlach
\cite{Gundlach_Chop1,Gundlach_Chop2} and Mart\'\i n-Garc\'\i a and
Gundlach \cite{critcont}. High-precision numerical results are shown
in Fig.~\ref{fig:choptuon}. Unlike a CSS ansatz, a DSS ansatz does not
reduce the equations to ODEs: because the symmetry is discrete, one
still has to solve a system of equations in 1+1 dimensions, with
periodic boundary conditions. The advantage of an exact DSS ansatz
over fine-tuning in the initial data is rather that the time evolution
method requires adaptive mesh refinement techniques in order to follow
the critical solution over many orders of magnitude, while the DSS
ansatz incorporates the scale-invariance as an explicit periodicity. A
DSS ansatz was also used for the spherically symmetric $SU(2)$
Yang-Mills field \cite{Gundlach_EYM}. A possible choice of coordinates
for a spherically symmetric critical solution is shown in
Fig.~\ref{fig:preCH}. 


A CSS or DSS solution is not known to be a critical solution until it
has been shown that it has only one growing perturbation mode. An
example for this is provided by the spherically symmetric massless
complex scalar field.  Hirschmann and Eardley \cite{HE1} found a
regular CSS solution but later discovered that it has three unstable
modes \cite{HE2}.

One important aspect of known type II critical solutions is that they
have no equivalent in the limit of vanishing gravity. It is only the
coupling to gravity that allows regular CSS and DSS solutions to exist.

What kind of regularity must a type II critical solution possess?  It
arises from the time evolution of smooth initial data, and nothing
prevents us from making the initial data analytic.  The critical
solution should therefore itself be analytic at all points that cannot
be causally influenced by the singularity. In particular, the past
light cone of the singularity should not be less regular than a
generic point. But self-similar spherically symmetric matter fields in
flat spacetime are singular either at the center of spherical symmetry
(to the past of the singularity), or at the past characteristic cone
of the singularity. Adding gravity makes solutions possible that are
regular at both places. These regular solutions are isolated, and are
therefore determined by imposing regularity at the center and the past
light cone.

We consider again the spherically symmetric scalar field, not only
because it has been our main example throughout, but also because the
critical solution is DSS, which is more general than CSS. The general
solution in flat spacetime is 
\begin{equation} 
\phi(r,t) = {f(t+r)-g(t-r)\over r},
\end{equation}
where $f(z)$ and $g(z)$ are two free functions of one variable ranging
from $-\infty$ to $\infty$. $f$ describes ingoing and $g$ outgoing
waves. Regularity at the center $r=0$ for all $t$ requires $f(z)=g(z)$
for $f(z)$ a smooth function. Physically this means that ingoing waves
move through the center and become outgoing waves. In order to study
self-similar solutions, we transform the general solution to the
self-similarity coordinates $x$ and $\tau$ introduced in
Eq. (\ref{x_tau}). For $t<-r$, that is in the past of the point
$t=r=0$, it can be written as
\begin{equation}
\label{DSSflat}
\phi(x,\tau) = {1-x\over x}F\left[\tau-\ln(1-x)\right]
-{1+x\over x}G\left[\tau-\ln(1+x)\right],
\end{equation}
where $F$ is related to $f$ and $G$ to $g$. Continuous self-similarity
$\phi=\phi(x)$ requires $F$ and $G$ to be constant. Discrete
self-similarity requires them to be periodic in their argument with
period $\Delta$. Regularity at the center $r=0$ for $t<0$, which is
$x=0$, requires $F=G$. Regularity at the past light cone $t=-r$ for
$t<0$, which is $x=1$, requires $F=0$. We see that a CSS or DSS solution
cannot be regular both at the past center and the past light cone,
unless it is the trivial solution $\phi=0$. Extending this argument to
two more coordinate patches, we can show that the solution can only be
regular in one of the four places: past center, past light cone, future
light cone and future center.

The presence of gravity changes this singularity structure
qualitatively. We shall now see that self-similar solutions can in
principle be regular both at the past center and past light cone.  We
can use the spacetime metric (\ref{SchwarzschildDSS}) with $x$ and
$\tau$ given by (\ref{x_tau}). For our discussion we only need the
scalar field equations. We introduce as first-order matter variables
the scale-invariant null derivatives of the scalar field,
\begin{equation}
\label{UVdef}
U,V = r\left({a\over
\alpha}\phi_{,t}\mp\phi_{,r} \right),
\end{equation}
which describe outgoing and ingoing waves. In the curved spacetime
wave equation we use the Einstein equations (\ref{dalpha_dr}) and
(\ref{da_dt}) to replace the metric derivatives $a_{,t}$ and
$\alpha_{,r}$ by stress-energy terms that are quadratic in $U$ and $V$. We
obtain 
\begin{eqnarray} 
\nonumber
U_{,x}&=&{f[(1-a^2)U+V]-{xU_{,\tau}}\over x(f+x)}, \\
V_{,x}&=&{f[(1-a^2)V+U]+{xV_{,\tau}}\over x(f-x)},
\label{DSS_wave}
\end{eqnarray}
where $f\equiv \alpha/a$.  The denominator of the $V_{,x}$ equation
vanishes at the past light cone $f-x=0$. We use the residual gauge
freedom $t\to t'(t)$ of that form of the metric to fix the past
lightcone of the point $r=t=0$ at $r=-t$, so that we have $f=1$ at
the past lightcone, which is at $x=1$. (Note that in this convention
$\alpha(0,t)\ne 1$, while Refs. \cite{Choptuik92,Gundlach_Chop2} use
the convention $\alpha(0,t)=1$.) $x=1$ is therefore a singular point
of the PDE system that we are trying to solve. (Similarly, the
denominator of the $U_{,x}$ equation vanishes at the future light
cone.)

We now consider the behavior of solutions at the past light cone. In
flat spacetime, where $f=a=1$, the second of Eqs. (\ref{DSS_wave})
reduces to the linear equation
\begin{equation}
\label{Vflat}
V_{,x}={U+{xV_{,\tau}}\over x(1-x)}.
\end{equation}
The general solution at the singular point $x=1$ can be written as the
sum of a regular and a singular solution. From (\ref{Vflat}) we see
that in the limit $x\to 1_-$, $V$ in the singular solution has the
form
\begin{equation}
V\simeq F\left[\tau-\ln(1-x)\right],
\end{equation}
where $F(z)$ is periodic with period $\Delta$. This is singular
because it oscillates infinitely many times as $x\to 1_-$. (We have
written this as an approximation because we have derived it using only
the $V$ equation instead of the complete wave equation, neglecting the
term $U$ in (\ref{Vflat}). In flat spacetime this solution is in fact
exact, compare (\ref{DSSflat}).) In the presence of gravity, however,
the numerator of the Eq. (\ref{DSS_wave}) contains the additional term
proportional to $(1-a^2)$, which is due to the curvature of spacetime
induced by the stress-energy of the scalar field. We can use this
additional term to impose the condition that the numerator of the $V$
equation vanishes at the light cone. We then find that both numerator
and denominator of this equation vanish as $O(x-1)$, and the solution
becomes analytic in a neighborhood of the light cone.

The condition that the numerator of the curved-space $V_{,x}$ equation
vanishes at $x=1$,
\begin{equation}
\label{reg_condition}
(1-a^2)V+U+V_{,\tau}=0,
\end{equation}
(recall that by gauge fixing $f=1$ at $x=1$) is an ODE for $V$ at the
past light cone, with periodic boundary conditions in $\tau$, for
given $U$ and $a$ at the past lightcone. 

We obtain a 1+1 dimensional mixed hyperbolic-elliptic boundary value
problem on the coordinate square $0\le x\le1$, $0\le \tau<\Delta$,
with regularity conditions at $x=0$ and $x=1$, and periodic boundary
conditions in $\tau$. (In a CSS ansatz in spherical symmetry, all
fields depend only on $x$, and one obtains an ODE boundary value
problem.) Well-behaved numerical solutions of these problems have been
obtained, and agree well with the critical solution found in collapse
simulations \cite{Gundlach_Chop2,critcont}. It remains an open
mathematical problem to prove existence and (local) uniqueness of
these numerical solutions.


\subsection{Analytical approaches}
\label{section:analyticapproaches}


Spherically symmetric CSS (but not DSS) solutions in GR have also been
studied independently of critical collapse. A large body of research
on spherically symmetric self-similar perfect fluid solutions with the
equation of state $p=k\rho$ predates Choptuik's discoveries and the
work of Evans and Coleman on critical fluid collapse
\cite{CarrColey,Bogoyavlenskii,FoglizzoHenriksen,BicknellHenriksen,OriPiran,OriPiran2,LakeZannias}.
In these papers, the Einstein equations are reduced to an ODE system
by the self-similar spherically symmetric ansatz, which is then
discussed as a dynamical system. Particularly interest has been paid
to the possible continuations through the sound cone of the
singularity (usually referred to as the sonic point), and the strength
and global or local nakedness of the singularity.

Perhaps the most interesting result of these studies is that naked
singularities appear to be generic not only for dust, but even for a
perfect fluid with pressure. For $k<0.036$ a CSS solution exists which
is analytic at the past sound cone and which is purely ingoing. For
$k<0.0105$ it contains a naked singularity \cite{OriPiran}. Harada and
Maeda have found that for this range it is also an attractor
\cite{HaradaMaeda,Harada}. The critical solution for this range
therefore probably sits at the boundary between dispersion and naked
singularity formation (rather than black hole formation.) The
Newtonian limit of the self-similar perfect fluid exists, and
corresponds to $k\to 0$ \cite{OriPiran}. Therefore the results of
Harada and Maeda extend to the Newtonian limit \cite{HaradaMaeda2}.

The critical solutions of perfect fluid collapse were not singled out
in these surveys of CSS solutions until after they had been discussed
in collapse simulations by Evans and Coleman. The position of the
critical solution in the general classification remained confused for
several years. It seems to have been settled now
\cite{CarrColeyGoliathNilssonUggla,CarrColey2}.The Evans-Coleman
solution is the unique solution that is analytic at the center and at
the sound cone, is ingoing near the center, and outgoing everywhere
else. The new classification uses a dynamical systems approach to the
ODEs that arise from the assumptions of spherical symmetry and
homotheticity. It combines results found in coordinates adapted to the
fluid, and coordinates adapted to the homotheticity. Generalizing
previous work \cite{OriPiran,Nolan}, Carr and Gundlach
\cite{CarrGundlach} have constructed the conformal diagrams for all
these solutions, including the critical solutions.

Scalar field spherically symmetric CSS solutions have been examined in
\cite{GoldwirthPiran,Brady_CSS_scalar}, with the purpose of studying
the formation of naked singularities from regular initial data. A
discrete subset of these solutions has the required regularity at the
center and the past light cone of the singularity. These solutions are
not critical solutions because they have many growing perturbation
modes.

A number of authors have attempted to throw light on critical collapse
with the help of analytic CSS solutions, even though they are not the
known critical solution. The 1-parameter family of exact self-similar
real massless scalar field solutions first discovered by Roberts
\cite{Roberts} is presented in Section \ref{section:singularity}. It
has been discussed in the context of critical collapse in
\cite{Brady_Roberts,Oshiro_Roberts}, and later
\cite{WangOliveira,Burko}. The solution can be given in double null
coordinates as
\begin{eqnarray}
\label{Roberts1}
ds^2 & = & - du\,dv + r^2(u,v) \,d\Omega^2, \\ r^2(u,v) & = & {1\over
4}\left[(1-p^2)v^2 - 2vu + u^2\right], \\
\label{Roberts3}
\phi(u,v) & = & {1\over 2} \ln {(1-p)v - u\over (1+p) v - u},
\end{eqnarray}
with $p$ a constant parameter. (Units $G=c=1$.) Two important
curvature indicators, the Ricci scalar and the Hawking mass, are
\begin{equation}
R={p^2 uv\over 2r^4}, \quad m = - {p^2 uv\over 8r}.
\end{equation}
The center $r=0$ has two branches, $u=(1+p)v$ in the past of $u=v=0$,
and $u=(1-p)v$ in the future.  For $0<p<1$ these are timelike
curvature singularities. The singularities have negative mass, and the
Hawking mass is negative inside the past and future light cones. One
can cut these regions out and replace them by Minkowski space without
creating a singular stress-energy tensor. The resulting spacetime
resembles the critical spacetimes arising in gravitational collapse in
some respects: it is self-similar, has a regular center $r=0$ at the
past of the curvature singularity $u=v=0$ and is continuous at the
past light cone. It is also continuous at the future light cone, and
the future branch of $r=0$ is again regular.

Roberts solutions with $p>1$ can be considered as black holes if they
are truncated in a suitable manner, and to leading order around the
critical value $p=1$, their mass is then $M\sim(p-1)^{1/2}$. The
pitfall in this approach is that only perturbations within the
self-similar family are considered, so the formal critical exponent
applies only to this one, very special, family of initial data. But
the $p=1$ solution has many growing perturbations which are
spherically symmetric (but not self-similar), and is therefore not a
critical solution in the sense of being an attractor of codimension
one. This was already clear because it did not appear in collapse
simulations at the black hole threshold, but in a tour de force Frolov
has calculated the perturbation spectrum analytically for both
spherical and nonspherical perturbations \cite{Frolov,Frolov3}. The
eigenvalues of spherically symmetric perturbations fill a sector of
the complex plane, with ${\rm Re}\lambda\le1$. Interestingly, all
nonspherical perturbations decay.

Frolov \cite{Frolov4} has suggested approximating the Choptuik
solution as the critical ($p=1$) Roberts solution, which has an
outgoing null singularity, plus its most rapidly growing (spherical)
perturbation mode, pointing out that this perturbation oscillates in
$\tau$ with a period $4.44$, but ignoring the fact that it also grows
exponentially. Hayward \cite{Hayward} and Clement and Fabbri
\cite{ClementFabbri1,ClementFabbri2} have also proposed critical
solutions with a null singularity, and have attempted to construct
black hole solutions from their linear perturbations. This is probably
irrelevant to critical collapse, as the critical spacetime does not
have an outgoing null singularity. The singularity is naked but first
appears in a point. The future light cone of that point is not a null
singularity but a Cauchy horizon with finite curvature. Beyond the
Cauchy horizon the solution is not unique. At least in the case of the
perfect fluid critical solution, the continuation may be so that the
singularity is a single spacetime point, an ingoing null singularity,
or a timelike singularity \cite{CarrGundlach}.

Other authors have attempted analytic approximations to the Choptuik
solution.  Pullin \cite{Pullin_Chop} has suggested describing critical
collapse approximately as a perturbation of the Schwarzschild
spacetime. Price and Pullin \cite{PricePullin} have approximated the
Choptuik solution by two flat space solutions of the scalar wave
equation that are matched at a ``transition edge'' at constant
self-similarity coordinate $x$. The nonlinearity of the gravitational
field comes in through the matching procedure, and its details are
claimed to provide an estimate of the echoing period $\Delta$.

Horowitz and Hubeny \cite{HorowitzHubeny} have pointed out a relation
between mass scaling in critical collapse and the quasi-normal modes
of black holes in string theory that they qualify as probably a
numerical coincidence. Consider black hole solutions in supergravity
in $D$ spacetime dimensions on an anti-deSitter (AdS) background in
which $d$ spacetime dimensions are extended. (The physical cases are
$D=10,11$ and $d=4$.) Such a solution has two independent scales: the
horizon radius $r_+$ and the cosmological radius $R$ that is related
to the cosmological constant as $\Lambda=R^{-2}$. The imaginary part
(damping rate) $\omega_{\rm Im}$ of the dominant quasinormal mode is
proportional to the Hawking temperature $T=T(R,r_+)$ for large black
holes ($r_+\gg R$). For small black holes ($r_+\ll R$), when the
cosmological constant can in some ways be neglected, it is
approximately proportional to the horizon radius, $\omega_{\rm
Im}\simeq \lambda r_+$. The numerical coincidence is that in the
physical case $d=4$ the proportionality constant $\lambda\simeq 2.67$
is similar to the growth rate of the one growing mode of the Choptuik
critical solution, $\lambda=1/\gamma\simeq 2.67$. If this is not just
a coincidence, the reason is certainly not understood, even on the
level of dimensional analysis. Furthermore, the proportionality breaks
down as $r_+\to 0$, and in $d=6$, the proportionality constant is not
the Choptuik exponent.

Motivated by this coincidence, Birmingham \cite{BirminghamAdS} has
calculated the quasinormal modes of the BTZ black hole in
2+1-dimensional AdS spacetime, and has found that $\omega_{\rm Im}=
(1/\gamma) r_+$ is exact in this case if one sets $\gamma=1/2$, the
value obtained by Birmingham and Sen (see above).


\subsection{2+1 spacetime dimensions}
\label{section:2+1}


General relativity in 2+1 spacetime dimensions is quite different from
3+1 dimensions. Spacetime in 2+1 dimensions is flat everywhere where
there is no matter, so that gravity is not acting at a distance in the
usual way. There are no gravitational waves. Black holes can only be
formed in the presence of a negative cosmological constant, so that
the spacetimes are asymptotically anti-de Sitter rather than
asymptotically flat. A negative cosmological constant in 2+1
dimensions has three important effects. It locally introduces a length
scale scale $l=(-\Lambda)^{-1/2}$ into the field equations. It also
changes the global structure of the spacetime. Null infinity becomes a
timelike surface, which is at infinite distance from the center and
infinite circumference radius $\bar r$, but a null ray can go out from
the center to null infinity and return while a finite proper time $\pi
l$ passes at the center. The only consistent boundary conditions for
the massless wave equations there are Dirichlet boundary
conditions. This means that all outgoing scalar waves are reflected
back to the center in a finite time. Finally, the presence of a
cosmological constant allows black hole solutions
\cite{BTZ,BTZ2}. (See \cite{Carlip} for a review.) The static,
circularly symmetric solutions with negative cosmological constant can
be written as
\begin{equation}
ds^2=-(-M+\bar r^2/l^2)\,dt^2 + (-M+\bar r^2/l^2)^{-1}\,dr^2 + \bar
r^2 \,d\theta^2,
\end{equation}
where the constant $M$ takes values $-1\le M<\infty$. $M=-1$ is
anti-de Sitter space, with a regular center. Solutions with $-1<M<0$
have point particle naked singularities at the center. Solutions with
$M\ge 0$ are black holes. The black hole horizon is at $\bar
r=l\sqrt{M}$. 

Birmingham and Sen \cite{BirminghamSen} have considered the formation
of a black hole from the collision of two point particles of equal
mass in 2+1 gravity with a negative cosmological constant. The initial
data are parameterized by the impact parameter and the speed of the
particles, and the black hole mass can be calculated in closed
form. Near the threshold of black hole formation, the black hole mass
is $M\simeq \sqrt{P}$ where $P$ is a known function of the two
parameters. The critical exponent $1/2$ is therefore universal within
this system. However, its phase space is only 2-dimensional.  Peleg
and Steif \cite{PelegSteif} have investigated the collapse of a dust
ring in 2+1 with $\Lambda<0$, where the space of initial data is also
2-dimensional, and also find $\gamma=1/2$. In both these examples no
CSS solution is involved, and the notion of black hole mass is quite
different from that in 3+1 dimensions.

Scalar field collapse in spherical symmetry in 2+1 dimensions with a
negative cosmological constant was investigated by Pretorius and
Choptuik (PC) \cite{PretoriusChoptuik}, and independently by Husain
and Olivier (HO) \cite{HusainOlivier}. The cosmological constant
introduces reflecting (Dirichlet) boundary conditions for the scalar
matter field, so that all matter must eventually fall into the black
hole. In that sense, the black hole mass is simply the asymptotic mass
of the spacetime, and there is no black hole threshold.  However, if
one is interested in 2+1 dimensional scalar field collapse mainly as a
toy model for 3+1 dimensions, then it is natural to use initial data
with compact support, to define the black hole threshold as the
formation of an apparent horizon before the scalar waves are reflected
off scri for the first time, and to define the {\it initial} black
hole mass as $M_{\rm AH}=\bar r^2_{\rm AH}/l^2$. (An apparent horizon in this
situation is a surface where the gradient of $\bar r$ is null.) One is
then playing a game similar to that in 3+1. Furthermore, if one can
achieve $M\ll 1$, then the cosmological constant should be negligible
on the relevant spacetime scales, and the dynamics should be
approximately scale-invariant, allowing in principle for type II
critical phenomena. This is indeed the case.

PC evolve the spacetime on spacelike slices that reach from the center
all the way out to (timelike) null infinity, using a free evolution
scheme, and double null coordinates to describe the metric. They
explicitly implement Dirichlet boundary conditions at null
infinity. HO evolve the spacetime on outgoing null cones, using a
fully constrained evolution scheme, and Bondi coordinates to describe
the metric. 

PC evolve several families of initial data whose length scale is $\bar
r_0\simeq 0.32\,l$. Therefore the effects of the cosmological constant
are already suppressed by a factor $\bar r^2/l^2\simeq 0.1$. PC find
type II critical phenomena with a universal CSS critical
solution. They find that the maximum value of the Ricci scalar scales
with the initial data amplitude $P$ as $R_{\rm max}\sim
(P-P_*)^{2\gamma}$, where $\gamma=1.20\pm 0.05$. They also roughly
observe apparent horizon mass scaling $M_{\rm AH}\sim
(P-P_*)^{2\gamma}$, with approximately the same critical exponent.

HO evolve one family of initial data on a scale $\bar r_0\simeq l$,
and find apparent horizon mass scaling with $\gamma\simeq 0.81$. Their
accuracy appears to be much lower than that of PC.

Garfinkle \cite{Garfinkle2+1} has found a 1-parameter family of exact
spherically symmetric CSS solutions for a massless scalar field with
$\Lambda=0$. The requirement that the solution is analytic restricts
the real parameter $q$ to positive integer values. He finds that the
$q=4$ solution is a good match to the critical solution found by PC in
numerical evolutions inside its past light cone. Outside the lightcone
the coincidence appears to be less accurate. Furthermore, the
lightcone of Garfinkle's solution is an apparent horizon, and all
spheres outside it are closed trapped surfaces. This seems to
contradict the assumption that the critical surface is at the
threshold of black hole formation. 

Garfinkle and Gundlach \cite{GarfinkleGundlach2+1} have calculated the
perturbation spectrum of Garfinkle's solutions in closed form with the
assumption that the perturbations are analytic. They find that the
$q=4$ solution has three growing modes, not one. However, the dominant
mode has $\lambda=7/8$ which, if it was the only growing mode, would
suggest a critical exponent of $\gamma=1/\lambda=8/7\simeq 1.14$, in
good agreement with PC. Hirschmann, Wang and Wu
\cite{HirschmannWangWu} impose an (unmotivated) additional
boundary condition at the past light cone on the perturbations, which
suppresses two of the three growing modes of Garfinkle and
Gundlach. However, the mode left over has $\lambda=1/4$, which disagrees
with the critical exponent of PC.

Cl\'ement and Fabbri \cite{ClementFabbri1,ClementFabbri2} have found
another 1-parameter family of exact spherically symmetric CSS scalar
field solutions for $\Lambda=0$ and have generalized them to a family
of numerical $\Lambda<0$ solutions that are asymptotically CSS near a
center. They interpret these as toy models for the critical solution,
and have calculated their perturbation spectrum.


\subsection{Th GR time evolution as a dynamical system}
\label{section:dynsim}


It has been pointed out by Argyres \cite{Argyres}, Koike, Hara and
Adachi \cite{KoikeHaraAdachi,KoikeHaraAdachi2} and others that the
time evolution near the critical solution can be considered as a
renormalization group flow on the space of initial data. For simple
parabolic or hyperbolic differential equations, one can in fact define
a discrete renormalization (semi)group acting on their solutions
\cite{Goldenfeld,BricmontKupiainen,ChenGoldenfeld,ChenGoldenfeldOOno}:
evolve initial data over a certain finite time interval, then rescale
the final data. Solutions which are fixed points under this
transformation are scale-invariant, and are often attractors or
critical points. For a general review of renormalization group ideas
in physics, see \cite{Wilson}.

One nice distinctive feature of GR in contrast to these simple models
is that one can use the coordinate freedom in GR to incorporate the
rescaling into the time evolution, by means of a converging shift
vector, and to make rescaling by a constant factor an evolution
through a constant time interval, by an appropriate choice of the
lapse. (This leads us to coordinates of the type $(\tau,x)$.) By a
suitable choice of the lapse and shift we can therefore turn the GR
time evolution into a dynamical system with the property that its
fixed points are CSS spacetimes and its limit cycles are DSS
spacetimes, the critical solutions for type II critical
phenomena. Alternatively, we can choose the lapse and shift so that
fixed points are stationary solutions, and its limit cycles periodic
solutions, that is, the critical solutions for type I critical
phenomena. (This requires coordinates of the type $(t,r)$.)  There are
a number of important problems associated with such a formulation.


\subsubsection{The choice of lapse and shift}


The phase space of GR is the space of pairs of 3-metrics and
extrinsic curvatures (plus any matter variables) that obey the
Hamiltonian and momentum constraints. In the following we restrict
ourselves to asymptotically flat data, that is, the space of initial
data for an isolated self-gravitating system. The evolution equations
of the dynamical system are in principle the ADM equations, but these
contain the lapse and shift as free fields that can be given arbitrary
values. In order to obtain an autonomous dynamical system, one needs a
general prescription that provides a lapse and shift for given initial
data.

The lapse and shift can be thought of as infinitesimal generators of
the coordinate freedom of GR while the spacetime is being evolved from
Cauchy data. In a relaxed notation, one can write the ADM equations as
$(\dot h,\dot K)=F (h,K,\alpha,\beta)$, where $h_{ij}$ is the
3-metric, $K_{ij}$ the extrinsic curvature, $\alpha$ the lapse and
$\beta^i$ the shift, and $F$ is a nonlinear second-order differential
operator which is linear in $\alpha$ and $\beta^i$. The lapse and
shift can be set freely, independently of the initial data. They
influence only the coordinates on the spacetime, not the spacetime
itself. We need to specify a prescription $(\alpha,\beta)=F (h,K)$ and
substitute it into the ADM equations to obtain $(\dot h,\dot
K)=F(h,K)$, which is an (infinite-dimensional) dynamical system.

We are faced with the general question: given initial data in GR,
is there a prescription for the lapse and shift such that, if these
are in fact data for a self-similar solution, the resulting time
evolution actively drives the metric to the special form
(\ref{DSS_coordinates}) that explicitly displays the self-similarity?
An algebraic prescription for the lapse suggested by Garfinkle
\cite{Garfinkle2} did not work, but maximal slicing with zero shift
does work if combined with a manual rescaling of space
\cite{GarfinkleMeyer}.

In a more systematic approach, Garfinkle and Gundlach
\cite{GarfinkleGundlach} have suggested several combinations of lapse
and shift conditions that not only leave CSS spacetimes invariant, but
also turn the Choptuik DSS spacetime into a limit cycle. Among these,
the combination of maximal slicing with minimal strain shift has been
suggested in a different context but for related reasons
\cite{SmarrYork}. Maximal slicing requires the initial data slice to
be maximal (${K_a}^a=0$), but other prescriptions, such as freezing
the trace of $K$ together with minimal distortion, allow for an
arbitrary initial slice with arbitrary spatial coordinates.

The main difficulty remaining is that all these coordinate conditions
are elliptic equations that require boundary conditions, and will turn
CSS spacetimes into fixed points (or DSS into limit cycles) only given
correct boundary conditions. Roughly speaking, these boundary
conditions require a guess of how far the slice is from the
accumulation point $t=t_*$, and answers to this problem only exist in
spherical symmetry.


\subsubsection{The phase space variables}


Turning a CSS spacetime or a stationary spacetime into a fixed point
of the dynamical system not only requires an appropriate choice of the
lapse and shift, but also of the phase space variables $Z(x^i)$. These
depend on the symmetry one wants to capture (CSS/DSS or
stationarity/periodicity). For stationary or periodic spacetimes, the
usual choice of variables $h$ and $K$ will do. We now give the
prescriptions for variables adapted to CSS and DSS solutions.

The Cauchy data of the gravitational field are the three-metric where
$i,j,k$ range over the three spatial coordinates. If we assume that
$h_{ij}(x^k)$ and $K_{ij}(x^k)$ are induced on the Cauchy slice from a
spacetime coordinate system of the form (\ref{CSS_coordinates}), so
that $x^\mu=(x^i,\tau)$, then they are of the form
\begin{equation}
h_{ij}(x,\tau)=l^2e^{-2\tau}\bar h_{ij}(x), \qquad
 K_{ij}(x,\tau)=le^{-\tau}\bar K_{ij}(x).
\end{equation}
We simply turn this around to define the variables $\bar h_{ij}$ and
$\bar K_{ij}$. They are of course restricted by the Hamiltonian and
momentum constraints.

It is often useful to assign dimensions to these variables as
follows. The coordinates $x^i$ and $\tau$ are dimensionless,
$le^{-\tau}$ has dimension length, and $g_{\mu\nu}$ has dimension
$l^2$. From this it follows that $h_{ij}$ and $K_{ij}$ have dimensions
$l^2$ and $l$. $\bar h_{ij}$ and $\bar K_{ij}$ are
dimensionless. Note that we need the value of $\tau$ in order to
reconstruct the physical variables from the barred variables. In each
case, engineering dimension goes with scaling dimension. Energy
density has dimension $l^{-2}$ in gravitational units, where length,
time and mass have the same dimension $l$. Therefore suitable
variables $Z$ for a perfect fluid are
\begin{equation}
\qquad le^{-\tau} u^i, \qquad \bar\rho\equiv l^2 e^{-2\tau}\rho.
\end{equation}
(The 4-velocity components $u^i$ is $u^i=dx^i/d\sigma$
where $x^i$ are the dimensionless self-similarity variables and
$\sigma$ is proper time with dimension $l$. In the coordinates $t$ and
$r$ we have introduced $u^r/u^t=v(x)$ would be a more natural choice.)
Scale-invariant variables for the scalar field are
\begin{equation}
\phi, \qquad \bar\Pi\equiv l e^{-\tau}\Pi,
\end{equation}
where $\Pi$ is the canonical momentum of $\phi$. (If we wanted to
write the wave equation in first-order form in space as well as in
time, we would have to add $\partial\phi/\partial x_i$ to this list.)
In spherical symmetry, we have $le^{-\tau}=f(x)r$ for some function
$f(x)$, for example $le^{-\tau}=r/x$, and this can be used to replace
any explicit appearance of $e^{-\tau}$ by $r$, so that we might use
$r\Pi$ and $r\partial\phi/\partial r$. This trick does not work
outside spherical symmetry.


\subsubsection{Other issues}


A specific difficulty of formulating GR as a dynamical system is
that even with a prescription for the lapse and shift in place, a
given spacetime does not correspond to a unique trajectory in phase
space. Rather, for each initial slice through the same spacetime one
obtains a different slicing of the entire spacetime. A possibility for
avoiding this ambiguity would be to restrict the phase space further,
for example by restricting possible data sets to maximal or constant
extrinsic curvature slices.

Another problem is that in order to talk about attractors and
repellers on the phase space we need a notion of convergence, that is
a distance measure. One requirement is that a spacetime with
gravitational waves going out to infinity should converge to Minkowski
space in this measure as time evolves, even though the radiation does
not disappear but just disperses. The problem is even more difficult
with fluid matter, which never reaches null infinity.  Another problem
in this context is that the critical solution, because it is
self-similar, is not asymptotically flat: it does not appear to be in
the phase space (of asymptotically flat data) of which it is to be a
critical point. The critical solution arises only in a region up to
finite radius as the limiting case of a family of asymptotically flat
solutions. At large radius, it is matched to an asymptotically flat
solution which is not universal but depends on the initial data (as
does the place of matching.)  Again, a distance measure in which any
near-critical solution approaches the critical solution in an
intermediate linear regime must take this into account, by putting
more emphasis on the behavior at the center than at infinity.


\subsection{Critical phenomena and naked singularities}
\label{section:singularity}


It has been conjectured, under the name of ``cosmic censorship'', that
naked singularities do not arise in the evolution of regular initial
data for reasonable kinds of matter coupled to GR. A naked singularity
is a curvature singularity from which information can travel to a
distant observer. For a general review of cosmic censorship, see
\cite{Wald_censorship}. Choptuik's results imply that any formulation
of cosmic censorship must be restricted to ``{\it generic} smooth
initial data for reasonable matter do not form naked singularities''.

The argument is as follows. As we shall see in this section, the
critical spacetime itself has a naked singularity. The numerical time
evolutions of Choptuik and others suggest very strongly that in
spherically symmetric situations, the critical spacetime can be
approximated arbitrarily well by fine-tuning any generic parameter of
the initial data to the black hole threshold. In dynamical systems
terms, the critical spacetime is an attractor in phase space whose
basin of attraction is the black hole threshold: the co-dimension-one
hypersurface in phase space that separates collapsing from dispersing
initial data. A region of arbitrarily high curvature is therefore seen
from infinity as fine-tuning is improved. Data exactly on the black
hole threshold would produce a naked singularity, with infinite
curvature. Critical collapse therefore provides a set of smooth
initial data for naked singularity formation that has codimension one
in the phase space of spherically symmetric scalar field solutions.

As we shall discuss in more detail in Section \ref{section:generic},
this result has been extended in two directions: on the one side,
critical phenomena have been established for several other matter
models in spherical symmetry. On the other side, for scalar field and
perfect fluid matter (with certain equations of state), it has been
shown in linear perturbation theory around spherical symmetry, that
the critical solution is an attractor of co-dimension one (ie has
exactly one unstable perturbation mode) also in the full,
non-spherical theory. 

This second result, if it goes beyond infinitesimal perturbations,
means that one can fine-tune any generic parameter, whichever comes to
hand, as long as it parameterizes a smooth curve in the space of
initial data. The first result seems to indicate that critical
phenomena are fairly independent of the matter models in a dynamical
regime where the matter equations of motion are approximately
scale-invariant. The two results together mean that, in a hypothetical
experiment to create a Planck-sized black hole in the laboratory
through a strong explosion, one could fine-tune any one design
parameter of the bomb, without requiring control over its detailed
effects on the explosion.

We now take a closer look at the singularity of the critical solution.
It is straightforward to calculate the curvature of the general DSS
metric (\ref{DSS_coordinates}). From there, or directly from
(\ref{homothetic_curvature}), one can show that, unless the spacetime
is flat, all curvature invariants blow up as suggested by their
dimension, for example the Ricci scalar behaves as
$R=l^{-2}e^{2\tau}\bar R(x,\tau)$, where $\bar R(x,\tau)$ is periodic
in $\tau$. Similarly, the square of the Riemann tensor scales as
$e^{4\tau}$. $\tau=\infty$, for any $x$, is a therefore a strong
curvature singularity. The Weyl tensor with index position
${C^a}_{bcd}$ is conformally invariant, so that components with this
index position remain finite as $\tau\to\infty$. In this property it
resembles the initial singularity in Penrose's Weyl tensor conjecture
rather than the final singularity in generic gravitational
collapse. This type of singularity is called conformally
compactifiable \cite{Tod_pc} or isotropic \cite{Goodeetal}. Is the
singularity timelike, null or a point , and how can one parameterize
the ``data on the singularity''?  

Choptuik's, and Evans and Coleman's, numerical codes were limited to
the region $t<0$, in the Schwarzschild-like coordinates
(\ref{tr_metric}), with the origin of $t$ adjusted so that the
singularity is at $t=0$. Evans and Coleman conjectured that the
singularity is shrouded in an infinite redshift based on the fact that
$\alpha$ grows as a small power of $r$ at constant $t$. This is
directly related to the fact that $a$ goes to a constant $a_\infty>1$
as $r\to\infty$ at constant $t$, as one can see from the Einstein
equation (\ref{dalpha_dr}). This in turn means simply that the
critical spacetime is not asymptotically flat, but asymptotically
conical at spacelike infinity, with the Hawking mass proportional to
$r$. 

Hamad\'e and Stewart \cite{HamadeStewart} evolved near-critical scalar
field spacetimes on a double null grid, which allowed them to follow
the time evolution up to close to the future light cone of the
singularity. They found evidence that this light cone is not preceded
by an apparent horizon, that it is not itself a (null) curvature
singularity, and that there is only a finite redshift along outgoing
null geodesics slightly preceding it. This was confirmed in Gundlach's
construction of the Choptuik critical solution as a boundary value
problem \cite{Gundlach_Chop2}. 

\begin{figure}
\label{fig:critical_structure}
\epsfysize=8cm
\centerline{\epsffile{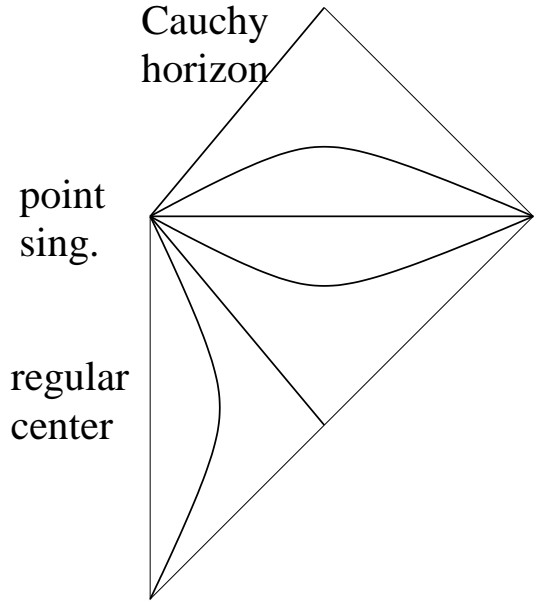}}
\caption{The global structure of spherically symmetric critical
spacetimes up to the Cauchy horizon. Infinity has been conformally
compactified. The curved lines are typical trajectories of the
homothetic vector field (in a CSS critical solution) or lines that are
invariant under the discrete isometry (in a DSS solution).}
\end{figure}

All spherically symmetric critical spacetimes appear to be
qualitatively alike as far as the singularity structure is concerned,
so that what we say about one is likely to hold for the others. These
solutions have a regular center for $t<0$, and a point-like
singularity at $t=r=0$. (As stated above, the Kretschmann scalar
scales as $t^{-4}$ at the center, and so diverges at $t=0$.) The
spacetime is regular, and in fact analytic, up to but not including
the point $r=t=0$ and its future light cone (the Cauchy horizon of the
spacetime). The conformal structure is shown in
Fig.~\ref{fig:critical_structure}. As we have discussed in Section
\ref{section:DSScalculations}, imposing analyticity at the center and
past lightcone (for the scalar field and other massless fields) or
past sound cone (for the perfect fluid) poses an ODE or PDE boundary
value problem. The reason to impose analyticity is that the critical
solution arises from generic initial data, including analytic initial
data. For the same reason, the critical solution should not be less
differentiable at the past light cone than elsewhere.


How singular is the Cauchy horizon? Hirschmann and Eardley \cite{HE1}
were the first to continue a candidate critical solution itself right
up to the future light cone. They examined a CSS complex scalar field
solution that they had constructed as a nonlinear ODE boundary value
problem, as discussed in Section \ref{section:DSScalculations}. The solution
they found is CSS and regular as required of a critical solution, but
later analysis \cite{HE2} showed that it has three growing modes and
so is not a critical solution. This should not matter for the global
structure that we are discussing here. The ansatz of Hirschmann and
Eardley for the self-similar complex scalar field is (we slightly
adapt their notation)
\begin{equation}
\phi(x,\tau) = f(x) e^{i\omega\tau}, \quad a=a(x), \quad
\alpha=\alpha(x),
\end{equation}
with $\omega$ a real constant. They continued the ODE evolution in the
self-similar coordinate $x$ through the coordinate singularity at
$t=0$ up to the future light cone by introducing a new self-similarity
coordinate $x$.  The self-similar ansatz reduces the field equations
to an ODE system. The past and future light cones, at $x=x_1$ and
$x=x_2$, are singular points of this system. At these ``points'' one
of the two independent solutions is regular and one singular. The
boundary value problem that originally defines the critical solution
corresponds to completely suppressing the singular solution at $x=x_1$
(the past light cone). The solution can be continued through this
point up to $x=x_2$. There it is a mixture of the regular and the
singular solution, and is approximately of the form
\begin{equation}
f(x)\simeq f_{\rm reg}(x) + (x_2-x)^{(i\omega+1)(1+\epsilon)} f_{\rm
sing}(x),
\end{equation}
with $f_{\rm reg}(x)$ and $f_{\rm sing}(x)$ regular at $x=x_2$, and
$\epsilon$ a small positive constant. The singular part of the scalar
field oscillates an infinite number of times as $x\to x_2$, but with
decaying amplitude. This means that the scalar field $\phi$ is just
differentiable, and that therefore the stress tensor is just
continuous. It is crucial that spacetime is not flat, or else
$\epsilon$ would vanish. For this in turn it is crucial that the
regular part $f_{\rm reg}$ of the solution does not vanish, as one
sees from the field equations.

The same effect can be seen when Choptuik's real scalar field solution
is continued to the future light cone
\cite{Gundlach_Chop2,Gundlach_Banach}. Consider the self-gravitating
wave equation (\ref{DSS_wave}) at the future light cone $f+x=0$. We
have fixed the gauge so that the future lightcone is at $x=x_f$. To
force the solution to be regular there, we would have to impose a
regularity condition similar to (\ref{reg_condition}), but this is not
possible: the solution is already completely determined by imposing
regularity at the center and at the past light cone. It can then be
shown \cite{critcont}that the solution near the future light cone has
the form
\begin{eqnarray}
U(\tau,x)&=&U_0(\tau)+|y|^\epsilon \check U_\epsilon(\tau)\, \hat
U_\epsilon(\hat\tau) 
+y\,U_1(\tau)+O\left(|y|^{1+\epsilon}\right), \\
V(\tau,x)&=&V_0(\tau)+y\,V_1(\tau)+O\left(|y|^{1+\epsilon}\right),
\\
a(\tau,x)&=&a_0(\tau)+y\,a_1(\tau)+O\left(|y|^{1+\epsilon}\right),
\end{eqnarray}
where
\begin{equation}
y\equiv x-x_f, \qquad \hat\tau\equiv \tau+H(\tau)-(1+\epsilon)\ln |y|,
\end{equation}
where all functions of $\tau$ or $\bar\tau$ are known and are periodic
with period $\tau$, and where $\epsilon$ is related to the
root-mean-square of $V_0(\tau)$ by
\begin{equation}
\epsilon={2\overline{V_0^2} \over 1-2\overline{V_0^2}}.
\end{equation}
This means that $\epsilon$ is positive for small nonvanishing
$V_0$. The situation is quite possible to the scalar field CSS
case. The term that determines the differentiability of the solution
is, roughly speaking, of the form $\phi\sim
\phi_0+|y|^{1+\epsilon}\cos\ln|y|$. Again the exponent $\epsilon$
turns out to be very small but positive, which means that $U$ is $C^0$
and $V$ is $C^1$. This in turn means that the scalar field $\phi$ is
$C^1$ and the curvature is $C^0$.  Note that $V_0$ describes a flux of
energy along the Cauchy horizon. If this flux is completely absent (or
if we solve the wave equation on flat spacetime) then $\epsilon=0$ and
the scalar field is only $C_0$. Therefore gravity allows the
self-similar scalar field to be more regular than it would be on flat
spacetime. 

The future light cone is a Cauchy horizon: the solution can be
continued through it, but the continuation is not unique. Physically,
one can think of the non-uniqueness as information emerging from the
singularity and propagating along the Cauchy horizon. Locally, one can
continue the solution through the Cauchy horizon to an almost flat
self-similar spacetime (even assuming self-similarity the solution is
not unique). It is not clear, however, if such a continuation can have
a regular center $r=0$ (for $t>0$), although this seems to have been
assumed in \cite{HE1}. It is also unknown what continuations are
possible if one drops the assumption of self-similarity to the future
of the Cauchy horizon. 

A complete kinematical description of spherically symmetric CSS
spacetimes by Carr and Gundlach (CG) \cite{CarrGundlach}. The main idea is
that the reduced 1+1 dimensional spacetime has two boundaries
$\tau=-\infty$ and $\tau=\infty$, and is fibrated by the integral
curves of the homothetic vector fields, or lines of constant $x$. The
manifold structure is therefore $-\infty<\tau<\infty$, $x_{\rm
min}<x<{\rm max}$ where the second interval may be open or closed at
each end. The boundary $\tau=\infty$ is the only curvature
singularity. Geometrically, it can be a single spacetime point (as in
the Roberts solution discussed above), or it can be a line consisting
of any number of timelike, spacelike and null pieces, corresponding to
a sequence of intervals in $x$. Because of this structure, each
possible singularity structure corresponds to a ``word'' made from a
small number of ``letters'', where each letter describes an interval
of the coordinate $x$, and the letters follow each other in the order
of increasing $x$. CG then draw the conformal diagram of all
spherically symmetric CSS spacetimes that are allowed dynamically with
perfect fluid matter, relying on the complete classification of Carr
and Coley \cite{CarrColey}. 

CG also discuss the global structure of the critical solutions for
spherical perfect fluid collapse with the equation of state
$p=\alpha\rho$, with $0<\alpha<1$ constant. They find that these
solutions are analytic from the regular center up to the Cauchy
horizon of the singularity. They then discuss the possible
continuations assuming that the spacetime remains CSS beyond the
Cauchy horizon. The critical solutions for $\alpha> 0.28$ have
no matter on the Cauchy horizon and are flat there: all the fluid
matter in the spacetime is moving outwards at the speed of the
light. The most natural continuation is therefore as a piece of vacuum
flat spacetime. Another possible continuation has an ingoing null
singularity covered by a spacelike singularity. The continuation is
filled with fluid particles that emerge from the null singularity and
run into the spacelike singularity, and so can be thought of as a
baby universe. The critical solutions for $\alpha< 0.28$ can
locally be continued through the Cauchy horizon as analytic CSS
solutions but this solution ansatz breaks down at a sonic point. To
obtain a geodesically complete continuation one has to assume the
existence of a shock somewhere to the future of the Cauchy
horizon. (The possible continuations are not discussed further.)

The possible DSS continuations of the scalar field critical solution
are discussed in \cite{critcont}. They include a continuation with a
regular center (as in the Roberts solution), and others. 

A region in a collapse solution starting from regular asymptotically
flat initial data can be made to approximate a region in the critical
solution arbitrarily well as one generic parameter of the initial data
is fine-tuned to the black hole threshold. In generic situations this
region of the critical solution can encompass all values of $x$ from
the regular center up to the Cauchy horizon, but not beyond. Therefore
the structure of the Cauchy horizon is of physical interest, but the
structure of possible continuations of the spacetime is more of a
mathematical curiosity. In particular, the question of ``what comes
out of the naked singularity'' in the limit of perfect fine-tuning to
the black-hole threshold cannot be answered consistently in classical
GR but is expected to involve quantum gravity effects.

\begin{figure}
\label{fig:collapse_to_critical}
\epsfysize=10cm
\centerline{\epsffile{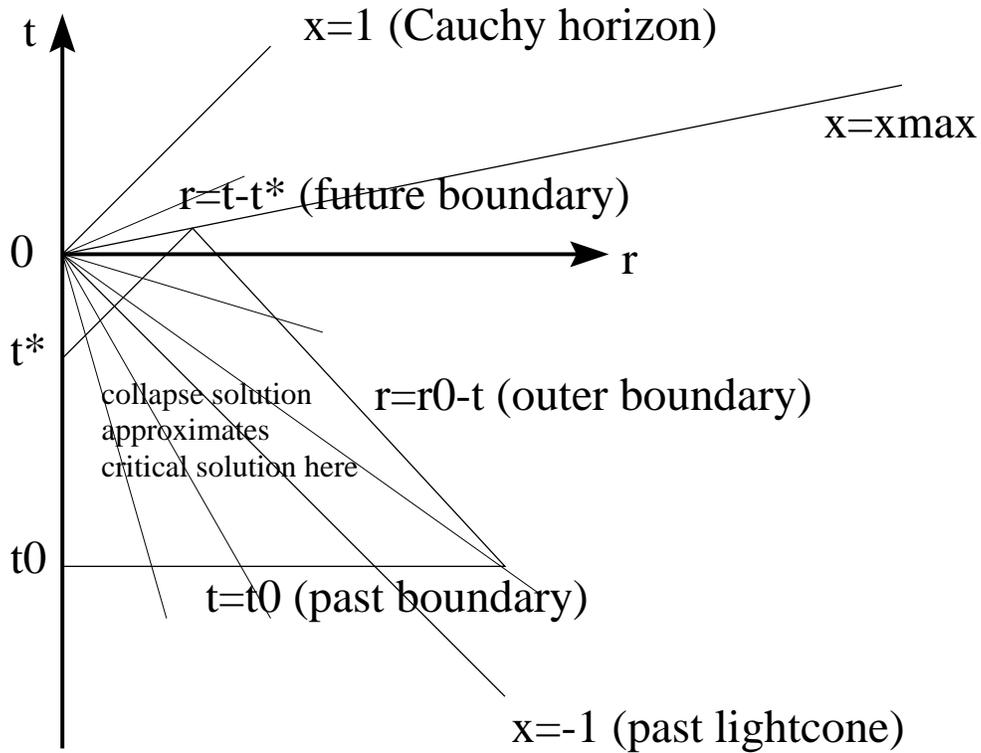}}
\caption{A generic near-critical collapse solution approximates the
critical solution in a spacetime region that is bounded very roughly
by a Cauchy surface $t=t_0$ in the past, the regular center $r=0$, an
outgoing light cone $r=t-t_*$, and an ingoing light cone
$r=r_0-t$. The parameters $r_0$ and $t_0$ depend on the initial data,
but depend only weakly on the fine-tuning parameter $p-p_*$. $t_*$ on
the other hand is proportional to $(p-p_*)^\lambda$. It can be seen
that as $p\to p_*$, the approximation region approaches the Cauchy
horizon of the critical solution but never crosses it it.}
\end{figure}

The following is a rough semiquantitative derivation of this fact,
illustrated in Fig.~\ref{fig:collapse_to_critical}. For simplicity of
presentation we treat $r$ and $t$ like the usual coordinates on flat
spacetime.  Let $t$ be proper time at the center, and let $t=0$ be the
singularity of the critical solution. Consider now a collapse solution
in the region where it is approximated well by the critical solution,
and adjust the origin of $t$ accordingly. Then the amplitude of the
one unstable mode of the critical solution is proportional to
$(p-p_*)(-t)^{-\lambda}$. The perturbation produces a significant
deviation from the critical solution at $t=t_*\sim
-(p-p_*)^\lambda$. The future lightcone of $r=0,t=t_*$, which is given
by $r=t-t_*$, is a heuristic future boundary to the region where the
collapse solution approximates the critical solution. (This assumes
that the growing mode is peaked at small $x$, which empirically is the
case.) The outer boundary of this region depends on the initial data,
and only weakly on $p-p_*$. We may approximate it as the ingoing null
cone $r= r_0-t$, and we may assume that $r_0>0$. It is easy to see
that $x$ within the approximation region takes its largest value at
its future tip, which is given by $r=(r_0+t_*)/2$ and
$t=(r_0-t_*)/2$. At that point $|x-1|\sim t_*\sim (p-p_*)^\lambda$,
and so we can see the critical spacetime arbitrarily close to its
Cauchy horizon at $x=1$.


In summary, the critical spacetimes that arise asymptotically in the
fine-tuning of gravitational collapse to the black-hole threshold have
a pointlike curvature singularity that is visible at infinity with a
finite redshift. The Cauchy horizon of the singularity is mildly
singular in the sense that the curvature is finite but not
differentiable. Beyond the Cauchy horizon, the continuation of the
spacetime is not unique. In these continuations, the curvature may
consist of the single point at the base of the Cauchy horizon, or it
may continue as an ingoing null singularity or a timelike
singularity. The non-uniqueness does not matter, as the critical
spacetime is relevant for critical collapse only up to its Cauchy
horizon, while its possible continuations are never seen in collapse.
In any case, there is a naked singularity the critical solutions
therefore provide counter-examples to any formulation of cosmic
censorship which states only that naked singularities cannot arise
from smooth initial data in reasonable matter models. The statement
must be that there is no {\it open ball} of smooth initial for naked
singularities.

So far we have only discussed the singularities of critical solutions,
that is of attractors of codimension one. A priori, however, regular
self-similar solutions could exist that are attractors, so that open
regions of initial data space would form naked singularities. This is
the case for the $S^3$ sigma model in spherical symmetry that is
discussed below, when the coupling to gravity is weak enough
($\eta<0.1$). The same is true for the perfect fluid with equation of
state $p=k\rho$, for $k<0.0105$ \cite{OriPiran,HaradaMaeda}. These two
examples are important as they further restrict the scope of any
future cosmic censorship theorem: Cosmic censorship must be formulated
to exclude such matter models. 


\section{Are critical phenomena generic?}
\label{section:generic}


The ultimate relevance of the critical phenomena discovered by
Choptuik obviously depends on how generic they are. Do they occur
whenever one fine-tunes to the black hole threshold, independently of
the family of initial data, their space symmetry, or the matter
content? The answer is not yet clear. Most numerical and analytical
studies have been limited to various matter models in spherical
symmetry. Two of these, the scalar field and the perfect fluid, have
been extended beyond spherical symmetry in perturbation theory. In
axisymmetry there is only a single published investigation. Nothing is
known in a generic 3D situation. A second line of generalization
begins from the fact that black holes can have angular momentum and
electric charge, as well as mass. What then happens if one takes a
family of initial data with angular momentum and/or charge and
fine-tunes to the black hole threshold? The answer is known for
charged but spherically symmetric initial data in one type of matter,
the charged scalar field. A prediction based on perturbation theory
has been made for the collapse of fluid initial data with angular
momentum, but to date no collapse simulations have been published
against which it could be checked. A third question concerning
genericity is if type II critical phenomena, where the critical
solution is scale-invariant, can arise if the field equations contain
a scale.


\subsection{Spherical symmetry}
\label{section:spherical}



\subsubsection{The scalar field}


Choptuik's results for the spherically symmetric scalar field have
been repeated by a number of other authors.  Gundlach, Price and
Pullin \cite{GPP2} could verify the mass scaling law with a relatively
simple code, due to the fact that it holds even quite far from
criticality. Garfinkle \cite{Garfinkle} used the fact that recursive
grid refinement in near-critical solutions is not required in
arbitrary places, but that all refined grids are centered on
$(r=0,t=t_*)$, in order to use a simple fixed mesh refinement on a
single grid in double null coordinates: $u$ grid lines accumulate at
$u=0$, and $v$ lines at $v=0$, with $(v=0,u=0)$ chosen to coincide
with $(r=0,t=t_*)$. Hamad\'e and Stewart \cite{HamadeStewart} have
written an adaptive mesh refinement algorithm based on a double null
grid (but using coordinates $u$ and $r$), and report even higher
resolution than Choptuik. Their coordinate choice also allowed them to
follow the evolution beyond the formation of an apparent horizon.

Christodoulou has also investigated the genericity of naked
singularities in the gravitational collapse of a spherically
symmetric scalar field \cite{Christodoulou5}. Christodoulou considers
a larger space of initial data that are not $C^1$. He shows that for
any data set $f_0$ in this class that forms a naked singularity there
are data $f_1$ and $f_2$ such that the data sets $f_0 + c_1 f_1 + c_2
f_2$ do not contain a naked singularity for any $c_1$ and $c_2$ except
zero. Here $f_1$ is data of bounded variation, and $f_2$ is absolutely
continuous data. Therefore, the set of naked singularity data is at
least codimension two in the space of data of bounded variation, and
of codimension at least one in the space of absolutely continuous
data. 

The perturbative result of Gundlach \cite{Gundlach_Chop2} claims that
the set of naked singularities (formed via Choptuik's critical
solution) is codimension exactly one in the set of smooth data. The
result of Christodoulou holds for any $f_0$, including initial data
for the Choptuik solution. The apparent contradiction is resolved if
one notes that the $f_1$ and $f_2$ of Christodoulou are not smooth in
(at least) one point, namely where the initial data surface is
intersected by the past light cone of the singularity in
$f_0$. (Roughly speaking, $f_1$ and $f_2$ start throwing scalar field
matter into the naked singularity at the exact moment it is born, and
therefore depend on $f_0$.) The data $f_0 + c_1 f_1 + c_2 f_2$ are
therefore not smooth.

Critical collapse of a massless scalar field in spherical symmetry in
5+1 spacetime dimensions was investigated in
\cite{GarfinkleCutlerDuncan}. Results are similar to 3+1 dimensions,
with a DSS critical solution and mass scaling with $\gamma\simeq
0.424$. Birukou et al \cite{Birukou, Birukou2} have developed a code
for arbitrary spacetime dimension. They confirm known results in 3+1
($\gamma\simeq 0.36$) and 5+1 ($\gamma\simeq 0.44$) dimensions, and
investigate 4+1 dimensions. Without a cosmological constant they find
mass scaling with $\gamma\simeq 0.41$ for one family of initial data
and $\gamma\simeq 0.52$ for another. They see wiggles in the $\ln M$
versus $\ln(p-p*)$ plot that indicate a DSS critical solution, but
have not investigated the critical solution directly. With a negative
cosmological constant and the second family, they find
$\gamma=0.49$. Clearly this needs more accurate investigation.

Scalar field collapse in spherical symmetry in 2+1 dimensions is
particularly interesting as it appears to admit an analytic approach
to critical phenomena. It is discussed in more detail in Section
\ref{section:2+1}.


\subsubsection{Other field theories}


Results similar to Choptuik's were subsequently found for a variety of
other models in spherical symmetry. All but one of these are field
theories restricted to spherical symmetry, so that the dynamical
variables are a small number of fields $Z(r,t)$. These include
varieties of the scalar field: real, complex, massless, massive, and
with nonlinear self-interaction, or coupled to the electromagnetic
field. Other natural field theories are Skyrme matter, non-linear
sigma models, Yang-Mills fields, or scalar-tensor gravity coupled to
some matter. In spherical symmetry, these can all be written as a
number (usually two) of free scalar fields coupled through terms
involving the fields and possibly their first derivatives. (For the
Yang-Mills field this is an artifact of the spherical symmetry.) A
perfect fluid is also a field theory in this sense, as long as no
shocks occur. When it is irrotational (which is automatic in spherical
symmetry) and barotropic, it can even be expressed as a scalar field
with a nonlinear term in first derivatives. From this point of view,
it is perhaps not surprising that all these models show generic
critical phenomena. As we have already mentioned, the perfect fluid
with equation of state $p=k\rho$, for $k<0.0105$, has a CSS solution
that is an attractor. It therefore produces naked singularities from
generic initial data.

An overview of all systems in which critical collapse was examined is
given in Table \ref{table:mattermodels}. The second column of this
table specifies the type of critical phenomena that is seen (compare
Sections \ref{section:typeI} and \ref{section:phasediagrams}). The
next column gives references to numerical evolutions of initial data,
while the last two columns give references to the semi-analytic
approach of first constructing the critical solution, and then its
perturbation spectrum. Some of these models are discussed in more
detail elsewhere because they introduce new phenomenology.

Related results not listed in the table concern spherically symmetric
dust collapse. Here, the entire spacetime, the Tolman-Bondi solution,
is given in closed form from the initial velocity and density
profiles. Excluding shell crossing singularities, there is a ``phase
transition'' between initial data forming naked singularities at the
center and data forming black holes. Which of the two happens depends
only the leading terms in an expansion of the initial data around
$r=0$ \cite{Christodoulou0,Jhingan}. Naked singularities are therefore
generic in dust collapse. For this reason, and because of the fact
that infinite density occurs generically even in the absence of
gravity, dust is usually considered an unphysical matter model as
far as the study of gravitational collapse is concerned. 


\subsubsection{Collisionless matter}
\label{section:vlasov}


Collisionless matter (obeying the Vlasov equation) model differs from
field theories by having a much larger number of matter degrees of
freedom. The matter content is described not by a field, or a finite
number of fields, on spacetime, but by a distribution on the point
particle phase space, which is spacetime $\times$ momentum
space. Remarkably, no type II scaling phenomena of the kind seen in
the scalar field were discovered in numerical collapse simulations of
collisionless matter in spherical symmetry, and it is not clear
whether or not type I critical phenomena have been seen.

When collisionless matter solutions are restricted to spherical
symmetry, individual particles move tangentially as well as radially,
and so individually have angular momentum, but the stress-energy
tensor averages out to a spherically symmetric one, with zero total
angular momentum. The matter is therefore described by a function
$f(r,t,p^r,L^2)$ of radius, time, radial momentum, and total angular
momentum. Note that this depends on two momentum variables as well as
on spacetime.

Two numerical simulations of critical collapse of collisionless matter
in spherical symmetry have been published to date. Rein, Rendall and
Schaeffer \cite{ReinRendallSchaeffer} find a mass gap at the black
hole threshold that depends on the initial matter -- black hole
formation turns on with a mass that is a large part of the ADM mass of
the initial data . No critical behavior of either type I or type II
was observed. Olabarrieta and Choptuik \cite{OlabarrietaChoptuik} find
evidence of a metastable static solution at the black hole threshold,
with type I scaling of its life time as in
Eq. (\ref{typeIscaling}). However, the critical exponent depends
weakly on the family of initial data, ranging from $5.0$ to $5.9$,
with a quoted uncertainty of $0.2$. They find partial universality in
the threshold solution: the matter distribution does not appear to be
universal, while the metric appears to be universal up to an overall
rescaling. This limited universality may just be compatible with the
results of Rein, Rendall and Schaeffer. It should be noted that the
parameter $p$ is chosen to be the particle rest mass $m$. To compare
with Rein, Rendall and Schaeffer, these evolutions must be rescaled to
set $m$ to a constant. Neither numerical investigation has either
enough numerical precision or has used enough different 1-parameter
families of data to settle the existence or absence of critical
phenomena in the Einstein-Vlasov system.

Mart\'\i n-Garc\'\i a and Gundlach \cite{vlasov1} have reviewed a
family of static spherically symmetric solutions for massive or
massless particles that is generic by function counting, and have
constructed a family of CSS spherically symmetric solution for
massless particles that is also generic by function counting. These
solutions are parameterized by one arbitrary function of two
variables, called $k(Y,Z)$ in the massless CSS case. The spacetime
metric, however, depends only on the integral $\bar k(Y)=\int
k(Y,Z)\,dZ$. Therefore there are infinitely many solutions with
different matter configurations but the same stress-energy tensor and
spacetime metric. The physical reason is the existence of an exact
symmetry: two massless particles with energy-momentum $p^\mu$ in the
solution can be replaced by one particle with $2p^\mu$. A similar
result holds for the perturbations. As the growth exponent $\lambda$
of a perturbation mode can be determined from the metric alone, this
means that there are infinitely many perturbation modes with the same
$\lambda$. If there is one growing perturbation mode, there are
infinitely many. This argument conclusively rules out the existence of
both type I and type II critical phenomena for massless particles.

Based on function counting, it appears possible that a similar result
holds approximately for massive particles that are moving
ultrarelativistically. If true, this would be harder to show as the
exact symmetry just mentioned is lost with massive particles. Applied
to a static solution, such a result would give theoretical support to
the observation of Olabarrieta and Choptuik of a critical solution
that is universal in the metric but not the matter. On the other hand,
applied to the perturbations of that solution, it would also rule out
(type I) critical phenomena. Therefore the situation with massive
particles is not completely understood.


\begin{table}

\caption{
\label{table:mattermodels}
Critical collapse in spherical symmetry}

\bigskip

\begin{tabular}{| l | c | c | c | c | c |}

\hline\hline

Matter & Type & Collapse  & Critical & Perturbations \\ 
& & simulations & solution & of crit. soln.\\

\hline\hline

Perfect fluid $p=k\rho$ & II & \cite{EvansColeman,NeilsenChoptuik} &
 CSS \cite{EvansColeman,Maison,NeilsenChoptuik} 
& \cite{Maison,KoikeHaraAdachi2,Gundlach_nonspherical,Gundlach_critfluid2} \\

\hline

Real scalar field: &&&& \\
 
-- massless, min. coupled & II & \cite{Choptuik91,Choptuik92,Choptuik94} &

DSS \cite{Gundlach_Chop1} 
& \cite{Gundlach_Chop2,MartinGundlach} \\

-- massive & I & \cite{BradyChambersGoncalves} 
& oscillating \cite{SeidelSuen} & \\

& II & \cite{Choptuik94}
& DSS \cite{HaraKoikeAdachi,GundlachMartin} 
& \cite{HaraKoikeAdachi,GundlachMartin} \\ 

-- conformally coupled & II &
\cite{Choptuik92} & DSS  &   \\

-- 4+1 & II & \cite{Birukou} & & \\

-- 5+1 & II & \cite{GarfinkleCutlerDuncan} & & \\

\hline

Massive complex scalar field & I (II) & \cite{HawleyChoptuik} &
\cite{SeidelSuen}& \cite{HawleyChoptuik} \\

Massless scalar electrodynamics & II & \cite{HodPiran_charge} &
DSS \cite{GundlachMartin} &
\cite{GundlachMartin} \\ 

\hline

2-d sigma model &&&& \\

-- complex scalar ($\kappa=0$) & II & \cite{Choptuik_pc} &
DSS \cite{Gundlach_Chop2}  &
\cite{Gundlach_Chop2} \\

-- axion-dilaton ($\kappa=1$) & II & 
\cite{HamadeHorneStewart} & 
CSS \cite{EardleyHirschmannHorne,HamadeHorneStewart} & 
\cite{HamadeHorneStewart} \\

-- scalar-Brans-Dicke ($\kappa>0$) & II & \cite{LieblingChoptuik,Liebling}
& CSS, DSS & \\

-- general $\kappa$ including $\kappa<0$ & II & & CSS, DSS \cite{HE3} & \cite{HE3} \\

\hline
 
$SU(2)$ Yang-Mills & I & \cite{ChoptuikChmajBizon} &
static \cite{BartnikMcKinnon} & \cite{LavrelashviliMaison} \\

& II & \cite{ChoptuikChmajBizon} & DSS \cite{Gundlach_EYM} 
& \cite{Gundlach_EYM} \\

& ``III'' & \cite{ChoptuikHirschmannMarsa} &
colored BH \cite{Bizon0,VolkovGaltsov} &
\cite{StraumannZhou,VolkovBrodbeckLavrelashviliStraumann} \\

$SU(2)$ Skyrme model & I & \cite{BizonChmaj} & static \cite{BizonChmaj}
& \cite{BizonChmaj} \\

& II & \cite{BizonChmajTabor}  

& static \cite{BizonChmajTabor}  & \\

$SO(3)$ Mexican hat & II & \cite{Liebling2} & DSS & \\

\hline

Vlasov & I? & \cite{ReinRendallSchaeffer,OlabarrietaChoptuik} &
\cite{vlasov1} & \\

\hline

\end{tabular}
\end{table}



\subsection{Perturbing around spherical symmetry}


The simplest way of taking critical collapse beyond the restriction to
spherical symmetry is to take a known critical solution in spherical
symmetry, and perturb it using nonspherical perturbations. Gundlach
\cite{Gundlach_nonspherical} has studied the generic non-spherical
perturbations around the critical solution found by Evans and Coleman
\cite{EvansColeman} for the $p={1\over3}\rho$ perfect fluid in
spherical symmetry. There is exactly one spherical perturbation mode
that grows towards the singularity (confirming the previous results
\cite{KoikeHaraAdachi,Maison}). There are no growing nonspherical
modes at all. 

The main significance of this result, even though it is only
perturbative, is to establish one critical solution that really has
only one unstable perturbation mode within the full phase space.  As
the critical solution itself has a naked singularity (see Section
\ref{section:singularity}), this means that there is, for this
matter model, a set of initial data of codimension one in the full
phase space of GR that forms a naked singularity.
This result also confirms the role of critical collapse as the most
``natural''way of creating a naked singularity.

Gundlach has extended his analysis to the perfect fluid with
$p=k\rho$, with the constant $k$ in the physical range $0<k<1$
\cite{Gundlach_critfluid2}.  For an intermediate range of values of
$k$, $1/9<k<0.49$, the result is the same as for $k=1/3$. At high $k$,
an additional mode is reported to appear with polar parity and angular
dependence $l=2$. Suppressing this mode would require fine-tuning an
additional 10 parameters, and so the spherically symmetric CSS solution is not a
critical solution outside spherical symmetry. (The unstable mode is a
complex conjugate pair of modes, and degenerate for $m=-2,-1,\dots
2$.) At low $k$, a single additional unstable mode appears with axial
parity and $l=1$ angular dependence. The appearance of this mode can
be shown analytically, and is more intuitively plausible: as $l=1$
axial perturbations are related to (infinitesimal) rotation, this mode
is interpreted as a centrifugal force disrupting collapse. Its
significance for critical collapse is discussed in Section
\ref{section:angmom}. The numerical perturbation results become
somewhat uncertain as $k$ approaches either end of the range, due to
increasing numerical error.

Gundlach and Mart\'\i n-Garc\'\i a \cite{MartinGundlach} have
constructed the nonspherical perturbations of the Choptuik critical
solution for the massless scalar field, which is spherically symmetric
and DSS. They find that all non-spherical perturbations decay.


\subsection{Axisymmetry}
\label{section:axisymm}


Every aspect of the basic scenario: CSS and DSS, universality and
scaling applies directly to a critical solution that is not
spherically symmetric, but all the models we have described are
spherically symmetric. The only exception is a numerical investigation
of critical collapse in axisymmetric pure gravity by Abrahams and
Evans \cite{AbrahamsEvans}. This was the first paper on critical
collapse to be published after Choptuik's initial paper, but has
remained the only one to go beyond spherical symmetry. Because of its
importance, we summarize its contents in some detail.

The physical situation under consideration is axisymmetric vacuum
gravity. The numerical scheme uses a 3+1 split of the spacetime. The
ansatz for the spacetime metric is
\begin{equation}
\label{axisymmetric}
ds^2=-\alpha^2\,dt^2 + \phi^4\left[e^{2\eta/3}(dr+\beta^r\,dt)^2 +
r^2e^{2\eta/3}(d\theta+\beta^\theta\,dt)^2 +
e^{-4\eta/3}r^2\sin^2\theta\,d\varphi^2\right],
\end{equation}
parameterized by the lapse $\alpha$, shift components $\beta^r$ and
$\beta^\theta$, and two independent coefficients $\phi$ and $\eta$ in
the 3-metric. All are functions of $r$, $t$ and $\theta$. The fact
that $dr^2$ and $r^2\,d\theta^2$ are multiplied by the same
coefficient is called quasi-isotropic spatial gauge. The variables for
a first-order-in-time version of the Einstein equations are completed
by the three independent components of the extrinsic curvature,
$K^r_\theta$, $K^r_r$, and $K^\varphi_\varphi$.  This ansatz limits
gravitational waves to one polarization out of two, so that there
are as many physical degrees of freedom as in a single wave
equation. In order to obtain initial data obeying the constraints,
$\eta$ and $K^r_\theta$ are given as free data, while the remaining
components of the initial data, namely $\phi$, $K^r_r$, and
$K^\varphi_\varphi$, are determined by solving the Hamiltonian
constraint and the two independent components of the momentum
constraint respectively.  There are five initial data variables, and
three gauge variables. Four of the five initial data variables, namely
$\eta$, $K^r_\theta$, $K^r_r$, and $K^\varphi_\varphi$, are updated
from one time step to the next via evolution equations. As many
variables as possible, namely $\phi$ and the three gauge variables
$\alpha$, $\beta^r$ and $\beta^\theta$, are obtained at each new time
step by solving elliptic equations. These elliptic equations are the
Hamiltonian constraint for $\phi$, the gauge condition of maximal
slicing (${K_i}^i=0$) for $\alpha$, and the gauge conditions
$g_{\theta\theta}=r^2 g_{rr}$ and $g_{r\theta}=0$ for $\beta^r$ and
$\beta^\theta$ (quasi-isotropic gauge).

For definiteness, the two free functions $\eta$ and $K^r_\theta$ in
the initial data were chosen to have the same functional form they
would have in a linearized gravitational wave with pure $l=2,m=0$
angular dependence. Of course, depending on the overall amplitude of
$\eta$ and $K^r_\theta$, the other functions in the initial data will
deviate more or less from their linearized values, as the non-linear
initial value problem is solved exactly. In axisymmetry, only one of
the two degrees of freedom of gravitational waves exists. In order to
keep their numerical grid as small as possible, Abrahams and Evans
chose the pseudo-linear waves to be purely ingoing. (In nonlinear GR,
no exact notion of ingoing and outgoing waves exists, but this ansatz
means that the wave is initially ingoing in the low-amplitude limit.)
This ansatz (pseudo-linear, ingoing, $l=2$ angular dependence),
reduced the freedom in the initial data to one free function of
advanced time, $I^{(2)}(v)$. A specific suitably peaked function was
chosen, and only the overall amplitude was varied.

Limited numerical resolution (numerical grids are now two-dimensional,
not one-dimensional as in spherical symmetry) allowed Abrahams and
Evans to find black holes with masses only down to $0.2$ of the ADM
mass. Even this far from criticality, they found power-law scaling of
the black hole mass, with a critical exponent $\gamma\simeq
0.36$. Determining the black hole mass is not trivial, and was done
from the apparent horizon surface area, and the frequencies of the
lowest quasi-normal modes of the black hole.  There was tentative
evidence for scale echoing in the time evolution, with $\Delta\simeq
0.6$, with about three echos seen. This corresponds to a scale range
of about one order of magnitude. By a lucky coincidence, $\Delta$ is
much smaller than in all other examples, so that several echos could
be seen without adaptive mesh refinement. The paper states that the
function $\eta$ has the echoing property $\eta(e^\Delta r,e^\Delta
t)=\eta(r,t)$. If the spacetime is DSS in the sense defined above, the
same echoing property is expected to hold also for $\alpha$, $\phi$,
$\beta^r$ and $r^{-1}\beta^\theta$, as one sees by applying the
coordinate transformation (\ref{x_tau}) to (\ref{axisymmetric}).

In a subsequent paper \cite{AbrahamsEvans2}, universality of the
critical solution, echoing period and critical exponent was
demonstrated through the evolution of a second family of initial data,
one in which $\eta=0$ at the initial time. In this family, black hole
masses down to $0.06$ of the ADM mass were achieved.  Further work on
critical collapse far away from spherical symmetry would be
desirable. An attempt with a current 3D numerical relativity code to
repeat the results of Abrahams and Evans was not successful
\cite{Alcubierre_Brill}. 


\subsection{Approximate self-similarity and universality classes}
\label{section:universalityclasses}


The presence of a length scale in the field equations is incompatible
with exact self-similarity. It can give rise to static (or
oscillating) asymptotically flat critical solutions and so to type I
critical phenomena at the black hole threshold. Depending on the
initial data and on how the scale appears in the field equations, it
can also become asymptotically irrelevant as a self-similar solution
reaches ever smaller spacetime scales. This behavior was already
noticed by Choptuik in the collapse of a massive scalar field,
$V(\phi)=m^2\phi^2$, or one with an arbitrary potential $V(\phi)$
\cite{Choptuik94} and confirmed by Brady, Chambers and Gon\c calves
\cite{BradyChambersGoncalves}. As the scalar field $\phi$ repeats
itself on ever smaller spacetime scales, the kinetic energy term
$(\nabla\phi)^2$ that appears in the stress-energy tensor diverges,
while the potential energy term $V(\phi)$ remains bounded because
$\phi$ is bounded.

In order to capture the notion of an asymptotically self-similar
solution, one may set the arbitrary scale $l$ in the definition
(\ref{x_tau}) of $\tau$ to the length scale set by the field
equations, for example $1/m$ in the example of the massive scalar
field.  Introducing self-similarity variables $x$ and $\tau$, and
suitable dimensionless first-order variables $Z$ (such as $a$,
$\alpha$, $\phi$, $r\phi_{,r}$ and $r\phi_{,t}$ for the spherically
symmetric scalar field), one can write the field equations as a first
order system
\begin{equation}
F\left(Z,Z_{,x},Z_{,\tau},e^{-\tau}\right)=0.
\end{equation}
Every appearance of $m$ in the original equations gives rise to an
appearance of $e^{-\tau}$ in the dimensionless equations for $Z$. If
the field equations contain only positive integer powers of $m$, one
can make an ansatz for the critical solution of the form
\begin{equation}
\label{asymptotic_CSS}
Z_*(x,\tau) = \sum_{n=0}^\infty e^{-n\tau} Z_n(x).
\end{equation}
This is an expansion around a scale-invariant solution $Z_0$ (obtained
by setting $m\to 0$, in powers of (scale on which the solution
varies)/(scale set by the field equations).

After inserting the ansatz into the field equations, each $Z_n(x)$ is
calculated recursively from the preceding ones. For large enough
$\tau$ (on spacetime scales small enough, close enough to the
singularity), this expansion is expected to converge. 
The leading order term $Z_0$ in the expansion of the self-similar
critical solution $Z_*$ obeys the equation
\begin{equation}
F\left(Z_0,Z_{0,x},Z_{0,\tau},0\right)=0.
\end{equation}
Clearly, this leading order term is independent of the overall scale
$l$, and can be calculated without reference to the following
terms. In the case of a DSS solution, $\Delta$ is also fixed exactly
in solving for $Z_0$ as an eigenvalue problem. (No corrections to
$\Delta$ arise at higher orders.) A similar ansatz can be made for the
linear perturbations of $Z_*$, and solved again recursively. One can
calculate the leading order perturbation term on the basis of $Z_0$
alone, and obtain the exact value of the critical exponent $\gamma$ in
the process. 

This procedure was carried out by Gundlach \cite{Gundlach_EYM} for the
Einstein-Yang-Mills system, and by Gundlach and Mart\'\i n-Garc\'\i a
\cite{GundlachMartin} for massless scalar electrodynamics. Both
systems have a single length scale $1/e$ (in units $c=G=1$), where $e$
is the gauge coupling constant. As another example, Liebling
\cite{Liebling2} has investigated the restriction to spherical
symmetry of a triplet of scalar fields with a Mexican hat
potential. The reduction to spherical symmetry gives rise to an
effective matter action
\begin{equation}
\int d^4x \sqrt{g}\left[ -{1\over 2}(\nabla\phi)^2 - {\phi^2\over r^2} -
{\lambda\over 4}(\phi^2-\eta^2)^2\right].
\end{equation}
Liebling finds that the sigma model-like term $\phi^2/r^2$ is
relevant, while the quartic potential is not. Similarly, the leading
order term $Z_0$ is the same in the critical solutions in the
spherically symmetric $SU(2)$ Yang-Mills \cite{ChoptuikChmajBizon} and
$SU(2)$ Skyrme models \cite{BizonChmajTabor}, because the terms that
distinguish them are irrelevant on small scales. This notion of
universality classes is fundamentally the same as in statistical
mechanics. Other examples include modifications to the perfect fluid
equation of state that do not affect the limit of high density. 

The critical exponent $\gamma$ depends only on $Z_0$, and is
therefore also independent of $l$. There is a region in the space of
initial data where in fine-tuning to the black hole threshold the
scale $l$ becomes irrelevant, and the behavior is dominated by the
critical solution $Z_0$. In this region, the usual type II critical
phenomena occur, independently of the value of $l$ in the field
equations. In this sense, all systems with a single length scale $l$
in the field equations are in one universality class
\cite{HaraKoikeAdachi,GundlachMartin}.  The massive scalar field, for
any value of $m$, or massless scalar electrodynamics, for any value of
$e$, are in the same universality class as the massless scalar field.

It should be stressed that universality classes with respect to a
dimensionful parameter arise in regions of phase space (which may be
large). Another region of phase space may be dominated by an
intermediate attractor that has a scale proportional to $l$. This is
the case for the massive scalar field with mass $m$: in one region of
phase space, the black hole threshold is dominated by the Choptuik
solution and type II critical phenomena occur, in another, it is
dominated by metastable oscillating boson stars, whose mass is $1/m$
times a factor of order 1 \cite{BradyChambersGoncalves}.

If there are several scales $l_0$, $l_1$, $l_2$ etc. present in the
problem, a possible approach is to set the arbitrary scale in
(\ref{x_tau}) equal to one of them, say $l_0$, and define the
dimensionless constants $\lambda_i=l_i/l_0$ from the others.  The size of
the universality classes depends on where the $\lambda_i$ appear in the
field equations. If a particular $\lambda_i$ appears in the field equations
only in positive integer powers, the corresponding $\lambda_i$ appears only
multiplied by $e^{-\tau}$, and will be irrelevant in the scaling
limit. All values of this $\lambda_i$ therefore belong to the same
universality class. As an example, adding a quartic self-interaction
$\lambda\phi^4$ to the massive scalar field gives rise to the
dimensionless number $\lambda/m^2$, but its value is an irrelevant in
the scaling limit. All self-interacting scalar fields are in fact in
the same universality class. Similarly, massive scalar
electrodynamics, for any values of $e$ and $m$, may form a single
universality class in a region of phase space where type II critical
phenomena occur. 

There are also examples of dimensionless parameters that are relevant,
so that they label 1-parameter families of universality classes, where
each universality class is in turn parameterized by one or several
irrelevant parameters. Such relevant dimensionless parameters include
the parameter $k$ in the perfect fluid equation of state $p=k\rho$,
the target space curvature $\kappa$ of the 2-dimensional sigma model,
or the conformal coupling of the scalar field (whereas the potential
is irrelevant.)


\subsection{Black hole charge}


Given the scaling power law for the black hole mass in critical
collapse, one would like to know what happens if one takes a generic
1-parameter family of initial data with both electric charge and
angular momentum (for suitable matter), and fine-tunes the parameter
$p$ to the black hole threshold. Does the mass still show power-law
scaling? What happens to the dimensionless ratios $L/M^2$ and $Q/M$,
with $L$ the black hole angular momentum and $Q$ its electric charge?

Gundlach and Mart\'\i n-Garc\'\i a \cite{GundlachMartin} have studied
scalar massless electrodynamics in spherical symmetry. Clearly, the
real scalar field critical solution of Choptuik is a solution of this
system too. Less obviously, it remains a critical solution within
massless (and in fact, massive) scalar electrodynamics in the sense
that it still has only one growing perturbation mode within the
enlarged solution space. Some of its perturbations carry electric
charge, but as they are all decaying, electric charge is a subdominant
effect. The charge of the black hole in the critical limit is
dominated by the most slowly decaying of the charged modes. From this
analysis, a universal power-law scaling of the black hole charge
\begin{equation}
Q\sim (p-p_*)^\delta
\end{equation}
was predicted. The predicted value $\delta\simeq 0.88$ of the critical
exponent (in scalar electrodynamics) was subsequently verified in
collapse simulations by Hod and Piran \cite{HodPiran_charge}. (The
mass scales with $\gamma\simeq 0.37$ as for the uncharged scalar
field.) General considerations using dimensional analysis led Gundlach
and Mart\'\i n-Garc\'\i a to the general prediction that the two
critical exponents are always related, for any matter model, by the
inequality
\begin{equation}
\delta\ge2\gamma.
\end{equation}
This has not yet been verified in any other matter model.


\subsection{Black hole angular momentum}
\label{section:angmom}


Gundlach's results on non-spherically symmetric perturbations around
spherically symmetric critical collapse of a perfect fluid
\cite{Gundlach_nonspherical} allow for initial data, and therefore
black holes, with infinitesimal angular momentum.  All nonspherical
perturbations decrease towards the singularity for equations of state
$p=k\rho$ with $k$ in the range $1/9<k<0.49$. This means generally
that the spherically symmetric critical solution is still a critical
solution for small deviations from spherical symmetry. More
specifically, infinitesimal angular momentum is carried by the axial
parity perturbations with angular dependence $l=1$. Using a
perturbation analysis similar to that applied to charge in scalar
electrodynamics, Gundlach \cite{Gundlach_angmom} (but see the
correction in \cite{Gundlach_critfluid2}) has derived the angular
momentum scaling law
\begin{equation}
\label{CSS_L}
L \sim (p-p_*)^\mu
\end{equation}
which is valid in the range $1/9<k<0.49$. The angular momentum
exponent $\mu(k)$ is related to the mass exponent $\gamma(k)$ by
\begin{equation}
\label{muofk}
\mu(k) = \Big(2+\lambda_1(k)\Big)\gamma(k)={5(1+3k)\over 3(1+k)} \gamma(k) .
\end{equation}
where $\lambda_1(k)$ is the growth rate of the dominant $l=1$ axial
perturbation mode (the spinup mode).  In particular for the value
$k=1/3$, where $\gamma\simeq 0.3558$, $\mu=5\gamma/2\simeq 0.8895$.

It is remarkable that $\lambda_1(k)$ (and in fact the entire $l=1$
axial spectrum) can be calculated exactly from the assumptions of CSS,
spherical symmetry, and regularity at the center alone. The dust limit
$k=0$ is also the Newtonian limit (in the sense that the gas particles
move much more slowly than light), and in this limit $\lambda_1=1/3$
\cite{HanawaMatsumoto}. A flat Friedmann universe is also CSS. The CSS
$l=1$ result would therefore indicate that the flat Friedmann perfect
fluid solutions have an infinite number of spinup modes that grow away
from the big bang singularity. In a cosmological context, these modes
are ruled out by boundary conditions at infinity, while such boundary
conditions are not relevant in the critical collapse context.

For $k>0.49$ the situation is unclear, as the spherically symmetric
critical situation does not appear to survive deviations from
spherical symmetry. For $0<k<1/9$, the spherically symmetric critical
solution has a single extra nonspherical unstable mode. This is
precisely an axial $l=1$ mode, threefold degenerate for $m=-1,0,1$,
which is related to infinitesimal angular momentum. The presence of
two growing modes of the critical solution would be expected to give
rise to interesting phenomena \cite{Gundlach_scalingfunctions}. Near
the critical solution, the two growing modes compete. Which one grows
to a nonlinear amplitude first depends both on their growth rates
$\lambda$ and on their initial amplitudes. Near the black hole
threshold, the mass $M$ of the final black hole will therefore depend
not only on the distance to the black hole threshold, but also on the
amount of angular momentum in the initial data. This dependence on two
parameters is again universal, and is encoded in two critical
exponents and one universal function of one argument:
\begin{equation}
M \simeq |\bar p|^{1\over\lambda_0} \cases{ F_M^+
(\delta), & $\bar p> 0$ \cr F_M^- (\delta), & $\bar p < 0$ ,  }
\end{equation}
where 
\begin{equation}
\delta=|\bar p|^{-{\lambda_1\over \lambda_0}}\bar q.
\end{equation}
Here $F_M^\pm$ are two universal functions, $\lambda_0$ is the growth
rate of the unstable spherical mode, and $\lambda_1$ is growth rate of
the unstable axial $l=1$ mode. $\bar p$ and $\bar q$ are
generalizations of the function $P$ defined above in
Eq. (\ref{Pdef}). Roughly speaking, they measure the self-gravity and
angular momentum of the initial data. (In general, $\bar q$ will
therefore be a vector. Here we have assumed axisymmetry for
simplicity.) Their exact definition is as follows. $\bar q$ is a
regular function on the space of initial data such that whenever a
black hole is formed, its sign changes when $\bar q$ changes sign.
Furthermore, $\bar p$ is another regular function in phase space with
the property that when $\bar q=0$, a black hole is formed if and only
if $\bar p>0$. We could equally well formulate this result in terms of
2-parameter families of initial data. A similar result holds for the
specific angular momentum $a=L/M$ of the black hole:
\begin{equation}
a \simeq |\bar p|^{1\over\lambda_0} \cases{ F_a^+
(\delta), & $\bar p> 0$ \cr F_a^- (\delta), & $\bar p < 0$ . }
\end{equation}
Because $\delta$ changes sign when the initial angular momentum
changes sign, it is clear from symmetry that $F_M$ is an even function
and $F_a$ is an odd function. Furthermore, one can set $F_M^+(0)=1$ by
convention. It is likely that adding a small amount of angular
momentum will not make subcritical initial data supercritical. If that
is correct, one would have $F_a^-=F_M^-=0$ for all
$\delta$. Furthermore, it appears likely that enough angular momentum
will make any data subcritical. This would mean that $F_a^+=F_M^+=0$
for $|\delta|$ above a threshold value.

These results are formal, but have real predictive power: in principle
the universal scaling functions $F_{M,a}^{\pm}$ can be determined from
a single 2-parameter family, and can then be verified in an infinite
number of other 2-parameter families. It is interesting that although
these results are based on linear perturbation theory, they make
predictions also for situations where the initial angular momentum is
so small that it can be treated as a small perturbation, but where the
final black hole is rapidly spinning nevertheless. Physically, such a
situation would arise because a small part of the initial mass
combines with a large part of the initial angular momentum, through a
spinup as the system contracts. 

If these predictions are correct, they extend the analogy between type
II critical phenomena and critical phase transitions to two
parameters. In the ferromagnet example $T_*-T$ and $\bf B$ would be
the equivalents of $\bar p$ and $\bf \bar q$, and $1/\xi$ and $\bf m$
would be the equivalents of $M$ and $\bf a$.

Note that these results hold for the perfect fluid with equation of
state $p=k\rho$ only in the range $0.0105<k<1/9$. For smaller values
of $k$ the black hole end state is replaced by a stable naked
singularity end state. For larger values of $k$, $\lambda_1$ is
negative. This means that fine-tuning to the black hole threshold
sends $\delta\to 0$, and angular momentum becomes a subdominant effect
(like electric charge in scalar electrodynamics). The black hole
angular momentum is then given by the simple power-law scaling
(\ref{CSS_L}). 

For the massless scalar field, angular momentum is also a decaying
perturbation of the spherically symmetric DSS solution.  A critical
exponent $\mu\simeq 0.76$ for the angular momentum was derived for the
massless scalar field in \cite{GarfinkleGundlachMartin} using
second-order perturbation theory. 

It should be stressed that at the time of writing there is no
nonlinear evolution study of the black hole threshold in the presence
of angular momentum, for any system.


\section{Phenomenology}
\label{section:phenomenology}



\subsection{Critical phenomena without gravity}
\label{section:withoutgravity}


Critical phenomena that arise at a threshold in initial data space and
involve universality, self-similarity and scaling are not restricted
to general relativity. Bizo\'n, Chmaj and Tabor
\cite{BizonChmajTabor2} (BCT) and Liebling, Hirschmann and Isenberg
(LHI) \cite{LieblingHirschmannIsenberg} have studied the round $S^3$
sigma model (\ref{S3sigma}) in Minkowski spacetime. Without gravity,
there are of course no curvature singularities and no black holes, but
singular behavior exists in the form of the blowup of matter fields.
Weak initial data remain regular and disperse. Strong initial data
blow up at the center in a finite time. There are two different kinds
of critical behavior:

The model admits a family of CSS solutions $\phi_n(r/t)$ for
$n=0,1,\dots$. The solution $\phi_n$ has $n$ unstable modes. $\phi_0$
is an attractor in the space of solutions that blow up. The blowup
therefore always takes the form $\phi_0$ in a region of phase
space. (This behavior carries over to a sufficiently weak coupling to
gravity, giving rise to generic naked singularities
\cite{BizonWassermann}.) $\phi_1$, which has one growing mode, is the
threshold, or critical, solution between blowup and dispersion in some
region of phase space. This was shown by both groups using both time
evolutions of several one-parameter families of initial data, and
perturbation analysis around the $\phi_n$.

The model also admits a scalable static solution $\phi_{\rm
static}(kr)$, where $k$ can take any positive value. This solution
does not decay as $r\to\infty$, and so has infinite energy. Therefore
it cannot be accessed from finite energy initial data, such as
Gaussian data. LHI find that this solution also sits at the collapse
threshold in the sense that if a small perturbation $\delta\phi(r)$ is
added to the initial data it will blow up, while adding
$-\delta\phi(r)$ results in dispersion. Nevertheless, BCT find using
perturbation theory that the static solution has in fact infinitely
many growing modes. They conjecture that almost all of these do not
matter in evolutions as they have support peaked at ever larger
$r$. LHI explain their findings using a similar argument. They
conjecture that as the initial data are fine-tuned to the blowup
threshold from the blowup side, a perturbation of $\phi_{\rm static}$
runs out to ever larger $r$ before turning around and causing the
blowup. For data below the threshold, this perturbation never turns
around.

More recently, Liebling \cite{Liebling3D} has generalized this model
to a 3+1 spacetime dimensions. The restriction of the $S^3$ sigma
model to spherical symmetry can be written as
\begin{equation}
\nabla^2 \chi=-{\sin 2\chi\over r^2}
\end{equation}
where $\chi=\chi(r,t)$ and $\nabla^2$ is the Minkowski wave operator
restricted to spherical symmetry. The generalized model is described
by the same wave equation for a single variable $\chi(x,y,z,t)$, where
$\nabla^2$ is now the 3+1-dimensional Minkowski wave operator. While
the full $S^3$ sigma model contains three coupled nonlinear fields,
this simplified version contains one nonlinear field with a
position-dependent potential. In particular, $\chi$ is still
constrained to vanish at the origin. Numerical evolutions shown that
the spherically symmetric critical solutions remains a critical
solution for nonspherical data in the generalized model, even with
angular momentum. This work is also remarkable for implementing
adaptive mesh refinement in 3+1 dimensions. 

Bizo\'n and Tabor \cite{BizonTabor} have examined self-similar
spherically symmetric solutions of the $SO(d)$ Yang-Mills field on
flat $(d+1)$-dimensional spacetime (which includes $SU(2)$ in the
usual 4-dimensional spacetime) for $d=4,5$. They find generic blowup
from smooth initial data. In $d=5$, this happens via an attracting CSS
solution. The critical solution between blowup and dispersion is a CSS
solution with one growing mode. In $d=4$ no CSS solutions exist, but
generic blowup seems to proceed via self-similarity of the second
kind. (The preferred slicing required for this is provided by the
assumption of spherical symmetry).

Type I critical phenomena can also arise without gravity. Honda and
Choptuik \cite{HondaChoptuik} consider the time evolution of a
spherically symmetric scalar field with a symmetric double well
potential $V(\phi)$ which has minima at $\phi=\pm\phi_0$. Their
initial data go from $\phi=\phi_0$ at $r=0$ to $\phi=-\phi_0$ at
$r=\infty$, with a smooth transition at $r\simeq r_0$. For generic
initial data, these initial data first contract and then disperse to
infinity, but for a discrete set of values of $r_0$, they form
strictly periodic soliton solutions. These solutions have only one
unstable perturbation mode, and so by fine-tuning $r_0$ to one of its
critical values, one can make these solitons survive ever longer, with
the lifetime scaling as (\ref{typeIscaling}). 


\subsection{CSS or DSS?}
\label{section:cssdss}


The critical solution in type II critical phenomena is either CSS or
DSS, depending on the matter model. No general criterion for why it is
one rather than the other is known. Therefore it is particularly
interesting to investigate continuous parameterized families of matter
models in which both types of symmetry occur, depending on the value
of the parameter. 


\subsubsection{Scalar field and stiff fluid}


One example of a 1-parameter family of matter models is the spherical
perfect fluid with equation of state $p=k\rho$ for arbitrary
$k$. Maison \cite{Maison} constructed the regular CSS solutions and
its linear perturbations for a large number of values of $k$. In each
case, he found exactly one growing mode, and was therefore able to
predict the critical exponent. As Ori and Piran before
\cite{OriPiran,OriPiran2}, he claimed that there are no regular CSS
solutions for $k> 0.89$. Recently, Neilsen and Choptuik
\cite{NeilsenChoptuik,NeilsenChoptuik2} have found CSS critical
solutions for all values of $k$ right up to $1$, both in collapse
simulations and by making a CSS ansatz. 

This raises an interesting question. Any solution with stiff
($p=\rho$) perfect fluid matter, where the velocity field is
irrotational, gives rise to an equivalent massless scalar field
solution. The fluid 4-velocity is parallel to the scalar field
gradient, (and therefore the equivalent scalar field solution has
timelike gradient everywhere), while the fluid density is related to
the modulus of the scalar field gradient. Furthermore, a spherically
symmetric velocity field is automatically irrotational, and so the
equivalence should apply. But the critical solution observed by
Neilsen and Choptuik in the stiff fluid is CSS, while the massless
scalar field critical solution is DSS; the two critical solutions are
different. How is this possible if the critical solution for each
system is supposed to be unique? 

This question has been answered in \cite{BradyChoptuikGundlachNeilsen}. One
needs to distinguish three different solutions (1) a CSS scalar field
solution that has exactly growing mode, and is therefore a candidate
critical solution, (2) the (CSS) critical solution observed in stiff
fluid collapse simulations, and (3) the (DSS) critical solution
observed in scalar field collapse simulations. The scalar field
gradient in solution (3) is timelike in some regions and spacelike in
others. In the regions with spacelike gradient it cannot be
interpreted as a stiff fluid solution. In particular, it is spacelike
in parts of the solution that are actually seen in near-critical
collapse, and so it can be a critical solution for scalar field
collapse, but not for perfect fluid collapse.  On the other hand,
solution (1) has an apparent horizon, which is a spacelike 3-surface
including the CSS singularity. All centered 2-spheres to the future of
that surface are closed trapped surfaces. Therefore solution (1), when
matched to an asymptotically flat exterior solution, contains a black
hole, and so it is not on the threshold of black hole formation. It
cannot be the fluid critical solution either.

However, the numerically observed fluid critical solution, solution
(2), agrees with the scalar field CSS solution, solution (1), in the
region where the scalar field gradient is timelike. In the region
where the scalar field gradient in solution (1) is spacelike, the
solutions differ completely. However, as the scalar field gradient
becomes null, the equivalent fluid density goes to zero, and
physically this means that the continuous fluid approximation breaks
down. To continue the spacetime, one has to introduce additional
physics ideas. Neilsen and Choptuik in fact modified their numerical
equations at low fluid densities in order to force the fluid density
to remain positive even for the stiff equation of state. The physical
effect is that the matter is blasted out at almost the speed of light,
leaving a very low density region behind. 


\subsubsection{The 2-dimensional sigma model}


Hirschmann and Eardley (from now on HE) \cite{HE3} looked for a
natural way of adding a non-linear self-interaction to the complex
scalar field without introducing a scale. (Dimensionful coupling
constants are discussed in Section \ref{section:universalityclasses}.)
An $n$-dimensional sigma model is a field theory whose fields $X^A$
are coordinates on an $n$-dimensional target manifold with metric
$G_{AB}(X)$. The action is the generalized kinetic energy
\begin{equation}
\int d^4x \sqrt{g} \ g^{\mu\nu}X^A_{,\mu}X^B_{,\nu}G_{AB}(X).
\end{equation}
The model discussed by HE has a 2-dimensional
target manifold with constant curvature. Using a single complex
coordinate $\phi$, its action can be written as 
\begin{equation}
\int d^4x \sqrt{g} {|\nabla\phi|^2\over
(1-\kappa|\phi|^2)^2}.
\end{equation}
This is then minimally coupled to gravity. (The constant target space
curvature is set by the real parameter $\kappa$.  Note that for
$\kappa=0$ the target space is flat, and $\phi$ is just a complex
scalar field. Moreover, for $\kappa > 0$ there are some rather unobvious
field redefinitions which make this model equivalent to a real
massless scalar field minimally coupled to Brans-Dicke gravity, with
the Brans-Dicke coupling given by
\begin{equation}
\omega_{\rm BD}=-{3\over2}+{1\over 8\kappa}.
\end{equation}
The value $\kappa=1$ ($\omega_{\rm BD}=-11/8$) also corresponds to an
axion-dilaton system arising in string theory
\cite{EardleyHirschmannHorne}.)

For each $\kappa$, HE constructed a CSS solution and its perturbations
and concluded that it is the critical solution for $\kappa>0.0754$,
but has three unstable modes for $\kappa<0.0754$. For $\kappa<-0.28$,
it acquires even more unstable modes. The positions of the mode
frequencies $\lambda$ in the complex plane vary continuously with
$\kappa$, and these are just values of $\kappa$ where a complex
conjugate pair of frequencies crosses the real axis. The results of HE
confirm and subsume collapse simulation results by Liebling and
Choptuik \cite{LieblingChoptuik} for the scalar-Brans-Dicke system,
and collapse and perturbative results on the axion-dilaton system by
Hamad\'e, Horne and Stewart \cite{HamadeHorneStewart}. Where the CSS
solution fails to be the critical solution, a DSS solution takes
over. In particular, for $\kappa=0$, the free complex scalar field,
the critical solution is just the real scalar field DSS solution of
Choptuik.

Liebling \cite{Liebling} has found initial data sets that find the CSS
solution for values of $\kappa$ (for example $\kappa=0$) where the
true critical solution is DSS. The complex scalar field in these data
sets is of the form $\phi(r)=\exp i\omega r$ times a slowly varying
function of $r$, for arbitrary $r$, while its momentum $\Pi(r)$ is
either zero or $d\phi/dr$. Conversely, data sets that are purely real
find the DSS solution even for values of $\kappa$ where the true
critical solution is the CSS solution, for example for
$\kappa=1$. These two special families of initial data maximize and
minimize the $U(1)$ charge. Small deviations from these data find the
sub-dominant ``critical'' solution for some time, then veer off and
find the true critical solution. 


\subsubsection{The 3-dimensional sigma model}
\label{section:3dsigma}


The transition from a DSS to a CSS critical solution can also occur
such that the period of the DSS solution diverges at the transition.
Aichelburg and collaborators \cite{Aichelburg1} have investigated the
3-dimensional sigma model whose target manifold is 3-dimensional with
unit positive curvature, that is, $S^3$ with a round metric on
it. Writing this metric as $d\phi^2+\sin^2\phi\
(d\theta^2+\sin^2\theta\,d\varphi^2)$, spherical symmetry can be
imposed by identifying $\theta$ and $\varphi$ with the angles of the
same name in spacetime. This is only the first of a family of
non-trivial identifications of the angles in the physical and the
target spaces. These identifications are labelled by a degree $m$. The
effective action in spherical symmetry is then
\begin{equation}
\label{S3sigma}
\eta \int d^4x \sqrt{g} \left(|\nabla\phi|^2+{m(m+1)\sin^2\phi\over r^2}\right)
\end{equation}
where $\phi$ depends only on $r$ and $t$. In the following we consider
only the trivial map with degree $m=1$. The parameter $\eta$ is
dimensionless in units $c=G=1$, and determines the strength of
coupling to gravity. 

For $0\le \eta<0.1$, a family of regular CSS solutions $\phi_n$
exists. In particular, for $\eta=0$ we have the 3-dimensional sigma
model on flat spacetime discussed in Section
\ref{section:withoutgravity}. $\phi_n$ has $n$ growing modes. $\phi_0$
is therefore an attractor, so that generic solutions form a naked
singularity. $\phi_1$ functions as a critical solution between naked
singularity formation (via $\phi_0$ and dispersion).  For all values
of $0.1<\eta<\infty$, type II critical phenomena are found at the
black hole threshold. The critical solution is DSS for
$0.2<\eta<\infty$. The scaling of the black hole mass shows the
expected periodic modulation on top of the power law. The scale period
$\Delta$ depends on $\eta$. It approaches a limiting value as
$\eta\to\infty$, and rises as $\eta$ decreases.  For $0.1<\eta<0.14$,
the critical solution is CSS.

Interesting new behavior occurs in the intermediate range
$0.14<\eta<0.2$ that lies between clear CSS and clear DSS. With
decreasing $\eta$, the overall DSS includes episodes of approximate
CSS \cite{Aichelburg2}. The length of these episodes (measured in the
log-scale time $\tau$) increases with decreasing $\eta$, while the
length of the non-CSS epoch remains approximately constant. As
$\eta\to 0.17$ from above, the duration of the CSS epochs, and hence
the overall DSS period $\Delta$, diverges. For $0.14<\eta<0.17$, time
evolutions of initial data near the black hole threshold no longer
show overall DSS, but they still show CSS episodes. Black hole mass
scaling is unclear in this regime.

The episodic CSS solution can be understood in terms of the following
dynamical systems picture \cite{Aichelburg3,Aichelburg4}. A regular
CSS solution exists for all $\eta<0.5$, and for $\eta>0.1$ it has
exactly one growing mode \cite{BizonWassermann}. We can picture this
in a three-dimensional toy-model. Let $x=y=z=0$ be the CSS fixed
point. Let the $x$-axis be its one growing mode, and let the $y$ and
$z$ axes be two of its decaying modes. Beginning near but not at the
CSS point, the solution moves away from it approximately along the $x$
axis, describes a loop in the $xy$ plane and returns approximately
along the $y$ axis, thus closing the loop. We then have a DSS
solution, with the CSS epoch corresponding to the DSS solution point
moving very slowly in the vicinity of the exact CSS fixed point. 

Let $(x_0,y_0)$ be the point of closest approach of the DSS loop to
the CSS point. As the solution point moves away from the CSS point,
$x\simeq x_0\exp\lambda\tau$, where $\lambda$ is the Lyapunov exponent
of the CSS solution. At $\tau\simeq \ln(x_1/x_0)/\lambda$, the
solution has reached some value $x_1$, after which linear perturbation
theory breaks down. A fixed $\tau$-interval $b$ later the solution
curve returns to the linear regime near the $y$ axis, and returns to
its starting point with exponentially decreasing $y$. In the limit
$\eta\to\eta_c$ the DSS solution touches the CSS solution(s). To
leading order $x_0$ is therefore proportional to
$\eta-\eta_c$. Therefore, $\Delta$ is expected to diverge as
\begin{equation}
\Delta\simeq -a\ln(\eta-\eta_c)+b
\end{equation}
for some constants $a$ and $b$. There are in fact not one but two CSS
fixed points, which differ only by the overall sign of the field
$\phi$, and the DSS solution oscillates between them. Therefore
$a=2/\lambda$. A good fit with the numerical simulations was achieved
with $\eta_c=0.17$. This is consistent with the fact that an exact DSS
solution could be found by making a DSS ansatz only for
$\eta>0.1726$. In terms of the phase space picture, the DSS loop is
broken for $\eta<0.17$. This is called a ``heteroclitic loop
bifurcation''. The situation for $0.14<\eta<0.17$ is unclear. It is
possible that the only genuine attractive fixed point in the black
hole threshold is the CSS solution, but that the DSS solution is
replaced by what could be called a ``strange attractor'' made up of
solutions that are almost but not quite DSS.

It should be stressed that the critical exponent of the DSS solution
is not related to the Lyapunov exponent $\lambda$ of the CSS
solution. In the toy model, assume that the phase point is not exactly
in the $xy$ plane, but slightly above it. When $x$ and $y$ return to
their closest approach to the CSS solution, $z$ has increased by a
factor $\exp\lambda_{\rm DSS}\Delta$, which defines the Lyapunov
exponent of the overall DSS solution. This factor cannot be derived
from perturbation theory around the CSS solution, but $\Delta$ and
$\lambda_{\rm DSS}$ can be determined in the standard manner of making
a DSS ansatz and studying its linear perturbations.


\subsection{Phase diagrams}
\label{section:phasediagrams}



\subsubsection{Type I and type II}
\label{section:typeIandII}


The same system can show type I critical behavior, where black hole
formation turns on at a universal minimum mass, and type II critical
behavior, where it turns on at zero mass, and the black hole mass
shows a universal power law. 

One system where this happens is the spherical $SU(2)$
Einstein-Yang-Mills system
\cite{ChoptuikChmajBizon,BizonChmaj,BizonChmaj2,BizonChmaj3}.  Which
kind of behavior arises appears to depend on the qualitative shape of
the initial data. In type II behavior, the critical solution is DSS
\cite{Gundlach_EYM}. In type I, the critical solution is a static,
asymptotically flat solution which had been found before by Bartnik
and McKinnon \cite{BartnikMcKinnon}.

The type I critical solution can also have a discrete symmetry, that
is, they can be periodic in time instead of being static. This
behavior was found in collapse situations of the massive scalar field
by Brady, Chambers and Gon\c calves (from now on BCG)
\cite{BradyChambersGoncalves}. Previously, Seidel and Suen
\cite{SeidelSuen} had constructed periodic, asymptotically flat,
spherically symmetric self-gravitating massive scalar field solutions
they called oscillating soliton stars. By dimensional analysis, the
scalar field mass $m$ sets an overall scale of $1/m$ (in units
$G=c=1$). For given $m$, Seidel and Suen found a 1-parameter family of
such solutions with two branches. The more compact solution for a
given ADM mass is unstable, while the more extended one is stable to
spherical perturbations. BCG report that the type I critical solutions
they find are from the unstable branch of the Seidel and Suen
solutions.  They see a 1-parameter family of (type I) critical
solutions, rather than an isolated critical solution. BCG in fact
report that the black hole mass gap does depend on the initial data.
As expected from the discrete symmetry, they find a small wiggle in
the mass of the critical solution which is periodic in
$\ln(p-p_*)$. If type I or type II behavior is seen appears to depend
mainly on the ratio of the length scale of the initial data to the
length scale $1/m$.

In the critical phenomena that we have discussed so far, with an
isolated critical solution, only one number's worth of information,
namely the separation $p-p_*$ of the initial data from the black hole
threshold, survives to the late stages of the time evolution. Recall
that our definition of a critical solution is one that has exactly one
unstable perturbation mode, with a black hole formed for one sign of
the unstable mode, but not for the other. This definition does not
exclude an $n$-dimensional family of critical solutions. Each solution
in the family then has $n$ marginal modes leading to neighboring
critical solutions, as well as the one unstable mode. $n+1$ numbers'
worth of information survive from the initial data, and the mass gap
in type I, or the critical exponent for the black hole mass in type
II, for example, depend on the initial data through $n$ parameters. In
other words, universality exists in diminished form. The results of
BCG are an example of a 1-parameter family of type I critical
solutions. Recently, Brodbeck et al. \cite{Brodbecketal} have shown,
under the assumption of linearization stability, that there is a
1-parameter family of stationary, rotating solutions beginning at the
(spherically symmetric) Bartnik-McKinnon solution. This could turn out
to be a second 1-parameter family of type I critical solutions,
provided that the Bartnik-McKinnon solution does not have any unstable
modes outside spherical symmetry \cite{Rendallpc}. Stability has now been
confirmed for nonspherical perturbations in
\cite{Brodbeckodd,WinstanleySarbach}.

Bizo\'n and Chmaj have studied type I critical collapse of an $SU(2)$
Skyrme model coupled to gravity, which in spherical symmetry with a
hedgehog ansatz is characterized by one field $F(r,t)$ and one
dimensionless coupling constant $\alpha$. Initial data $F(r)\sim
\tanh(r/p)$, $\dot F(r)=0$ surprisingly form black holes for both
large and small values of the parameter $p$, while for an intermediate
range of $p$ the endpoint is a stable static solution called a
skyrmion. (If $F$ was a scalar field, one would expect only one
critical point on this family.) The ultimate reason for this behavior
is the presence of a conserved integer ``baryon number'' in the matter
model. Both phase transitions along this 1-parameter family are
dominated by a type I critical solution, that is a different skyrmion
which has one unstable mode. In particular, an intermediate time
regime of critical collapse evolutions agrees well with an ansatz of
the form (\ref{typeIintermediate}), where $Z_*$, $Z_0$ and $\lambda$
were obtained independently. It is interesting to note that the type I
critical solution is singular in the limit $\alpha \to 0$, which is
equivalent to $G \to 0$, because the known type II critical solutions
for any matter model also do not have a weak gravity limit.

Apparently, type I critical phenomena can arise even without the
presence of a scale in the field equations. A family of exact
spherically symmetric, static, asymptotically flat solutions of vacuum
Brans-Dicke gravity given by van Putten was found by Choptuik,
Hirschmann and Liebling \cite{ChopHirschLieb} to sit at the black
hole-threshold and to have exactly one growing mode. This family has
two parameters, one of which is an arbitrary overall scale.


\subsubsection{Triple points}
\label{section:triplepoints}


In analogy with critical phenomena in statistical mechanics, let us
call a graph of the black hole threshold in the phase space of some
self-gravitating system a phase diagram. The full phase space is
infinite-dimensional, but one can plot a two-dimensional
submanifold. In such a plot the black hole threshold is generically a
line, analogous to the fluid/gas dividing line in the
pressure/temperature plane.

Interesting phenomena can be expected in systems that admit more
complicated phase diagrams. The massive complex scalar field, for
example, admits stable stars as well as black holes and flat space as
possible end states. There are three phase boundaries, and these
should intersect somewhere. A generic two-parameter family of initial
data is expected to intersect each boundary in a line, and the three
lines should meet at a triple point.

Similarly, in a system where the black hole/dispersion phase boundary
is type I in one part of the phase space and type II in another, one
might expect these two lines to intersect in a suitable two-parameter
family of data. Is the black hole mass at the intersection finite or
zero? Is there a third line that begins where the type I and type II
lines meet?

Choptuik, Hirschmann and Marsa \cite{ChoptuikHirschmannMarsa} have
investigated this for a specific two-parameter family of initial data
for the spherically symmetric $SU(2)$ Yang-Mills field, using a
numerical evolution code that can follow the time evolutions for long
after a black hole has formed. There is a third type of phase
transition along a third line which meets the intersection of the type
I and type II lines. On both sides of this
``type III'' phase transition the final state is a Schwarzschild black
hole with zero Yang-Mills field strength, but the final state is
distinguished by the value of the Yang-Mills gauge potential at
infinity, which can take two values, corresponding to two distinct
vacuum states.  The critical solution is an unstable black hole with
Yang-Mills hair, which collapses to a hairless Schwarzschild black
hole with either vacuum state of the Yang-Mills field, depending on
the sign of its one growing perturbation mode. The critical solution
is not unique, but is a member of a 1-parameter family of hairy black
holes parameterized by their mass. As the ``triple point'' is
approached the mass of this black hole becomes arbitrarily small (and
what happens exactly at the triple point needs to be investigated
separately.)


\subsection{Astrophysical scenarios}



\subsubsection{Primordial black holes}


Any application of critical phenomena to astrophysics or cosmology
would require that critical phenomena are not an artifact of the
simple matter models that have been studied so far, and that they are
not an artifact of spherical symmetry. At present both these
assumptions seem reasonable.

Critical collapse also requires a fine-tuning of initial data to the
black hole threshold. Niemeyer and Jedamzik \cite{NiemeyerJedamzik}
have suggested a cosmological scenario that gives rise to such
fine-tuning. In the early universe, quantum fluctuations of the metric
and matter can be important, for example providing the seeds of galaxy
formation. If they are large enough, these fluctuations may even
collapse long before stars or galaxies form, giving rise to what is
called primordial black holes. Large quantum fluctuations are
exponentially more unlikely than small ones, $P(\delta)\sim
e^{-\delta^2}$, where $\delta$ is the density contrast of the
fluctuation. One would therefore expect the spectrum of primordial
black holes to be sharply peaked at the minimal $\delta$ that leads to
black hole formation. That is the required fine-tuning. In the
presence of fine-tuning, the black hole mass is much smaller than the
initial mass of the collapsing object, here the density
fluctuation. In consequence, the peak of the primordial black hole
spectrum might be expected to be at exponentially smaller values of
the black hole mass than expected naively. See also
\cite{NiemeyerJedamzik2,Yokoyama,GreenLiddle}.


\subsubsection{Realistic equations of state}


Critical collapse is not likely to be relevant in the universe in the
present epoch as there is no known mechanism for fine-tuning of
initial data. Furthermore, if one could fine-tune the gravitational
collapse of stars made of realistic matter (i.e. not scalar fields) it
seems more likely that type I critical phenomena would be observed,
i.e. there would be a universal mass gap. However, Novak \cite{Novak}
has evolved initial data obtained by using the density profile of a
static spherical neutron star with a realistic equation of state, but
giving it a non-zero velocity profile with ingoing velocity. At
sufficiently high speeds the star collapses to a black hole. Contrary
to expectation for astrophysical collapse situations formulated above,
at the threshold there is no mass gap, and instead a mass scaling with
a critical exponent is observed. This may be because the inward
velocity imparted to the star in the initial data is quite large -- if
the velocities become relativistic, type II phenomena might be
expected again because the scale set by the rest mass density becomes
irrelevant. 

Static neutron stars for a given equation of state form a family
parameterized by the central density. The graph of mass versus central
density has a maximum, and stars on the high density branch are
unstable. Novak uses density profiles from the stable branch. Below a
certain central density, that is far enough from the maximum of the
curve, the star cannot be made to collapse for any velocity. Above the
threshold, there is power-law scaling at the black hole threshold. The
same critical exponent was found for two different velocity profiles
and the same equation of state, and a different exponent for one
profile and a different equation of state, where the fine-tuned
parameter was the overall amplitude of the velocity profile.

In the fluid collapse simulations of Evans and Coleman
\cite{EvansColeman} and Neilsen and Choptuik \cite{NeilsenChoptuik}
the critical solution is CSS and smooth. By contrast, the mechanism of
dispersing almost all the mass and making only a small black hole in
Novak's simulations is a core bounce and the formation of a shock. If
a universal critical solution exists it cannot be smooth. Furthermore,
the equation of state is not barotropic. The initial data are at zero
temperature and therefore effectively barotropic, but heating occurs
in the shock. The zero temperature equation of state is not compatible
with self-similarity on the density scales used by Novak in his
initial data, although this may change during collapse. There is not
enough numerical detail in Novak's paper to settle this.


\subsection{Critical collapse in semiclassical gravity}


Type II critical phenomena provide a relatively natural way of
producing arbitrarily high curvatures, where quantum gravity effects
should become important, from generic initial data. Therefore, various
authors have investigated the relationship of Choptuik's critical
phenomena to quantum black holes. It is widely believed that black
holes should emit thermal quantum radiation, from considerations of
quantum field theory on a fixed Schwarzschild background on the one
hand, and from the purely classical three laws of black hole mechanics
on the other (see \cite{Wald_BH} for a review). But there is no
complete model of the back-reaction of the radiation on the black
hole, which should be shrinking. In particular, it is unknown what
happens at the endpoint of evaporation, when full quantum gravity
should become important. It is debated in particular if the
information that has fallen into the black hole is eventually
recovered in the evaporation process or lost.

To study these issues, various 2-dimensional toy models of gravity
coupled to scalar field matter have been suggested which are more or
less directly linked to a spherically symmetric 4-dimensional
situation (see \cite{Giddings_BH} for a review). In two space-time
dimensions, the quantum expectation value of the matter stress tensor
can be determined from the trace anomaly alone, together with the
reasonable requirement that the quantum stress tensor is
conserved. Furthermore, quantizing the matter scalar field(s) $f$ but
the metric as a classical field can be formally justified in the limit
in which the number $N$ of identical matter fields goes to
$\infty$. The two-dimensional gravity used is not the two-dimensional
version of Einstein gravity, which is trivial, but a scalar-tensor
theory of gravity. $e^\phi$, where $\phi$ is called the dilaton, in
the 2-dimensional toy model plays essentially the role of $r$ in 4
spacetime dimensions. There seems to be no preferred 2-dimensional toy
model, with arbitrariness both in the quantum stress tensor and in the
choice of the classical part of the model. In order to obtain a
resemblance of spherical symmetry, a reflecting boundary condition is
imposed at a timelike curve in the 2-dimensional spacetime. This plays
the role of the curve $r=0$ in a 2-dimensional reduction of the
spherically symmetric 4-dimensional theory.

How does one expect a model of semiclassical gravity to behave when
the initial data are fine-tuned to the black hole threshold?  First of
all, until the fine-tuning is taken so far that curvatures on the
Planck scale are reached during the time evolution, universality and
scaling should persist, simply because the theory must approximate
classical GR. Approaching the Planck scale from above,
one would expect to be able to write down a critical solution that is
the classical critical solution asymptotically at large scales, as an
expansion in inverse powers of the Planck length.  This ansatz would
recursively solve a semiclassical field equation, where powers of
$e^{\tau}$ (in coordinates $x$ and $\tau$) signal the appearances of
quantum terms.  Note that this is exactly the ansatz
(\ref{asymptotic_CSS}), but with the opposite sign in the exponent, so
that the higher order terms now become negligible as $\tau\to-\infty$,
that is away from the singularity on large scales. On the Planck scale
itself, this ansatz would not converge, and self-similarity would
break down.

Addressing the question from the side of classical GR, Chiba and Siino
\cite{ChibaSiino} write down ad-hoc semiclassical Einstein-scalar
field equations in spherical symmetry in 4 spacetime dimensions that
are inspired by a 2-dimensional toy model. They note that their
quantum stress tensor diverges at $r=0$. Ayal and Piran
\cite{AyalPiran} make an ad-hoc modification to these semiclassical
equations. They modify the quantum stress tensor by a function which
interpolates between 1 at large $r$, and $r^2/L_p^2$ at small
$r$. They justify this modification by pointing out that the resulting
violation of energy conservation takes place only at the Planck
scale. It takes place, however, not only where the solution varies
dynamically on the Planck scale, but at all times in a Planck-sized
world tube around the center $r=0$, even before the solution itself
reaches the Planck scale dynamically. With this modification, Ayal and
Piran obtain results in agreement with our expectations set out
above. For far supercritical initial data, black formation and
subsequent evaporation are observed. With fine-tuning, as long as the
solution stays away from the Planck scale, critical solution phenomena
including the Choptuik universal solution and critical exponent are
observed. (The exponent is measured as $0.409$, but should be
Choptuik's one of $0.374$ in this regime.) In an intermediate regime,
the quantum effects increase the critical value of the parameters
$p$. This is interpreted as the initial data partly evaporating while
they are trying to form a black hole.

Researchers coming from the quantum field theory side seem to favor a
model (the RST model) in which ad hoc ``counter terms'' have been
added to make it integrable. The matter is a conformally rather than
minimally coupled scalar field. The field equations are trivial up to
an ODE for a timelike curve on which reflecting boundary conditions
are imposed. The world line of this ``moving mirror'' is not clearly
related to $r$ in a 4-dimensional spherically symmetric model, but
seems to correspond to a finite $r$ rather than $r=0$. This may
explain why the problem of a diverging quantum stress tensor is not
encountered. Strominger and Thorlacius \cite{StromingerThorlacius}
find a critical exponent of $1/2$, but their 2-dimensional situation
differs from the 4-dimensional one in many aspects. Classically
(without quantum terms) any ingoing matter pulse, however weak, forms
a black hole. With the quantum terms, matter must be thrown in
sufficiently rapidly to counteract evaporation in order to form a
black hole. The initial data to be fine-tuned are replaced by the
infalling energy flux. There is a threshold value of the energy flux
for black hole formation, which is known in closed form. (Recall this
is a integrable system.) The mass of the black hole is defined as the
total energy it absorbs during its lifetime.  This black hole mass is
given by
\begin{equation}
M\simeq \left({\delta\over\alpha}\right)^{1\over2}
\end{equation}
where $\delta$ is the difference between the peak value of the flux
and the threshold value, and $\alpha$ is the quadratic order
coefficient in a Taylor expansion in advanced time of the flux around
its peak. There is universality with respect to different shapes of
the infalling flux in the sense that only the zeroth and second order
Taylor coefficients matter. See also \cite{Kiem,ZhouKirstenYang}.

Peleg, Bose and Parker \cite{PelegBoseParker,BoseParkerPeleg} study
the so-called CGHS 2-dimensional model. This (non-integrable) model does
allow for a study of critical phenomena with quantum effects turned
off. Again, numerical work is limited to integrating an ODE for the
mirror world line. Numerically, the authors find black hole mass
scaling with a critical exponent of $\gamma\simeq 0.53$. They find the
critical solution and the critical solution to be universal with
respect to families of initial data. Turning on quantum effects, the
scaling persists to a point, but the curve of $\ln M$ versus
$\ln(p-p_*)$ then turns smoothly over to a horizontal
line. Surprisingly, the value of the mass gap is not universal but
depends on the family of initial data. While this is the most
``satisfactory'' result among those discussed here from the classical
point of view, one should keep in mind that all these results are
based on mere toy models of quantum gravity.

Rather than using a consistent model of semiclassical gravity, Brady
and Ottewill \cite{BradyOttewill} calculate the quantum stress-energy
tensor of a conformally coupled scalar field on the fixed background
of the perfect fluid CSS critical solution and treat it as an
additional perturbation, on top of the perturbations of the fluid-GR
system itself. In doing this, they neglect the indirect coupling
between fluid and quantum scalar perturbations through the metric
perturbations. From dimensional analysis, the quantum perturbation has
a Lyapunov exponent $\lambda=2$. If this is larger than the positive
Lyapunov exponent $\lambda_0$, it will become the dominant
perturbation for sufficiently good fine-tuning, and therefore
sufficiently good fine-tuning will reveal a mass gap. For a spherically symmetric
perfect fluid with equation of state $p=k\rho$, one finds that
$\lambda_0>2$ for $k>0.53$, and vice versa. If $\lambda_0>2$, the
semiclassical approximation breaks down for sufficiently good
fine-tuning, and this calculation remains inconclusive.


\section{Conclusions}
\label{section:conclusions}



\subsection{Summary}


When one fine-tunes a one-parameter family of initial data to get
close enough to the black hole threshold, the details of the initial
data are completely forgotten in a spacetime region, and all
near-critical time evolutions look the same there. The only
information remembered from the initial data is how close one is to
the threshold. Either there is a mass gap (type I behavior), or black
hole formation starts at infinitesimal mass (type II behavior). In
type I, the universal critical solution is time-independent, or
periodic in time, and the better the fine-tuning, the longer it
persists. In type II, the universal critical solution is
scale-invariant or scale-periodic, and the better the fine-tuning, the
smaller the black hole mass, according to the famous formula
Eq. (\ref{power_law}).

These phenomena are best understood in dynamical systems terms. The
dispersion and black hole end states are then considered as competing
attractors. The black hole threshold is the boundary between their
basins of attraction. It is a hypersurface of codimension one in the
(infinite-dimensional) phase space. By definition initial data on the
black hole threshold evolve neither to a black hole nor to dispersion,
and so remain in the black hole threshold. The black hole-threshold is
therefore a dynamical system of its own, of one dimension fewer -- a
critical surface in dynamical systems terms. The critical solution is
simply an attractor in the critical surface.

In terms of the full dynamical system, the critical solution is an
attractor of codimension one. It has a single unstable linear
perturbation mode which drives it out of the critical
surface. Depending on the sign of this perturbation, the perturbed
critical solution tips over towards forming a black hole or towards
dispersion. In the words of Eardley, all one-parameter families of
data trying to cross the black hole threshold are funneled through a
single time evolution.  If the critical solution is time-independent,
its linear perturbations grow or decrease exponentially in time. If it
is scale-invariant, they grow or decrease exponentially with the
logarithm of scale.  The power-law scaling of the black hole mass
follows by dimensional analysis. Because we are really discussing a
field theory, many aspects of the dynamical systems picture remain
qualitative, but its basic correctness is demonstrated by 
quantitative calculations of critical solutions and critical
exponents.

The importance of type II behavior lies in providing a natural route
from large to very small scales, with possible applications to
astrophysics and quantum gravity. Natural here means that the critical
phenomena occur in many simple matter models and are apparently not
limited to spherical symmetry either. As far as any generic parameter
in the initial data provides some handle on the amplitude of the one
unstable mode, fine-tuning any one generic parameter creates critical
phenomena, and can in principle create arbitrarily large curvatures
visible from infinity from asymptotically flat regular initial data.

This property of type II critical behavior restricts what version
of cosmic censorship one can hope to prove. At least in some matter
models (scalar field, perfect fluid), fine-tuning any smooth
one-parameter family of smooth, asymptotically flat initial data,
without any symmetries, gives rise to a naked singularities. In this
sense the set of initial data that form a naked singularity is
codimension one in the full phase space of smooth asymptotically flat
initial data for well-behaved matter.

Finally, critical phenomena are the outstanding contribution of
numerical relativity to knowledge in GR to date, and they continue to
act as a motivation and a source of testbeds for numerical
relativity.


\subsection{Outlook}


Numerical work continues to establish the generality of critical
phenomena in gravitational collapse, or to find a counter-example
instead. In particular, future research will investigate highly
non-spherical situations. Given the twin facts that black holes can
have angular momentum and electric charge as well as mass, and that
angular momentum and electric repulsion are expected to oppose
gravitational attraction, it is particularly interesting to
investigate collapse with large angular momentum and/or electric
charge.

Going beyond spherical symmetry poses a formidable numerical
challenge. In fully 3D simulations it is difficult to obtain adequate
numerical resolution for any purpose, let along critical collapse with
its wide range of length scales. Axisymmetric simulations have been
almost abandoned by the numerical relativity community in favor of
fully 3D codes. The important results of Abrahams and Evans on
critical collapse in axisymmetric vacuum gravity have not yet been
repeated either with an axisymmetric or with a fully 3D
code. 

Nevertheless there is encouraging progress. Several groups are
independently developing codes that are accurate enough to resolve
critical phenomena in axisymmetric fluid collapse, and working on
adaptive mesh refinement for axisymmetric and 3D codes. 3D adaptive
mesh refinement has recently been demonstrated for critical phenomena
in flat spacetime nonlinear wave equation \cite{Liebling3D}, and
axisymmetric adaptive mesh refinement has been demonstrated for
axisymmetric scalar field collapse \cite{Pretorius_thesis}.

The fundamental mathematical question in the field is why all simple
matter models investigated so far show critical phenomena. Put
differently, the question is why they admit a critical solution: an
attractor of codimension one at the black hole threshold. If the
existence of a critical solution is really a generic feature, then
there should be at least an intuitive argument, and perhaps a
mathematical proof, for this important fact.  

Collisionless matter suggests a possible restriction. It has been
shown analytically \cite{vlasov1}, that there are no type I or type II
critical phenomena in the spherical Einstein-Vlasov system with
massless particles. On an intuitive level, the explanation seems to be
that collisionless matter, which is not a field theory, has infinitely
many more degrees of freedom than either gravity, or a field theory or
a perfect fluid describing matter. There is numerical evidence for
type I critical phenomena with massive particles
\cite{OlabarrietaChoptuik}, but it is not completely conclusive. The
intuitive function-counting argument that rules out critical phenomena
with massless particles appears to apply to massive particles as well,
but has not been made rigorous.

As critical solutions are interesting in part because they generate a
naked singularity from regular (even analytic) initial data in
reasonable matter models (or even in vacuum gravity), the nature of
this singularity is of particular interest. There is ongoing research
in this area, both numerical and analytic, including the stability of
the Cauchy horizon and the possible singularity structures beyond it.

Also on the mathematical side, the technical challenge remains to make
the intuitive dynamical systems picture of critical collapse more
rigorous, by providing a distance measure on the phase space, and a
prescription for a flow on the phase space (equivalent to a
prescription for the lapse and shift). The latter problem is
intimately related to the problem of finding good coordinate systems
for the binary black hole problem.

On the phenomenological side, it is conceivable that the scope of
critical collapse will be expanded to take into account new phenomena,
such as multicritical solutions (with several growing perturbation
modes), or critical solutions that are neither static, periodic, CSS
or DSS. More complicated phase diagrams than the simple black
hole-dispersion transition are already being examined, and the
intersections of phase boundaries are of particular interest.

A large number of people have contributed indirectly to this paper,
but I would particularly like to thank Pat Brady, Matt Choptuik, David
Garfinkle, Jos\'e M. Mart\'\i n-Garc\'\i a, Alan Rendall and Bob Wald
for helpful conversations, and an anonymous referee for a careful
reading.




\end{document}